\documentclass[fleqn,usenatbib,useAMS]{mnras}

\usepackage{graphicx}	
\usepackage{amsmath}	
\usepackage{amssymb}	
\usepackage[varg]{txfonts} 

\usepackage[T1]{fontenc}
\usepackage{ae,aecompl}

\let\boldgrk=\gkvecten
\let\boldgrksc=\gkvecseven

\def\gkthing#1{{\mathchoice%
	{\hbox{{\boldgrk\char#1}}}
	{\hbox{{\boldgrk\char#1}}}
	{\hbox{{\boldgrksc\char#1}}}
	{\hbox{{\boldgrksc\char#1}}}}}

\def\vtheta{\gkthing{18}}

{\newif\ifnotend
\notendtrue
\def\veclist{ABCDEFGHIJKLMNOPQRSTUVWXYZabcdefghijklmnopqrstuvwxyz.}
\def\top#1#2.{#1}
\def\tail#1#2.{#2.}
\loop\expandafter\xdef\csname v\expandafter\top\veclist\endcsname%
{{\noexpand\bf\expandafter\top\veclist}}
\edef\veclist{\expandafter\tail\veclist}
\if\veclist.\notendfalse\fi\ifnotend\repeat}
\def\Gyr{\,\mathrm{Gyr}}
\def\Myr{\,\mathrm{Myr}}
\def\kpc{\,\mathrm{kpc}}
\def\Kms{\,\mathrm{km\,s}^{-1}}
\def\FeH{\hbox{[Fe/H]}}
\def\FeA{\hbox{[$\alpha$/Fe]}}
\def\ex#1{\langle#1\rangle}
\def\msun{{\rm M}_\odot}
\def\d{{\rm d}}
\renewcommand{\[}{\begin{equation}}
\renewcommand{\]}{\end{equation}}

\newcommand{\kms}{\ km s$^{-1}$}

\newcommand{\Gaia}{\textit{Gaia}}

\newcommand{\rps}{r_{\rm ps}}
\newcommand{\psips}{\psi_{\rm ps}}

\def\HI{H${\scriptstyle\rm I}$}
\def\Msun{\ifmmode{\>\!{\rm M}_{\odot}}\else{M$_{\odot}$}\fi}
\def\Zsun{\ifmmode{\>\!{\rm Z}_{\odot}}\else{Z$_{\odot}$}\fi}
\def\XH{\ifmmode{\>X_{\textnormal{\sc h}}}\else{$X_{\textnormal{\sc h}}$}\fi}

\def\Usol {\ifmmode{U_\odot} \else {$U_\odot$}\fi} 
\def\Vsol {\ifmmode{V_\odot} \else {$V_\odot$}\fi} 
\def\Wsol {\ifmmode{W_\odot} \else {$W_\odot$}\fi} 
\def\Vsolar {\ifmmode{{\bf v}_\odot} \else {${\bf v}_\odot$}\fi} 
\def\vsolar {\ifmmode{v_\odot} \else {$v_\odot$}\fi} 
\def\VLSR {\ifmmode{{\bf \Theta}_{\rm 0}} \else {${\bf \Theta_{\rm 0}}$}\fi} 
\def\Vlsr {\ifmmode{{\bf v_{\rm LSR}}} \else {${\bf v_{\rm LSR}}$}\fi} 
\def\Rsolar{\ifmmode{R_0}\else {$R_0$}\fi}
\def\zsolar{\ifmmode{z_0}\else {$z_0$}\fi}
\def\phisolar{\ifmmode{\phi_0}\else {$\phi_0$}\fi}
\def\Omegasolar{\ifmmode{{\bf \Omega}_\odot}\else {${\bf \Omega}_\odot$}\fi}

\def\Vzmax{\ifmmode{\vert V_{z_{\rm max}}\vert}\else {$\vert V_{z_{\rm max}}\vert$}\fi}
\def\zmax{\ifmmode{\vert z_{\rm max}\vert}\else {$\vert z_{\rm max}\vert$}\fi}
\def\Az{\ifmmode{{\cal A}_z}\else {${\cal A}_z$}\fi}

\def\GBP {\ifmmode{G_{\rm BP}} \else {$G_{\rm BP}$} \fi}
\def\GRP {\ifmmode{G_{\rm RP}} \else {$G_{\rm RP}$} \fi}

\begin{document}

\title[Dissecting the stellar disc's phase space by age, action, chemistry and location]{The GALAH survey and \Gaia\ DR2: dissecting the stellar disc's phase space by age, action, chemistry and location}

\author[Bland-Hawthorn et al.]{Joss Bland-Hawthorn$^{1,2,3}$\thanks{Contact e-mail: \href{mailto:jbh@physics.usyd.edu.au}{jbh@physics.usyd.edu.au}},
Sanjib~Sharma$^{1,2}$,
Thor~Tepper-Garcia$^{1,2}$,
James~Binney$^{4}$, \newauthor
Ken~C.~Freeman$^{5}$,
Michael~R.~Hayden$^{1,2}$,
Janez~Kos$^{1}$,
Gayandhi~M.~De~Silva$^{1,6}$, \newauthor
Simon Ellis$^6$,
Geraint~F.~Lewis$^{1}$,
Martin~Asplund$^{2,5}$,
Sven~Buder$^{7,8}$, 
Andrew~R.~Casey$^{9,10}$, \newauthor
Valentina~{D'Orazi}$^{11}$,
Ly~Duong$^{5}$,
Shourya~Khanna$^{1,2}$,
Jane~Lin$^{5}$, 
Karin~Lind$^{7,12}$, \newauthor 
Sarah~L.~Martell$^{2,13}$,
Melissa~K.~Ness$^{14,15}$, 
Jeffrey~D.~Simpson$^{13}$,
Daniel~B.~Zucker$^{6}$,  \newauthor
Toma\v{z}~Zwitter$^{16}$,
Prajwal~R.~Kafle$^{17}$,
Alice~C.~Quillen$^{18}$,
Yuan-Sen~Ting$^{19,20,21}$, \newauthor
Rosemary~F.~G.~Wyse$^{22}$ 
and~the~GALAH team
\\
(Affiliations listed after the references)}

\pubyear{2018}

\maketitle

\begin{abstract}
We use the second data releases of the ESA \Gaia\ astrometric survey and the high-resolution GALAH spectroscopic survey to analyse the structure of our Galaxy's disc components.  With GALAH, we separate the $\alpha$-rich and $\alpha$-poor discs (with respect to Fe), which are superposed in both position and velocity space, and examine their distributions in action space. We examine the distribution of stars in the $zV_z$ phase plane, for both $V_\phi$ and $V_R$, and recover the remarkable ``phase spiral'' discovered by \Gaia. We identify the anticipated quadrupole signature in $zV_z$ of a tilted velocity ellipsoid for stars above and below the Galactic plane. By connecting our work with earlier studies, we show that the phase spiral is likely to extend well beyond the narrow solar neighbourhood cylinder in which it was found. The phase spiral is a signature of corrugated waves that propagate through the disc, and the associated non-equilibrium phase mixing. The radially asymmetric distribution of stars involved in the phase spiral reveals that the corrugation, which is mostly confined to the $\alpha$-poor disc, grows in $z$-amplitude with increasing radius. We present new simulations of tidal disturbance of the Galactic disc by the Sagittarius (Sgr) dwarf.  The effect on the $zV_z$ phase plane lasts $\gtrsim 2\Gyr$ {\it but a subsequent disc crossing wipes out the coherent structure}. We find that the phase spiral was excited $\lesssim 0.5\Gyr$ ago by an object like Sgr with total mass $\sim3\times10^{10}\:\msun$ (stripped down from $\sim5\times10^{10}\:\msun$ when it first entered the halo) passing through the plane.

\end{abstract}


\section{Introduction} 
\label{s:intro}

The ESA \Gaia\ astrometric mission \citep{perryman2001,prusti2016} has been
eagerly anticipated for many years by the stellar and Galactic communities
and the early results have not disappointed 
\citep[DR2:][]{Brown2018}.  Wide-field stellar surveys across the Galaxy are
fundamental to astrophysics because there are important measurements that can
only be made in the near field.  The remarkable precision of measured stellar
parameters by \Gaia\ after only two years of observations has triggered a
flurry of new discoveries and new fields of study
\citep[e.g.][]{antoja2018,helmi2018,babusiaux2018,eyer2018}.
Complementary spectroscopic surveys -- 
RAVE \citep{Steinmetz2006}, APOGEE \citep{Majewski2017}, 
\Gaia-ESO \citep{Gilmore2012}, LAMOST \citep{Deng2012} -- 
and are now
able to exploit the overlap of targets with \Gaia\ DR2 -- 
this is the golden
age of Galactic archaeology \citep{freeman2002}.

Our focus here is on the synergy between \Gaia\ and the Galactic Archaeology
with HERMES (GALAH)\footnote{\url{https://galah-survey.org/}} survey based at
the Anglo-Australian Telescope (AAT) in Australia. This survey brings a
unique perspective to Galactic archaeology by measuring accurate radial
velocities and up to 30 elemental abundances
for about a million stars \citep{DeSilva2015,Martell2017}.  The HERMES
instrument was designed and optimised for the GALAH survey specifically for
the pursuit of Galactic archaeology \citep{freeman2008,barden2010}.  The
GALAH selection criteria were designed to be as simple as possible to avoid
problems that adversely affect earlier surveys \citep{Sharma2011}. The
primary selection is based on a magnitude range of $12 < V < 14$ and a cut in
Galactic latitude, $\vert b\vert > 10\,\mathrm{deg}$. Thus GALAH probes
mainly the thin and thick discs of the Galaxy. The impact of \Gaia\ was a key
consideration from the outset, particularly with regard to choosing a bright
limiting magnitude to ensure good distances for all stars.

The second GALAH data release (GALAH DR2) features stellar parameters, radial
velocities and up to 23 elemental abundances for $342\,682$ stars
\citep{Buder2018a}. All of these stars have complementary data from the
\Gaia\ DR2 data archive. GALAH's high-quality radial velocities, with mean
errors of $0.1-0.2$\kms\ \citep{zwitter2018}, are more accurate than the \Gaia\
radial velocities, but comparable to or better than typical transverse
velocities derived from the proper motions \citep{Brown2018}. The
GALAH-\Gaia\ synergy\footnote{The power of this synergy is demonstrated by
\citet{Kos2018b} who reveal that four well known NGC ``open clusters'' first
identified in 1888 are projections and not physical systems. This requires the modulus of the velocity {\it vector} to be measured to better than 1\kms.} is
particularly strong for local dwarfs that dominate the survey within about 1 kpc. We exploit this advantage in the present study.

A major topic is the remarkable discovery of a phase-space signal in the
local stellar disc by the \Gaia\ team \citep{antoja2018}. In a volume element
defined by ($\Delta R, \Delta \phi, \Delta z$) = ($\pm 0.1$, $\pm 0.1$, $\pm
1$) kpc$^3$ centred on the Sun, \citet{antoja2018} detect a coherent spiral
pattern in the space spanned by $z$ and $V_z$. This phenomenon is indicative
of a system that is settling from a mildly disturbed state to a stationary
configuration \citep{LyndenBell1967} through the process of phase mixing
\citep[e.g.][\S4.10.2]{Binney2008}.

In Section~\ref{s:params}, we establish notation and define the required
coordinate systems. In Section~\ref{s:galah_gaia}, we characterise the GALAH
survey and use it to understand better the gross structure of the Galaxy's
discs. In Section~\ref{s:spiral_phase}, we analyse the phase spiral in detail
using data from \Gaia\ and GALAH taking advantage of the insights
provided by angle-action coordinates. In Section~\ref{s:v_ellip}, we
present a novel manifestation of the tilt of the velocity ellipsoid as one
moves away from the plane, and discuss the relation of the spirals seen in
$V_R$ and $V_\phi$.
In Section 6, some dynamical implications are considered prior to a search,
in Section 7, for the phase spiral signal in N-body simulations of Sgr
impacting the Galactic disc.  Section 8 sums up and
provides some pointers to future work.

\begin{figure*}
\includegraphics[width=15cm]{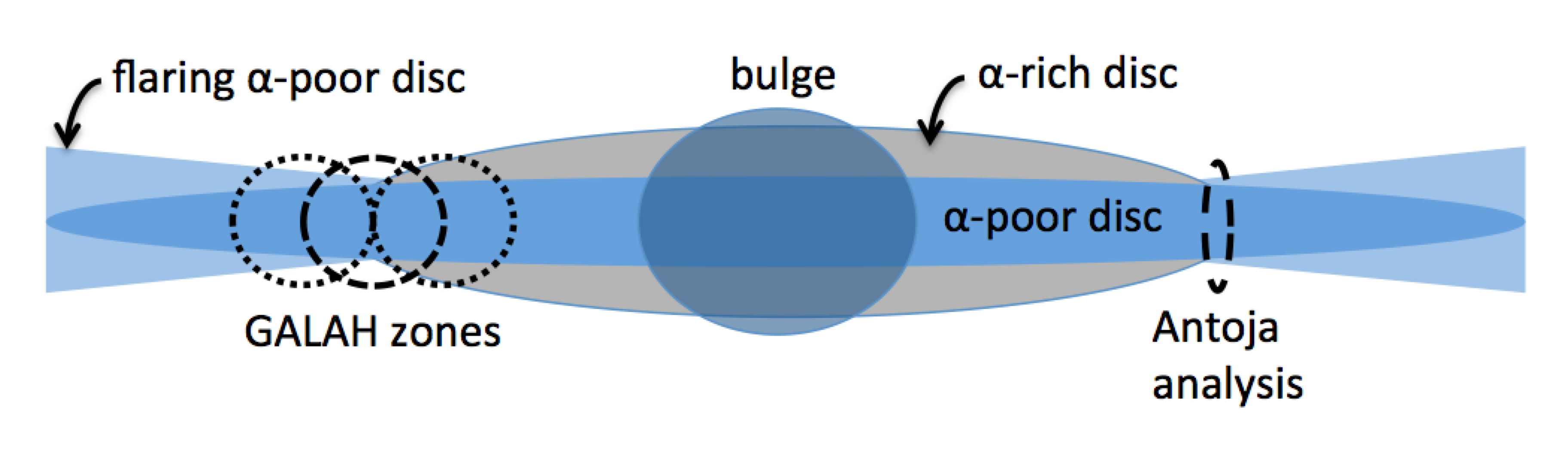}
\caption{Schematic diagram of the modern interpretation of the disc structure \citep[e.g.][]{Hayden2015}. The $\alpha$-rich, inner thick disc has a shorter scalelength than the $\alpha$-poor inner thin disc and terminates near the Solar Circle. Here, the $\alpha$-poor disc takes over and begins to flare at larger radii. It is appropriate now to speak
of the $\alpha$-rich and $\alpha$-poor discs (relative to Fe) rather than the thick and thin discs. The thin vertical ellipse shows the extent of the \citet{antoja2018} \Gaia\ analysis. The large dashed and dotted circles show the domain our analysis; GALAH's low-latitude limit ($\vert b\vert >10^\circ$) is not shown. Both the \Gaia\ and GALAH studies were performed in the solar neighbourhood. Anywhere in the Galaxy, the stellar metallicity [Fe/H] declines with both $R$ and $z$.}
\label{f:disc_sketch}
\end{figure*}

\section{Coordinates}
\label{s:params}

 \subsection{Terminology}

Since the seminal work of \cite{Gilmore1983}, we have become accustomed to the idea that the Galaxy's disc comprises two components. Traditionally these have been called the thin and the thick discs because \cite{Gilmore1983}
distinguished them by their contrasting vertical density profiles. In the last few decades, it has become clear that this terminology is unfortunate
because the real dichotomy is between stars that are poor and rich in
$\alpha$-chain elements (O, Ne, Mg, Si, S, Ar, Ca, Ti) with respect to iron.

\cite{Fuhrmann1998} first noted in a survey of local F and G stars that there are two distinct populations characterised by high and low [Mg/Fe] abundances, and these are associated with the geometrically thick and thin disks respectively. Many subsequent papers provided overwhelming evidence in support of this picture, most notably \cite{Bensby2003} and \cite{Bensby2014} with follow-up studies of more elements at higher sensitivity. \cite{Bensby2011} first suggested a shorter scalelength for the $\alpha$-rich disk compared to the dominant $\alpha$-poor disk that was confirmed by the APOGEE survey \citep{Bovy2015,Hayden2015}. In recent years, this chemical distinction 
has been further emphasized by enhanced C/N in the older $\alpha$-rich population, presumably due to the effects of dredge-up in old turn-off dwarfs and giants \citep{Masseron2015,Hawkins2015}.

Chemistry always provides a firmer foundation for galaxy dissection than kinematics or spatial distributions because a star's chemical composition is invariant even as its orbit evolves. (There are exceptions to every rule, e.g. diffusion in giants can lead to different element ratios.)
Moreover, the classical thin and thick discs are understood to overlap extensively in both configuration
and velocity space so, in much of phase space, stars cannot be assigned to one
disc or the other if only phase-space coordinates are known. Hence dissection by chemistry is the preferred option.

As the numbers of stars with good spectra has grown, it has become clear that
$\alpha$-rich stars form a structure that is quite different from that formed
by $\alpha$-poor stars. \autoref{f:disc_sketch} summarises our current
understanding of the Galaxy's discs. A disc comprising $\alpha$-rich stars
extends out to about the solar radius, having a scaleheight of order $1\kpc$.
A disc comprising $\alpha$-poor stars extends beyond the solar radius. Out to
that radius, it has a scaleheight of order $0.3\kpc$ but further out it
flares, so at large radii $\alpha$-poor stars can be found far from the plane.

In view of this picture, we shall refer to the $\alpha$-poor and $\alpha$-rich discs rather than to the thin and thick discs, consistent with the language used in the earlier papers cited above. Looking forwards, however, an improved nomenclature needs to recognize stellar populations that are even more depleted in [$\alpha$/Fe] at all [Fe/H] than the inconveniently labelled $\alpha$-poor disc. This realization dates back to at least \cite{Russell1992} in a seminal study of stellar abundances in the Magellanic Clouds. 

\citet{Hasselquist2017} and Hasselquist et al (2018, submitted) compile updated abundance data and show that there are {\it three} [$\alpha$/Fe] sequences in the ([Fe/H], [$\alpha$/Fe]) plane where the lowest [$\alpha$/Fe] sequence arises from the accretion of massive dwarfs. A better naming convention, suggested by L.E. Hernquist in discussions with the GALAH team, is as follows: (i) $\alpha_+$ for the [$\alpha$/Fe]-rich sequence, (ii) $\alpha_{\rm o}$ for the confusingly named [$\alpha$/Fe]-poor sequence, and (iii) $\alpha_-$ for the newly established lower sequence. The compact notation refers to the stellar populations and is not specific to a structural component. We illustrate the ``$\alpha$ notation'' in a later figure (\autoref{f:disc_sketch_ripple}) and propose to use this language in subsequent GALAH papers.

\subsection{Reference frame}

We employ right-handed frames of reference for both the heliocentric and Galactocentric systems. In the heliocentric system, the $x$ axis and unit vector {\bf i} point towards the Galactic Centre, the $y$ axis and unit
vector {\bf j} point in the direction of rotation, and the $z$ axis and unit
vector {\bf k} point towards the North Galactic Pole (NGP). In this frame, a star's velocity components are $(V_x,V_y,V_z)$.
Hence we place Sgr A* at
$(x,y,z)=(\Rsolar,0,\zsolar)\kpc$, where \Rsolar\ = $8.2\pm 0.1$ kpc and \zsolar\ = $25\pm 5$ pc \citep{blandhawthorn2016a} consistent with the new ESO Gravity measurement 
\citep{Abuter2018}, 
and the Sun's velocity with respect
to a co-located particle on a circular orbit is $\Vlsr = \Usol {\bf i}\> +
\Vsol {\bf j}\> + \Wsol {\bf k}$, 
with $(\Usol,\Wsol)=(11.1,7.25)\Kms$ \citep{Schoenrich2010}.

We employ Galactocentric cylindrical coordinates $(R,\phi,z)$ centred on Sgr
A* with $\phi$ increasing clockwise when viewed from the north and the Sun
located at $\phisolar=\pi$.  To convert velocities from the
heliocentric to the Galactocentric frame, we take the angular velocity of the
Sun to be $\Omegasolar=(\VLSR+\Vsol)/\Rsolar=30.24\Kms\kpc^{-1}$ from the
measure proper motion of Sgr A* \citep{Reid2004}.

\subsection{Angle-action variables}

Motion in the $zV_z$ plane is simplest when cast in terms of angle-action
coordinates \citep{Binney2008,Binney2018}. The actions $J_R$ and $J_z$ quantify
the amplitudes of a star's oscillation parallel and perpendicular to the
Galactic plane, respectively. In an axisymmetric potential, the third action
$J_\phi$ is the component of angular momentum around the symmetry
axis: $J_\phi\equiv L_z$. Each action $J_i$ is associated with an angle
variable $\theta_i$ such that $(\vtheta,\vJ)$ forms a complete set
of canonical coordinates for phase space. In the potential for which they are
defined, the actions are constants of motion, while the angle variables
increase linearly in time $\theta_i(t)=\theta_i(0)+\Omega_it$, where the
$\Omega_i$ are the star's three fundamental frequencies. We use the software package {\tt AGAMA} \citep{Vasiliev2018} to compute angles and actions for motion in the Galactic potential derived by \cite{Piffl2014}. Our actions were compared to those computed from {\tt galpy} \citep{Bovy2015} and the results are broadly similar given the observational uncertainties. 
The \citet{Piffl2014} potential is preferred because it is constrained to fit the well established, vertical density profile through the Sun's position \citep{Gilmore1983}, which is necessary for deriving an accurate oscillation period for a star away from the plane. We quote actions with dimensions $L^2T^{-1}$ as multiples of $\Rsolar\VLSR=1952\kpc\Kms$ for which $\VLSR=238\Kms$.
We recognize that the actions are not true invariants over the lifetime of the Galaxy, but they are demonstrated to be useful in the halo or the disc over many orbits \citep[e.g.][]{Solway2012}.

\section{Disc dissection with GALAH data}
\label{s:galah_gaia}

The GALAH survey exploits the High Efficiency and Resolution Multi-Element
Spectrograph (HERMES) at the Anglo-Australian Telescope \citep{Sheinis2015}.
This instrument employs the Two Degree Field (2dF) fibre positioner at
the f/3.3 Prime Focus to provide multi-object ($n \approx 400$),
high-resolution (${\cal R} \approx 28,000$) spectra of many stars in a single
shot. HERMES is a fibre-fed, multi-channel spectrograph optimised to carry
out Galactic archaeology from a 4m-class telescope \citep{DeSilva2015}.
HERMES has four optical spectrographs covering 471--490 nm, 564--587 nm,
647--673 nm and 758--788 nm for determining elemental abundances for up to 30
elements (up to 45 elements for the brightest stars). HERMES exploits a photonic comb to internally calibrate the fibre to fibre response across the full field of all four detectors \citep{BlandHawthorn2017,Kos2018c}.

Here we use the internal data release of $505\,571$ stars provided to the GALAH
team which includes GALAH DR2 \citep{Buder2018a} augmented with HERMES data
from parallel observations of open clusters, and K2 \citep{Wittenmyer2018} and
TESS fields \citep{Sharma2018}.  These new observations, which provide
improved calibrations of stellar masses and gravities, were reduced with the
same pipeline as DR2 \citep{kos2017}.  The additional numbers of stars
are $2\,498$, $97\,133$ and $42\,764$, respectively.  From this sample, we select stars
with the \Gaia\ DR2 relative parallax uncertainty of less than 20\% and
distance $|R-\Rsolar|< 1.0$ kpc and $|\phi-\phisolar|< 15^\circ$. These
criteria yield a sample of $192\,972$ stars.

\begin{figure}
\includegraphics
[width=0.5\textwidth]
{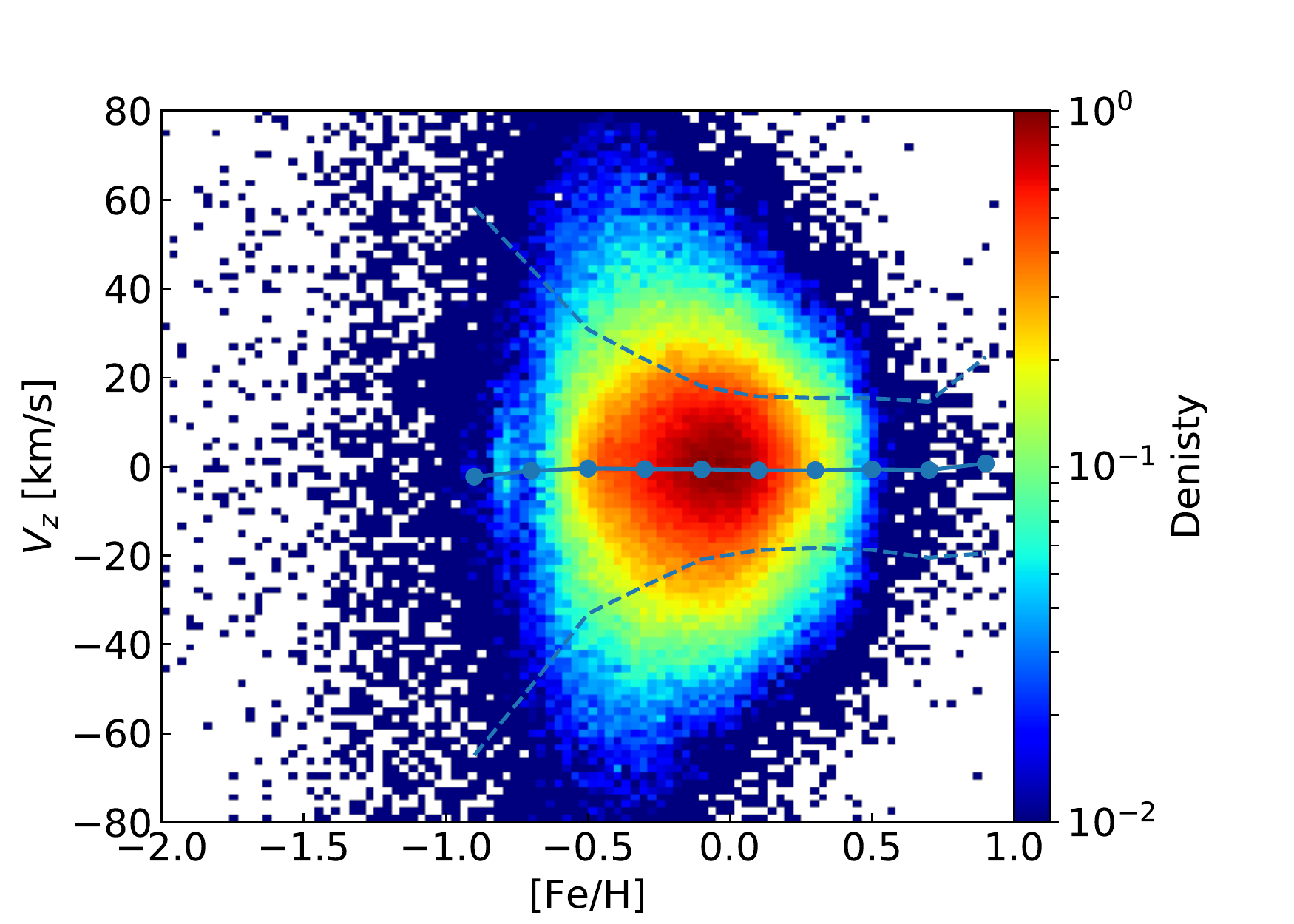}
\caption{Normalised density of GALAH stars in the plane [Fe/H] vs. $V_z$. This is mostly understood in terms of the well known age-metallicity and age-velocity dispersion relations \citep[e.g.][]{Sharma2014}. [Fe/H] is taken from the GALAH survey; $V_z$ is determined from \Gaia\ proper motions and GALAH radial velocities. The mean trend is indicated; the 1$\sigma$ error tracks show the progression from the thin disc to the thick disc and halo as [Fe/H] declines.}
  \label{f:galah_feh_vz}
\end{figure}

\begin{figure}
\includegraphics
[width=0.5\textwidth]
{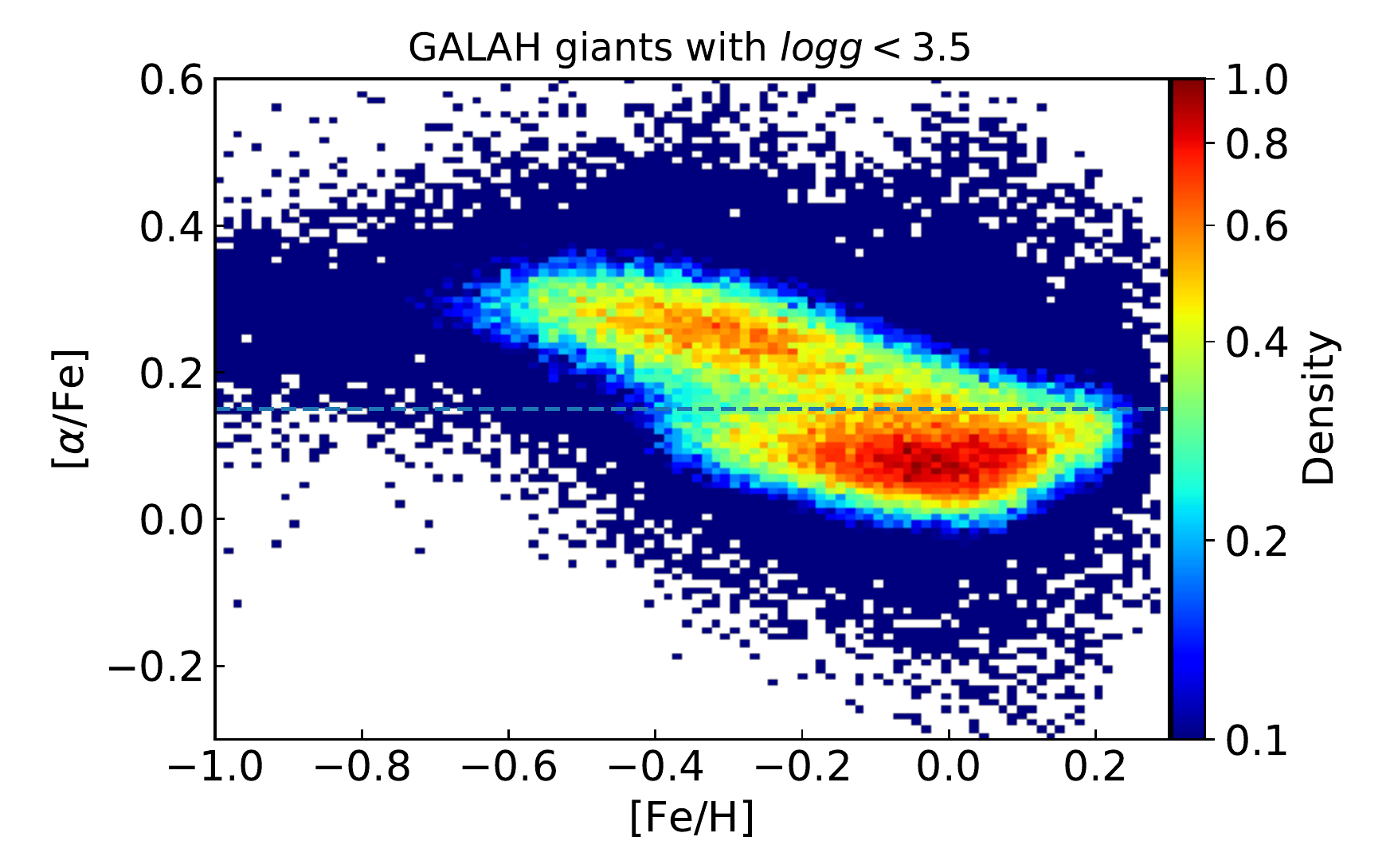}
\caption{Normalised density of giants in the [Fe/H] vs. [$\alpha$/Fe] plane using data from the GALAH survey ($\log g < 3.5$).  A clear separation of the high and low [$\alpha$/Fe] populations is visible. The dashed boundary line distinguishes the high and low [$\alpha$/Fe] populations \citep[cf.][]{Adibekyan2012}. The simple choice of boundary is vindicated in Section~\ref{s:action} where we show that the dynamical properties (actions) of the high and low [$\alpha$/Fe] populations are distinct.
}
\label{f:thin_thick_sep}
\end{figure}

\begin{figure*}
\includegraphics[width=0.8\hsize]
{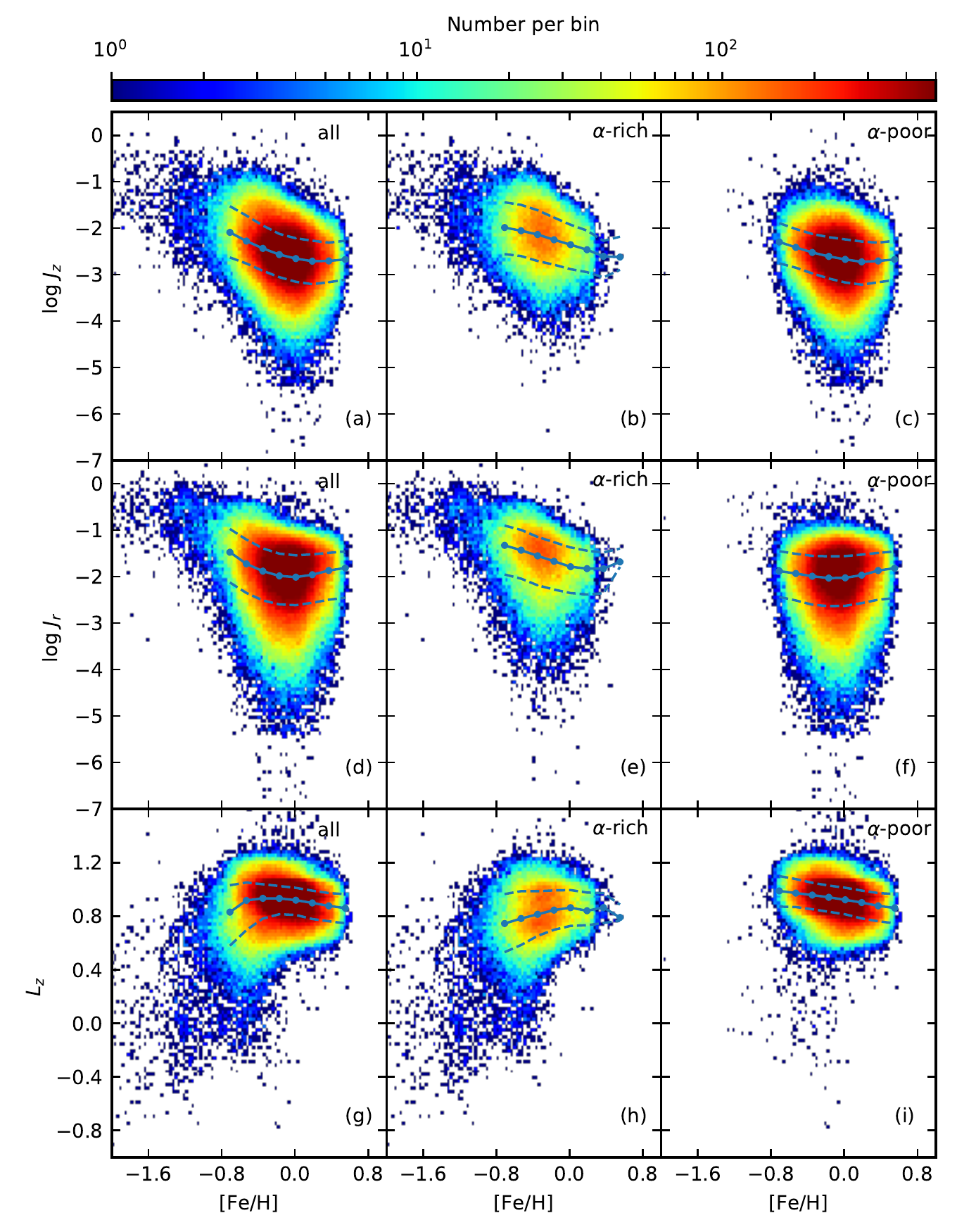}
\caption{Stellar metallicity vs. actions for stars that lie within 1 kpc of the Sun in both GALAH and \Gaia\ DR2 (\autoref{f:disc_sketch}); the normalised stellar densities are shown in colour. 
The actions are computed with the {\tt AGAMA} code using the \citet{Piffl2014} galactic potential, with the normalisation defined in Section 2. 
The left column includes all stars, the middle column is for the $\alpha$-rich disc, and the right column for the $\alpha$-poor disc. (a-c) [Fe/H] vs. $\log J_z$, (d-f) [Fe/H] vs. $\log J_R$, (g-i) [Fe/H] vs. $J_\phi \equiv L_z$. The mean trend and 1$\sigma$ dispersion tracks are also shown.
}
\label{f:gaia_actions_feh}
\end{figure*}

\subsection{Chemodynamic correlations}\label{s:gross_struct}
\autoref{f:galah_feh_vz} shows the density of GALAH stars in the plane of [Fe/H] vs. $V_z$ where the Sun lies at the density peak around (0,0). As [Fe/H]
becomes negative, the density decreases at first gradually and then at $\FeH\sim -0.7$ very steeply as the $\alpha$-rich disc gives way to the halo. At positive [Fe/H], the fall-off in density becomes steep at $\FeH\sim0.4$, significantly more metal-rich than the metallicity of the local ISM determined from the small abundance dispersion of local B stars \citep{Nieva2012}. This is a signature of radial migration \citep{Sellwood2002,SchoenrichB2009} which favours stars with low vertical dispersion $\sigma_z$ on near circular orbits \citep{Minchev10a,Daniel2018}. We provide more evidence of disc migration in Sec. 4.5.

The dashed lines in \autoref{f:galah_feh_vz} illustrate one standard deviation in
$V_z$ at a given [Fe/H]. This is remarkably flat at $\sim20\Kms$ for
$\FeH>-0.1$, but as [Fe/H] falls below $-1$, it increases with ever-increasing
rapidity.  The increase of $\sigma_z$ as [Fe/H] decreases arises from two effects: (a) the
stochastic acceleration of stars causes older, more metal-poor, populations within the thin disc to have larger velocity
dispersions \citep{AumerBS2016}, and (b) as [Fe/H] declines, the $\alpha$-rich disc is more dominant and its stars have larger $\sigma_z$ (hence its thickness).

\autoref{f:thin_thick_sep} shows the distribution in the $(\FeH,\FeA)$ plane
of GALAH stars that are classified as giants ($\log g<3.5$). The distribution
is strongly bimodal. We use the horizontal dashed line to classify stars as
members of either the $\alpha$-rich or $\alpha$-poor disc.  This is a simpler
definition than one used in many other studies, which divide the discs by a
line that runs down toward $\FeA\sim 0$ at the largest values of [Fe/H].
We adopt the horizontal line in \autoref{f:thin_thick_sep} because it reflects the bimodality that is unambiguously present here
and because GALAH captures significantly fewer $\alpha$-rich stars than
$\alpha$-poor ones, so it is better to contaminate the $\alpha$-poor sample
with a few $\alpha$-rich stars than allow significant contamination of the
$\alpha$-rich sample by outliers from the $\alpha$-poor disc.
A more complicated selection, such as those employed by \cite{Bensby2014} and \cite{Hayden2015}, has the potential to contaminate the sample with stars that have more thin disk kinematics.
Specifically, \cite{Haywood2013} and \cite{Hayden2017} show that the metal-rich intermediate $\alpha$ stars have kinematics similar to the thin disk.

\autoref{f:gaia_actions_feh} contrasts the action distributions of stars of different chemical compositions. From top to bottom, the panels of the left column (a), (d) and (g) show the distributions over $J_R$, $J_z$ and $L_z$ for all stars. The middle and right columns separate them into $\alpha$-rich and $\alpha$-poor stars, respectively.
For a star describing a perfect circular orbit confined to the Galactic plane at the Sun's radius, $L_z=J_\phi=1$, $J_R=0$ and $J_z=0$.
For metallicities below [Fe/H]$\approx$-1, the transition to the high energy, low-density stellar halo (low $L_z$) in all panels is clear.
Panel (h) reveals that, in the $\alpha$-rich disc, $L_z$ tends to grow with increasing [Fe/H] while panel (i) shows that the opposite correlation prevails in the $\alpha$-poor disc. The decrease of [Fe/H] with increasing $L_z$ in the $\alpha$-poor disc reflects the radial metallicity gradient (i.e. metal poor stars are from the outer disk and have large $L_z$) that is a general feature of galaxies discussed by many authors 
\citep{Chiappini2001}.

The $\alpha$-rich disc shows the {\it opposite} correlation, a remarkably clear trend discovered by \cite{Spagna2010} that is poorly understood.
\cite{SchoenrichM2017} argue that the effect arises because the $\alpha$-rich disc grew rapidly in radial extent on a $\sim$Gyr timescale coincident with the transition from Type II to Type Ia supernova enrichment. In contrast, the accreting proto-cloud that formed the thick disc may have undergone a rapid collapse over roughly the same timescale, with stars forming during the collapse phase (outside-in scenario).

\begin{figure}
\includegraphics[width=0.5\textwidth]{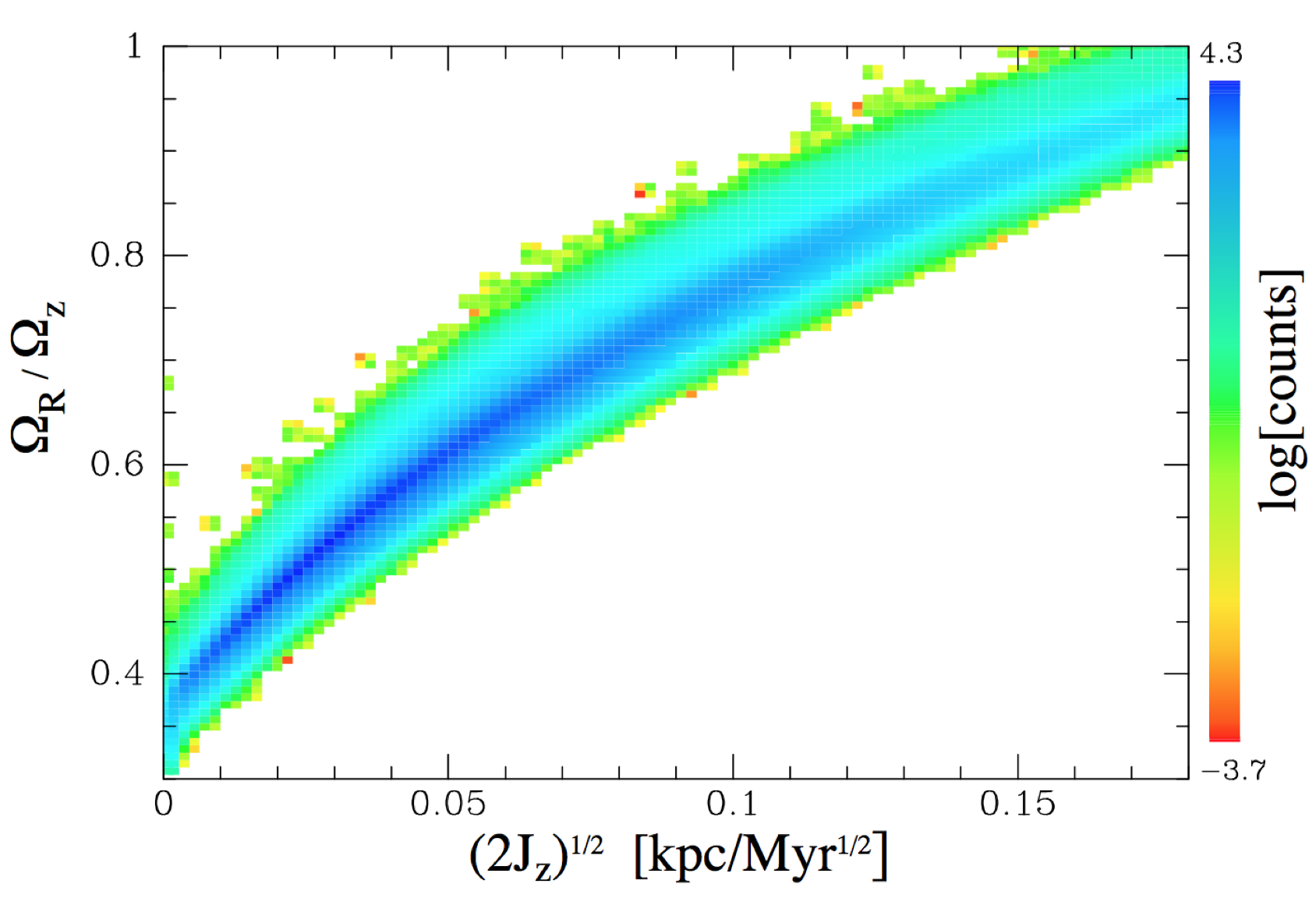}
\caption{The frequency ratio $\Omega_R/\Omega_z$ vs. $\sqrt{2 J_z}$ for orbits calculated in a realistic Galactic potential;
the ratio for most stars lies in the range $0.5-0.7$. The factor 2 in front of $J_z$ ensures that the area inside a curve of constant $J_z$ is $2\pi J_z$ in the $zV_z$ plane. Note that a range of frequencies operates in $R$ and $z$ for all values of $J_z$.}
\label{f:omega_Rz}
\end{figure}

We can relate the action-metallicity dependencies to the better known metallicity-dispersion relations in the following way. 
For all actions, we may write
\[
2\pi J_i=\oint \dot x_i\,\d x_i={1\over\Omega_i}\int_0^{2\pi}(\dot x)^2\,\d\theta_i={2\pi\over\Omega_i}\ex{v^2_i}.
\]
Thus the time average of a star's squared velocity component is related to the action $J_i$ through the associated frequency $\Omega_i$ such that $\ex{v_i^2}=\Omega_i J_i$. Passing from this result for time averages for individual stars to population averages over the stars that reach a given
place is non-trivial, but it generally follows that
\[
\sigma_i^2/\sigma_j^2=\ex{\Omega_i J_i}/\ex{\Omega_j J_j}
\]
where $\ex{.}$ is an appropriate average. When dividing the median tracks in \autoref{f:gaia_actions_feh}b,e for the $\alpha$-rich disc, we find that $\ex{J_z/J_R} \approx 0.1$ near solar metallicity, asymptoting to 0.25 in the metal-poor limit. In the $\alpha$-poor disc, the same trend is seen but the asymptotic limit is significantly lower at 0.2. More broadly, we may write
\[
\sigma_z/\sigma_R \approx \sqrt{\ex{\Omega_z/\Omega_R}}\sqrt{\ex{J_z/J_R}} .
\]
In \autoref{f:omega_Rz}, we note that $\ex{\Omega_R/\Omega_z}\approx 0.6\pm 0.1$ such that $\sigma_z/\sigma_R \approx 0.6\pm 0.1$ consistent with both the Gaia-ESO survey \citep[$\sigma_z/\sigma_R\approx 0.7$;][]{Guiglion2015}\footnote{There is a typographical error in \citet{Guiglion2015} stating that $\sigma_R/\sigma_z \approx 0.7$ in conflict with the data presented in their Fig. 11; their quoted \citet{Binney2014} result is also misstated.} and the RAVE survey \citep[$\sigma_z/\sigma_R\approx 0.6$;][]{Binney2014}.

In \autoref{f:gaia_actions_feh}f, the $J_R$
distribution of the $\alpha$-poor stars has a much sharper top-right corner
than the corresponding distribution of $J_z$ shown in panel (c).
This indicates that the most metal-rich stars extend to the largest values of
$J_R$ but they shun the largest values of $J_z$. This phenomenon has two
possible explanations. One is that spiral structure almost instantaneously
accelerates stars to significant random velocities within the plane but it
takes significant time for molecular clouds to convert in-plane motion to
vertical motion \citep{AumerBS2016}. The other possible explanation is that
the most metal-rich stars can enter the GALAH sample only by migrating
outwards from their birth radii, and stars with large $J_z$ are less likely
to migrate than stars with small $J_z$ \citep{Solway2012}. {\it It is interesting
to note that the upper envelopes of the $J_R$ and $J_z$ distributions of the
$\alpha$-rich stars are almost identical with the most metal-rich stars shunning the highest values of both actions.} 
We discuss the expected signatures of stellar
migration in more detail in Sec. 4.5.

In an upcoming paper, \citet{Hayden2019} derive new velocity dispersion-metallicity relations from the GALAH survey. They take stars (mostly dwarfs) within 500 pc of the Sun and measure the velocity dispersion profiles in all 3 components ($V_R$, $V_\phi$, $V_z$) for populations based on their metallicity and [$\alpha$/Fe] abundance. The velocity dispersions increase smoothly as [$\alpha$/Fe] increases for a fixed metallicity, [Fe/H]. The velocity dispersions also increase as metallicity decreases for stars at a fixed [$\alpha$/Fe]. These results mirror those presented here for the actions, with the kinematics of the $\alpha$-rich population becoming more like
the $\alpha$-poor population as [Fe/H] declines. The most metal-poor, $\alpha$-poor stars have larger velocity dispersions than those with higher metallicities. These stars are the scatter of blue points in \autoref{f:gaia_actions_feh}c,f,i that fall outside the main trends and make up the outer flaring disc.

\begin{figure}
\centering \includegraphics[width=0.5\textwidth]{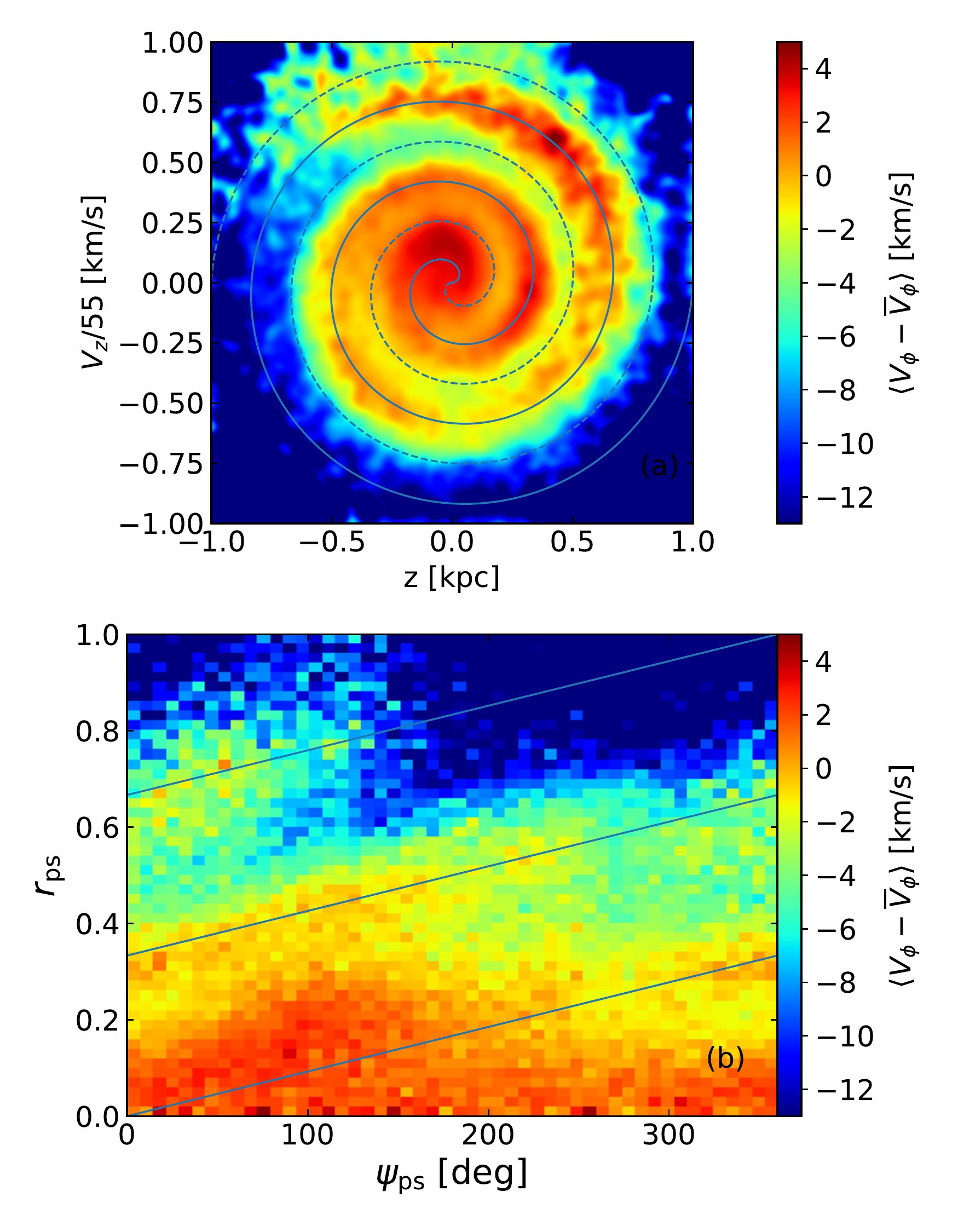}
\caption{
Kinematic properties of stars in the solar neighbourhood using data from \Gaia\ DR2. Stars were selected to be within $|R-\Rsolar|<1.0$ kpc and $|\phi-\phisolar|<15^\circ$. 
Top: map of $\ex{V_\phi(z,V_z)-\overline{V}_\phi}$ with the volume-weighted median value ($\overline{V}_\phi=228.3$\kms) subtracted. The archimedean spirals are defined in equation (3). 
Bottom: the unwound spiral pattern as a function of $\rps$ (dimensionless) and $\psips$ (in degrees). 
The three diagonal lines wrap around every $2\pi$ and thus correspond to three different $\rps$ intervals in the $zV_z$ plane. Note the underlying velocity gradient in $\rps$ due to the asymmetric drift.
}\label{f:gaia_maps4}
\end{figure}

\section{The phase spiral}
\label{s:spiral_phase}

\citet{antoja2018} selected stars from \Gaia\  DR2 that lie in the thin cylinder
$|R-\Rsolar|< 0.1\;{\rm kpc}$ with $|\phi-\phisolar|< 7.5^\circ$. To
revisit the phase spiral in more detail, we expand the volume by an order of
magnitude to $|R-\Rsolar|< 1.0\;{\rm kpc}$ and $|\phi-\phisolar|<
15^\circ$. This expansion ensures that a useful number of GALAH stars are included. As
in \citet{antoja2018}, we restrict consideration to stars that satisfy $\sigma_\varpi/\varpi<0.2$ so that distances can be readily inferred
from parallaxes \citep{Schoenrich2018}. The domain of our study and that of \cite{antoja2018} is shown in \autoref{f:disc_sketch}.

Following \cite{antoja2018}, the top left panel of \autoref{f:gaia_maps4} plots the mean value of $V_\phi$ in the $zV_z$ plane. The overall tendency is for the centre of the panel to be red, signalling large values of
$\ex{V_\phi}$, while the edges are blue because there $\ex{V_\phi}$ is lower.
This low-order structure reflects the familiar phenomenon of asymmetric drift:
stars that make large vertical excursions (large $|z|$ and/or large $|V_z|$)
tend to have guiding centres inside $R_0$ and are consequently visiting us
near apocentre. 
Superposed on the low-order trends in \autoref{f:gaia_maps4} is a prominent
one-armed spiral of stars with anomalously high $\ex{V_\phi}$.

The benefits of expanding the sample volume are clear, but there is also a penalty. 
The phase spiral in \autoref{f:gaia_maps4} shows the change of $\ex{V_\phi}$ with position in the $zV_z$ plane. Comparing the stellar kinematics in the red and yellow parts of the spiral, the 
change of $\ex{V_\phi}$ is due to a shift in the shape of the distribution 
of $V_\phi$, but only for stars within $\approx$25\kms\ of the peak in the $\ex{V_\phi}$ distribution.
Stars further from the peak are not affected: they have the same distribution in the red and yellow parts of the spiral. $\ex{V_\phi}$ is a proxy for the guiding center radius, so the shift in $\ex{V_\phi}$ is just 
a shift in the mean guiding center radius between stars in the red and yellow 
parts of the phase spiral. This shift corresponds to only 5\kms\ in $\ex{V_\phi}$ or 
roughly 200 pc in guiding centre radius. 
By increasing the radial interval from 200 pc to 2 kpc, the small shift in the mean guiding centre radius will be masked by including a much larger overall range of guiding center radii. On the other hand, the shot noise is reduced.

In \autoref{f:gaia_maps4}, a solid black line
has been drawn roughly along the crest of this spiral as follows.  A system
$(\rps,\psips)$ of polar coordinates are defined for the $zV_z$ plane by
\begin{align}
\rps &= \sqrt{(V_z/55\Kms)^2+(z/{\rm kpc})^2}\cr
\psips &= \tan^{-1}[(V_z/55\Kms)(z/{\rm kpc})^{-1}]
\end{align}
where $z$ is in units of kpc. 
The curve is then the Archimedean spiral
\begin{align}
\rps &= k (\psips+\eta)
\end{align}
where $k=1/(6\pi)$ is a constant and $\eta=0,\pi$ (solid, dashed) allows the spiral to be rotated.
The right-hand panel of \autoref{f:gaia_maps4} shows how, when $\ex{V_\phi}$
is plotted in $(\rps,\psips)$ space, the spiral
unravels into sloping straight lines that wrap from the right to the left
edge of the plot on account of the periodicity of $\psips$.

Within the $zV_z$ plane, stars move clockwise on oval curves: our adopted scaling $z/\kpc\sim V_z/55\Kms$ leads to circles consistent
with the discovery paper \citep{antoja2018}. The angular frequency
at which a star moves on its circle decreases as the circle's radius
increases, so an initially radially directed line of stars is steadily sheared into a spiral like that evident in \autoref{f:gaia_maps4}.

\begin{figure}
\centerline{\includegraphics[width=.97\hsize]{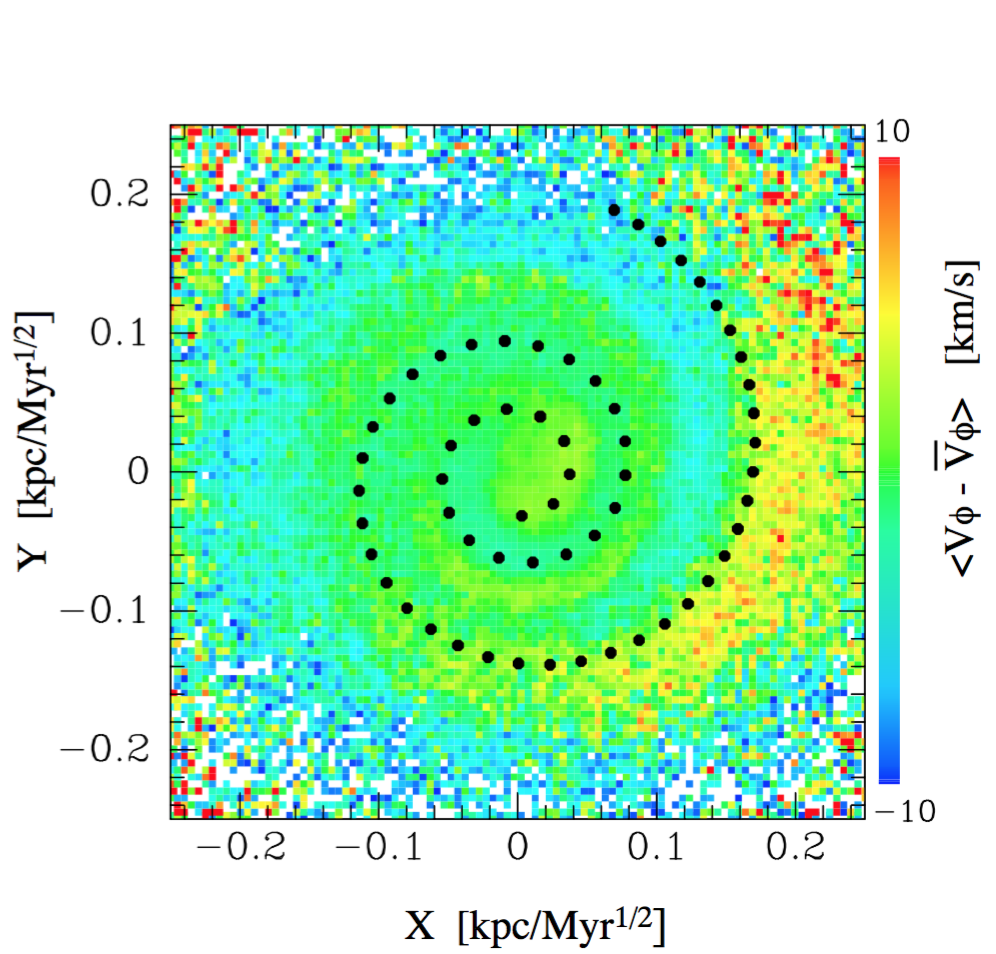}}
\centerline{\includegraphics[width=1.\hsize]{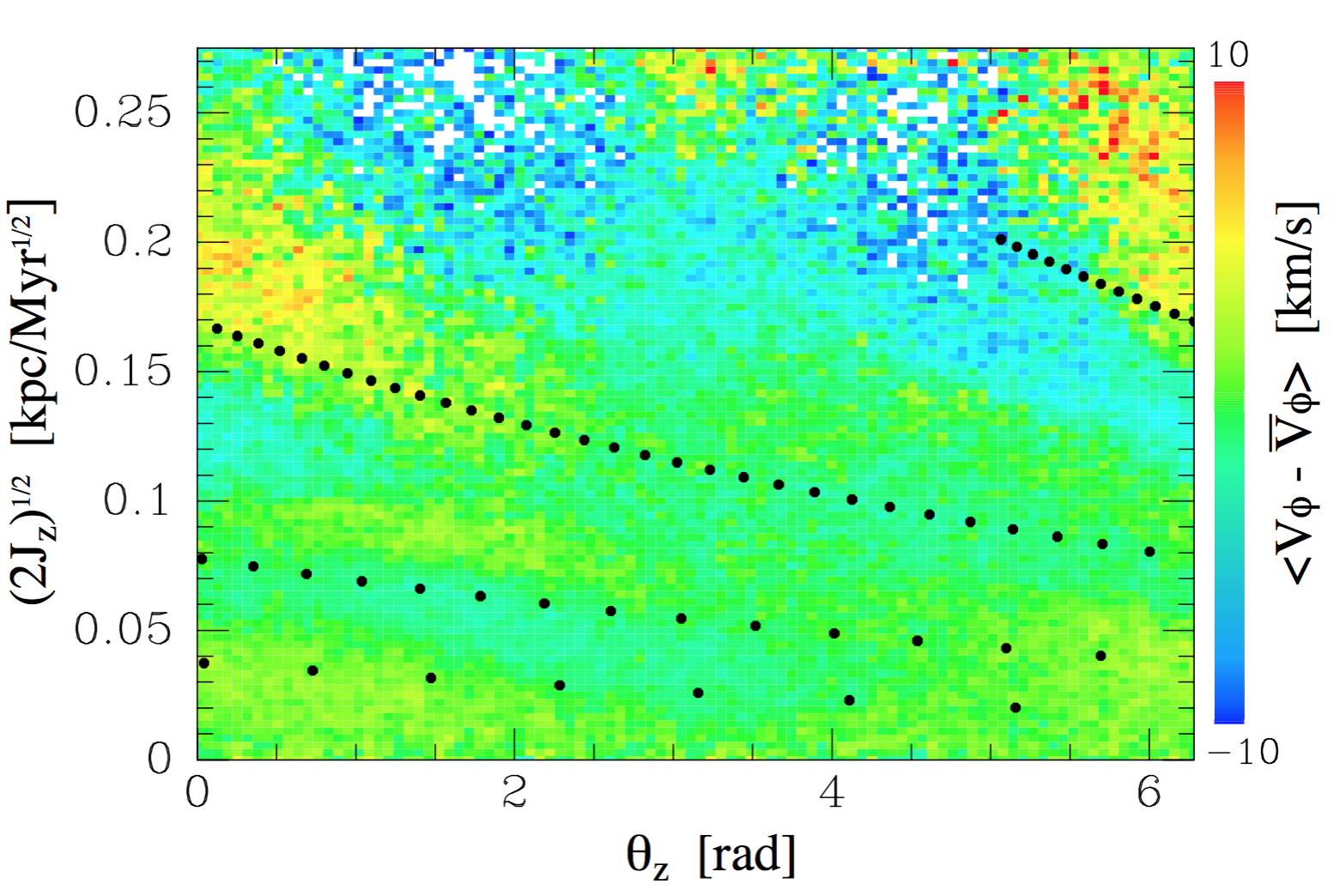}}
\caption{Top: the spiral in the angle-action coordinates defined by equation
(\ref{eq:defXY}). The colour scale shows
$\ex{V_\phi(z,V_z)-\overline{V}_\phi}$, where the overline means the average over
$\theta_z$ at the given $J_R$.  Bottom: the spiral unwrapped by  plotting same quantity in the
angle-action coordinates analogous to $(\rps,\psips)$. The dots show the
relation in equation $\ref{eq:Jspiral}$.
The action-angle figures convey the same information as the projections in \autoref{f:gaia_maps4}.
}\label{f:jjbJplane}
\end{figure}

Using angle-action coordinates, we can quantify this phase-wrapping process.
\autoref{f:jjbJplane} shows the plane defined by coordinates
\[\label{eq:defXY}
X\equiv\sqrt{2 J_z}\:\cos\,\theta_z\quad\quad
Y\equiv-\sqrt{2 J_z}\:\sin\,\theta_z\;,
\]
where the minus sign is included in the definition of $Y$ so stars circulate
clockwise as $\theta_z$ increases.\footnote{The factor 2 in
eq.~(\ref{eq:defXY}) ensures that the area inside a curve of constant $J_z$ is
$2\pi J_z$, as it is in the $zV_z$ plane.}
If the stellar $z$ oscillations were harmonic, these coordinates would be simply
linearly scaled versions of $z$ and $V_z$. In reality, they have a complex
relation to
$(z,V_z)$ because the $z$ oscillations are strongly anharmonic. In the $XY$ plane,
stars move exactly on circles of radius $\sqrt{2 J_z}$ at the angular velocity $\Omega_z$ that can be
computed as a function of the radius. The dots in \autoref{f:jjbJplane} trace
the curve
\[\label{eq:Jspiral}
\theta_z=\Omega_z(J_{\phi0},J_z)\,t_o+0.75
\]
where $J_{\phi0}$ is the angular momentum of a circular orbit at $\Rsolar$
and $t_o=515\Myr$. This curve delineates quite well the inside of the blue spiral.  The
colour scale in this figure shows the amount by which $\ex{V_\phi}$ deviates
from its average value around the relevant circle. The spiral, which
decreases in amplitude towards the centre, can now be traced all the way to
the centre.

\cite{Binney2018} argue that the spiral arises because, at a given value of $J_z$, the values of $\Omega_z$ vary systematically with $L_z$: stars
with smaller guiding-centre radii have higher frequencies. Hence, when some
event bunches stars in $\theta_z$, the stars with smaller values of $L_z$
move ahead of the stars with larger $L_z$ as they move around their common
circle in the $xy$ plane. The dots in \autoref{f:jjbJplane} mark the
locations reached by stars that lie at the centre of the $L_z$ distribution,
so the part of each circle on which $V_\phi$ is below average should lie at later times on the clock than the part at which $V_\phi$ is above average. This is precisely
what we see in the upper panel of the figure.

\begin{figure}[.]
\centering \includegraphics[width=0.48\textwidth]
{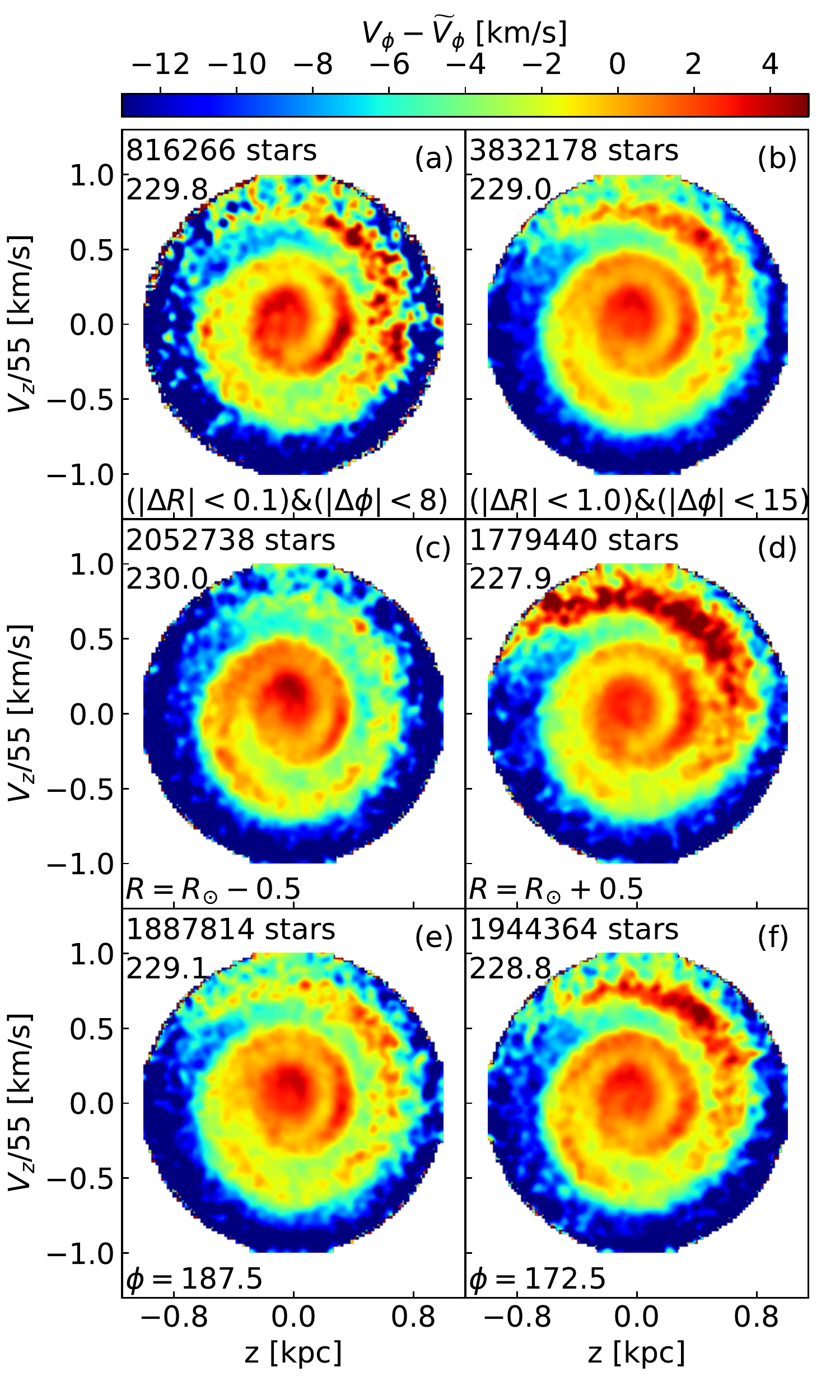}
\caption{Kinematic properties of stars in the solar neighbourhood in the $zV_z$ plane using data from \Gaia\ DR2 explored by volume and location. The colour code shows the deviation in volume-weighted $V_\phi$ with respect to the median value quoted in each panel along with the number of stars used.
(a) Stars with $|R-\Rsolar|<0.1$ kpc and $|\phi-\phisolar|<7.5^\circ$ presented in Antoja's original analysis for a narrow vertical cylinder. 
(b) Stars with $|R-\Rsolar|<1.0$ kpc and $|\phi-\phisolar|<15^\circ$ highlighting the same phase spiral over the 20$\times$ larger GALAH volume.
(c) Stars with $(|R-(\Rsolar-0.5)|<1.0$ kpc and $|\phi-\phisolar|<15^\circ$. (d) Stars with $|R-(\Rsolar+0.5)|<1.0$ kpc and $|\phi-\phisolar|<15^\circ$. Note the inner spiral in (c) is vertically elongated consistent with the smaller Galactic radius; the outer spiral in (d) is more prominent at larger radius.
(e) Stars with $|R-\Rsolar|<1.0$ kpc and $|\phi-(\phisolar+7.5^\circ)|<7.5^\circ$. 
(f) Stars with $|R-\Rsolar|<1.0$ kpc and $|\phi-(\phisolar-7.5^\circ)|<7.5^\circ$. The form of the spiral is remarkably invariant across most panels, but differents sections can be enhanced relative to others, e.g. note the asymmetry in azimuth and radius in the prominence of the outer spiral (d,f compared to c,e). ($\phi$ is quoted in degrees.)
\label{f:gaia_maps3}}
\end{figure}
\subsection{Slicing by location}\label{s:location}

\autoref{f:gaia_maps3} explores how the properties of the phase spiral vary with location and volume in the Galaxy. Panel (a) exploits the data used by \citet{antoja2018} that are confined to a narrow vertical cylinder local to the Sun. Panel (b) is the expanded sample around the Sun studied in
\autoref{f:gaia_maps4}. The volume of the latter sample is twenty times
larger than the former, yet the spiral pattern remains essentially the same.
The remaining panels, which also use larger volumes, examine samples in neighbouring spheres close to the
solar neighbourhood, two offset to larger and smaller radii (c,d), and two offset
in both directions in azimuth (e,f).

As illustrated in
\autoref{f:disc_sketch}, panel (c) examines a sample centred at
$R=\Rsolar-0.5$ kpc, while panel (d) explores a sample centred at $R=\Rsolar+0.5$ kpc. The spiral pattern is evident but shows clear differences from inner to outer disc, with the outer disc spiral being stronger at higher $\vert z\vert$. The inner part of the spiral in (c) is stretched vertically relative to (d) consistent with the stronger disc gravity (see below).

Panel (e) explores a sample with
Galactocentric longitude $|\phi-(\phisolar+7.5^\circ)|< 7.5^\circ$, while panel (f) explores a sample with $|\phi-(\phisolar-7.5^\circ)|< 7.5^\circ$. There is little evolution with azimuth over the inner spiral; the outer spiral varies with azimuth in addition to radius. The outer part of the phase spiral being more prominent at larger Galactocentric radius and smaller azimuthal angle has an important consequence that we return to below.

We also inspected the 2-kpc diameter volume centred at $R=\Rsolar\pm 1$
kpc; the phase spiral was evident but less well defined due to the smaller number of stars and the larger cumulative errors on measured parameters. 

\begin{figure}
\centering \includegraphics[width=0.48\textwidth]
{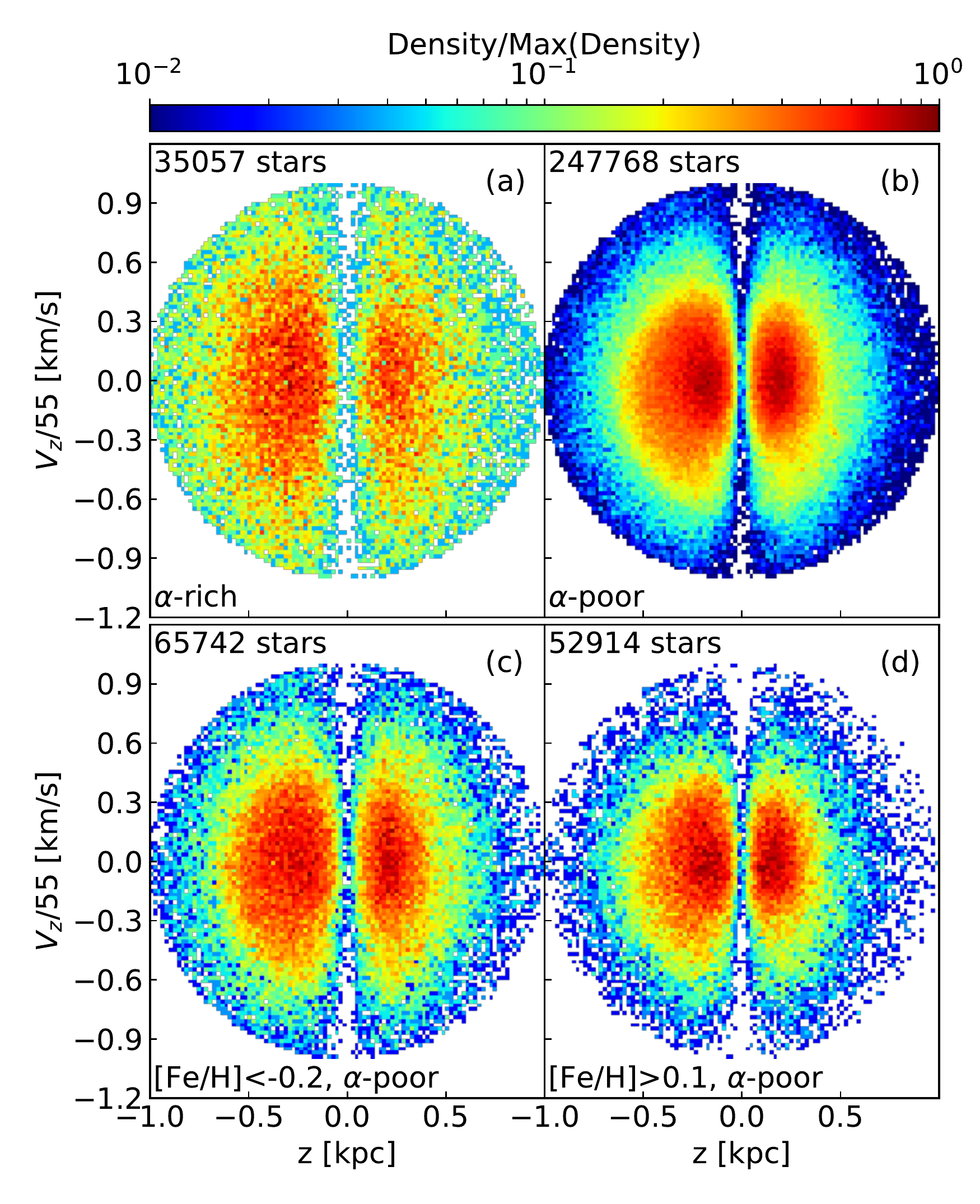}
\caption{The density of GALAH stars in the 
$zV_z$ plane as a function of chemistry: (a) $\alpha$-rich disc, (b) $\alpha$-poor disc, (c) $\alpha$-poor, metal-poor disc, (d) $\alpha$-poor, metal-rich disc. The number of stars used in each panel is indicated.}\label{f:zVzdensFeH}
\end{figure}

\begin{figure}
\centering \includegraphics[width=0.48\textwidth]
{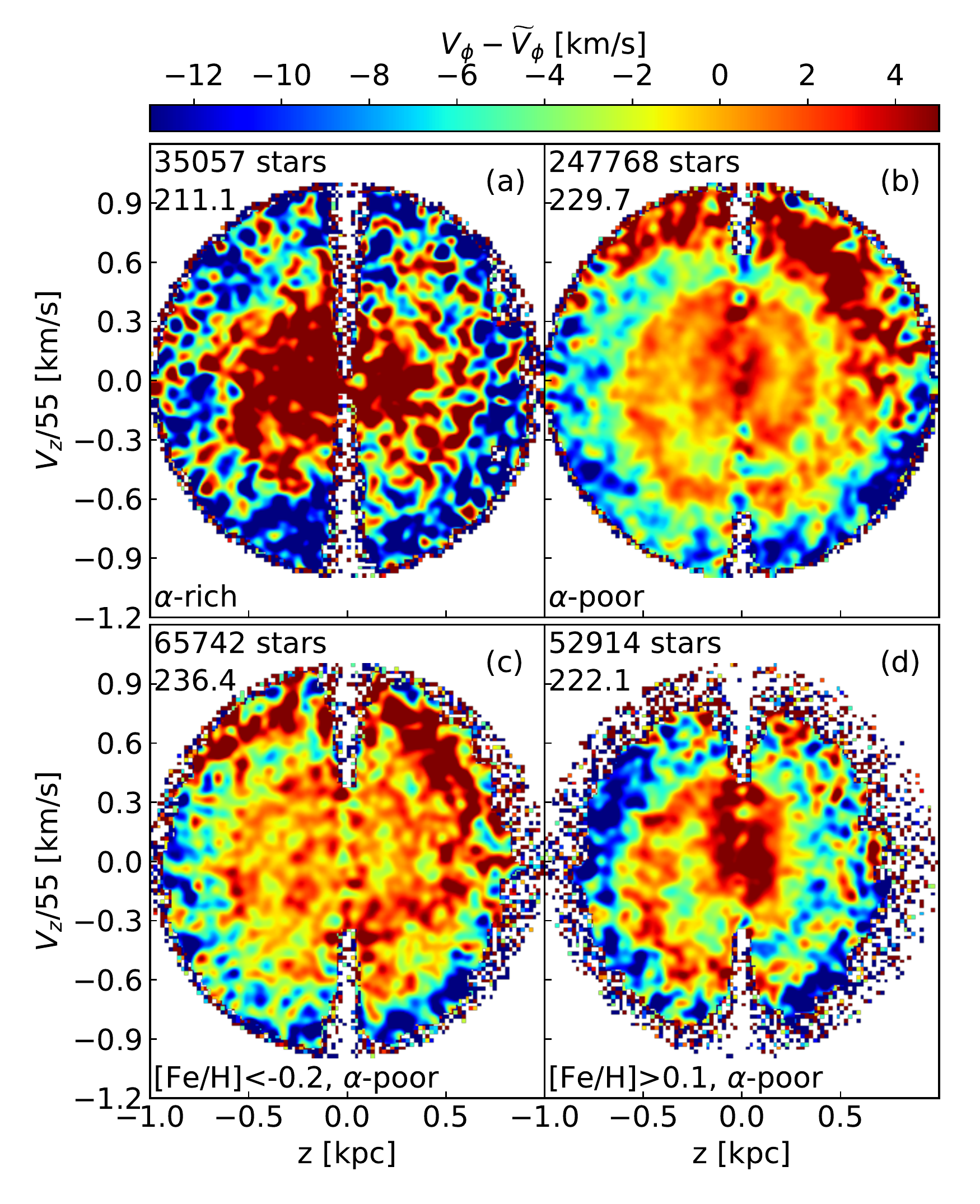}
\caption{Analogue of \autoref{f:zVzdensFeH} but showing $\ex{V_\phi}$ instead of the density of stars in the $zV_z$ plane. In panel (d), the inner phase spiral is prominent in the younger, $\alpha$-poor, metal-rich disc, but is not obviously evident in the older stars, i.e. the $\alpha$-rich disc and the $\alpha$-poor, metal-poor disc (a,c). The outer spiral is most prominent in the $\alpha$-poor, metal-poor disc (b,c).
\label{f:gaia_galah_maps3}}
\end{figure}

\subsection{Slicing by chemistry}
\label{s:chemistry}

In \autoref{f:zVzdensFeH}, we show how the distribution of stars in the $zV_z$ plane varies with metallicity. 
As a result of GALAH's selection criterion ($\vert b \vert > 10^\circ$), there is a deficit of stars at small $z$. On account of the age-metallicity
relation, metal-rich stars tend to be younger than metal-poor ones, so by the
age-velocity dispersion relation they are more strongly concentrated to the
centre of the $zV_z$ plane than metal-poor stars, and especially more than
$\alpha$-rich stars, which, as discussed in Section~\ref{s:gross_struct},
reach exceptionally large values of $J_z$. 

As \autoref{f:gaia_galah_maps3} shows, these differences by chemical
composition in the locations of stars within the $zV_z$ plane impact the
visibility of the phase spiral in different chemically defined populations.
Panel (a) shows the phase plane for $\alpha$-rich stars in the sense of
\autoref{f:thin_thick_sep}, while panel (b) shows the $\alpha$-poor stars
\citep[cf.][]{Adibekyan2012}.  Additionally, we subdivide the $\alpha$-poor
disc into (c) metal poor ([Fe/H]$<$-0.2) and (d) metal rich ([Fe/H]$>$0.1) bins. {\it The phase spiral shows a clear trend in metallicity across the $zV_z$ plane}: for metal-rich stars, the inner spiral is strong; for metal-poor stars, particularly in the $\alpha$-poor disc, the outer spiral is strong.

The $\alpha$-rich disc (a) shows much weaker evidence for the phase spiral than the $\alpha$-poor disc (b), but in
part this may reflect the order of magnitude smaller size of the sample.
To check if this fully accounts for the difference, we examined samples of
$\alpha$-poor stars that were restricted to  $31\,666$ stars by random
sub-sampling: even these attenuated samples of $\alpha$-poor stars 
showed the spiral more clearly than the
$\alpha$-rich sample. This indicates that the spiral pattern in the
$\alpha$-rich disc is intrinsically weaker than in the colder
$\alpha$-poor disc.

As we vary the metallicity of the $\alpha$-poor stars under study, two
effects are in play.  First, on account of the disc's radial metallicity
gradient, the mean guiding-centre radius of the sample will decrease as the
metallicity is increased, and we saw above (\autoref{f:gaia_maps3}) that
decreasing the radius weakens the outer spiral. Secondly, by the age-velocity dispersion
relation, the $zV_z$ footprint of the stars will move inwards as
metallicity increases (\autoref{f:zVzdensFeH}), which will again weaken the spiral
(\autoref{f:jjbJplane}).
In line with these expectations, in \autoref{f:gaia_galah_maps3},
the outer spiral is clearest for the most metal poor stars (c)
and becomes weaker as we increase the metallicity of the sample (d).

The trend in the spiral pattern with metallicity is smoothly varying with no
evidence for a chemically homogeneous or single-age population (e.g. star
cluster) dominating the phase spiral at fixed $V_\phi$. This rules out the idea that it is associated with one or more large disrupting star clusters.

\begin{figure}
\centering \includegraphics[width=0.48\textwidth]%
{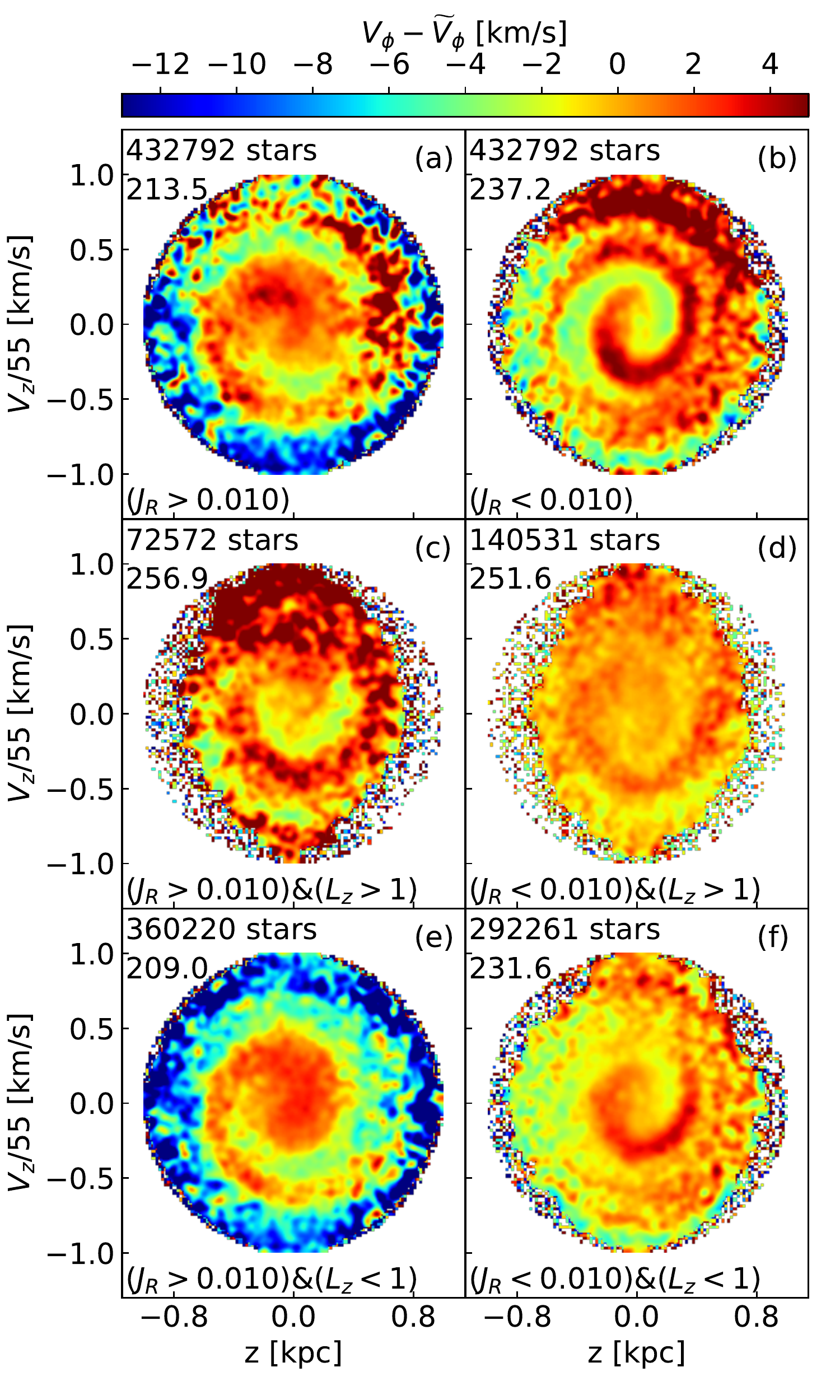}
\caption{Dissection by actions ($J_R$, $J_\phi \equiv L_z$) over the original Antoja volume ($|\Delta R|<0.1$ kpc). The left panels show stars with $J_R > \overline{J}_R$, where $\overline{J}_R$ is the median value over the volume. The right panels show stars with $J_R < \overline{J}_R$. The character of the phase spiral is very different for stars with eccentric (a) compared to more circular (b) orbits. Thus, we divide the distribution in action further:
(c) $J_R > \overline{J}_R$, $L_z > 1$; 
(d) $J_R < \overline{J}_R$, $L_z > 1$; 
(e) $J_R > \overline{J}_R$, $L_z < 1$; 
(f) $J_R < \overline{J}_R$, $L_z < 1$. 
Stars with $L_z > 1$ have guiding-centre radii outside the Solar Circle. Comparison of
panels (f) and (d) reveals that the tight inner spiral arises from stars with less eccentric orbits and guiding radii inside \Rsolar\ that reach apogalacticon in the solar neighbourhood. Comparison of panels (c) and (e) shows that stars on eccentric orbits from the outer disc are much less relaxed.
}\label{f:gaia_actions}
\end{figure}

\subsection{Slicing by actions}
\label{s:action}

In \autoref{f:gaia_actions}, we explore how the $zV_z$ phase plane varies when stars
are selected by the values of their actions $J_R$ and $L_z$. Stars are
split by whether their radial action is greater than (left panels: a, c, e) or less than
(right panels: b, d, f) the median value over the \Gaia\ volume, $\overline{J}_R=0.01$.
Stars with larger $J_R$ move on more eccentric orbits. Near the centre of the
phase plane, the spiral is much more clearly traced by stars with less eccentric orbits. In fact, panels (b) and (f) provide {\it the clearest manifestation to date that the phase spiral can be traced to the origin of the vertical phase plane.} Notice that the $V_\phi$ scales of the two panels
are quite different, so $V_\phi$ for the low-eccentricity stars is systematically larger than for the high-eccentricity stars.
This indicates that stars with large $J_R$ typically have guiding-centre
radii inside \Rsolar, a consequence of both the steep radial density gradient
within the disc and the outward decline in $\sigma_R$.

The lower row in \autoref{f:gaia_actions} shows the effect of further
splitting stars by their values of $L_z \equiv J_\phi$. The guiding centres of stars
with $L_z > 1$ lie outside $R_0$, and conversely for stars with $L_z<1$.
Comparison of panels (d) and (f) shows significant difference in the structure
of the innermost portion of the phase spiral -- stars with smaller $L_z$ form a
tighter spiral that reaches the centre sooner. 
Panel (c) shows that the stars on eccentric
orbits with large guiding centres are very far from relaxed, with a strong
bias towards values of $\psips\sim90^\circ$.

\begin{figure}
\centering \includegraphics[width=0.48\textwidth]
{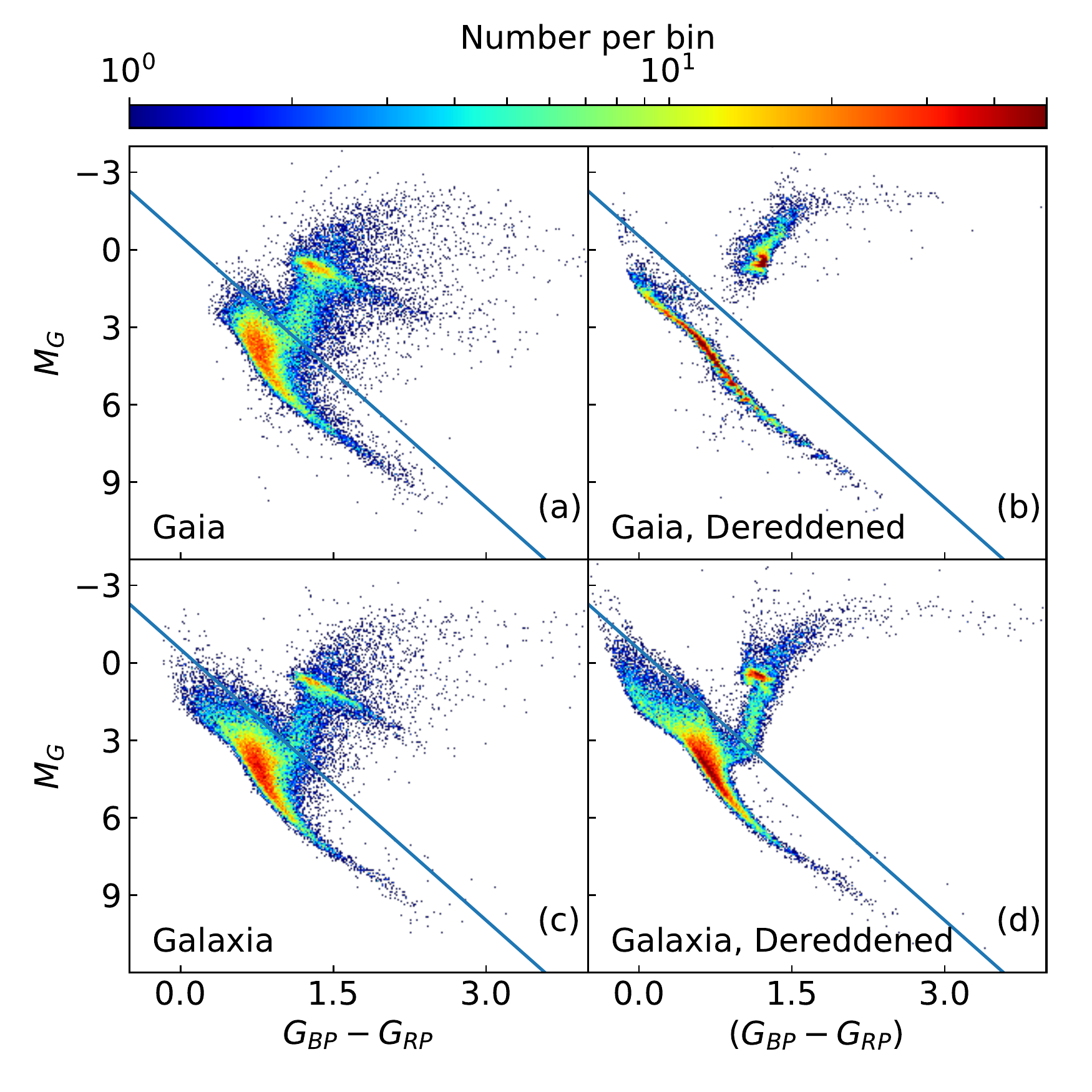}
\caption{
The colour-magnitude distribution of \Gaia\   
stars in the solar neighborhood and  that of a mock sample generated with the {\tt Galaxia} model for the Galaxy \citep{Sharma2011} which includes a sophisticated treatment of the dust extinction. 
The \Gaia\ sample is similar to that used by  \citet{antoja2018} and consists of stars with valid radial velocities, less than 20\% error on parallax, $|\Delta R <1.0|$ kpc, $|\Delta \phi|<15^\circ$, and $|\Delta z|<1$ kpc. The mock sample was generated to match the $G$ magnitude distribution and the selection criteria of the \Gaia\ sample.
Qualitatively, the raw counts in the colour-magnitude diagram uncorrected for dust extinction agree very well. The dereddened data and models agree less well because the dereddened 
\Gaia\ stars are a biassed sample; we do not use this sample in our analysis.
The diagonal blue line is how we separate giants (upper) from dwarfs (lower); see the main text.
\label{f:cmd_gaia_galaxia}}
\end{figure}
\begin{figure}
\centering \includegraphics[width=0.5\textwidth]
{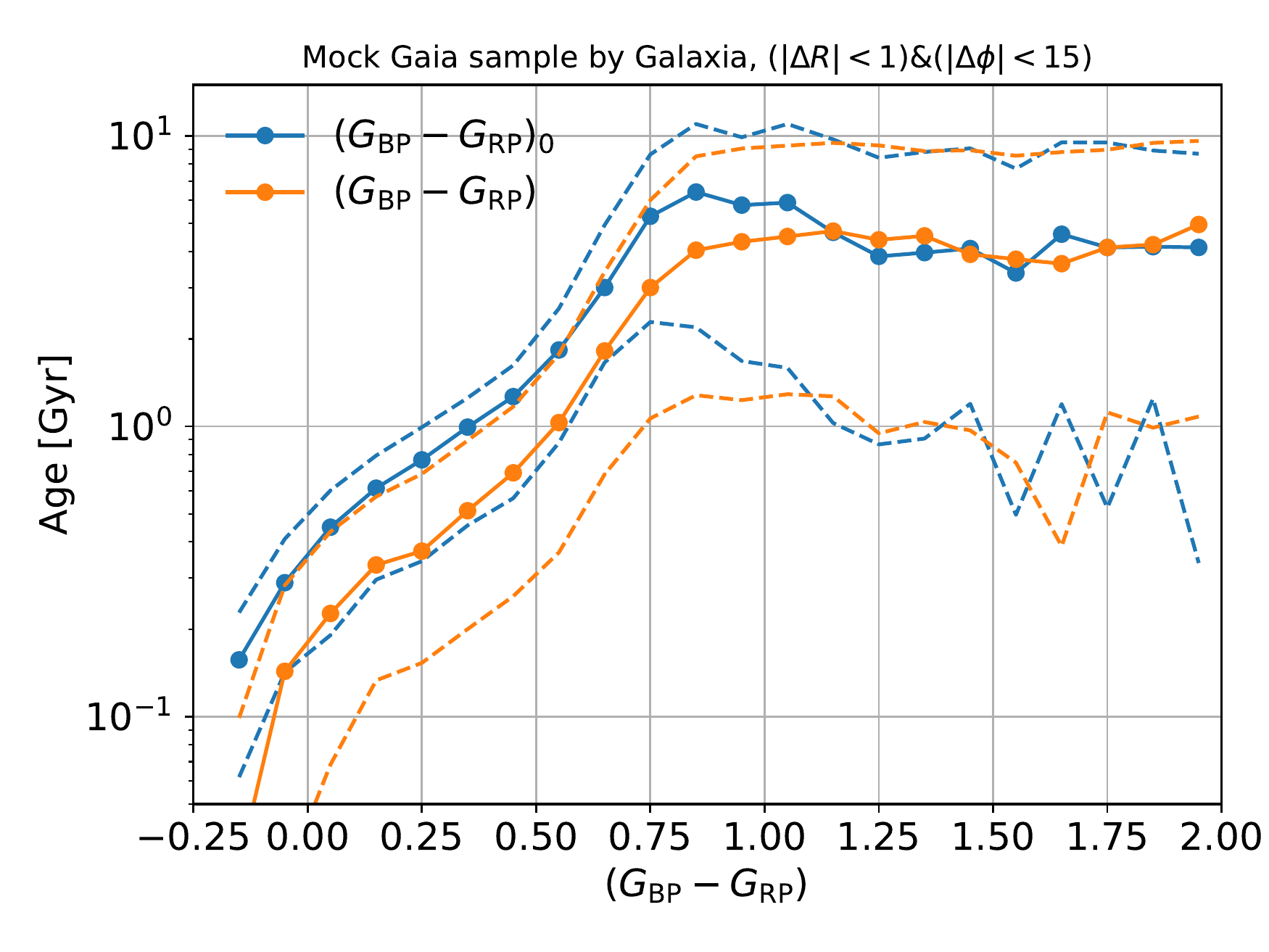}
\caption{
Given the assumed model and selection function for the Galaxy of \autoref{f:cmd_gaia_galaxia}, we derive a mean stellar age as a function of the \Gaia\ colour for the dwarf stars.
The lower solid curve is based on the raw \Gaia\ photometric data uncorrected for dust extinction; the 16 and 84 percentile ranges are shown as dashed lines. The upper curves are the \Gaia\ photometric data corrected for dust extinction. Age discrimination is impossible beyond about 3.5 Gyr.
\label{f:age_vs_bprp}}
\end{figure}

\begin{figure}
\centering \includegraphics[width=0.48\textwidth]
{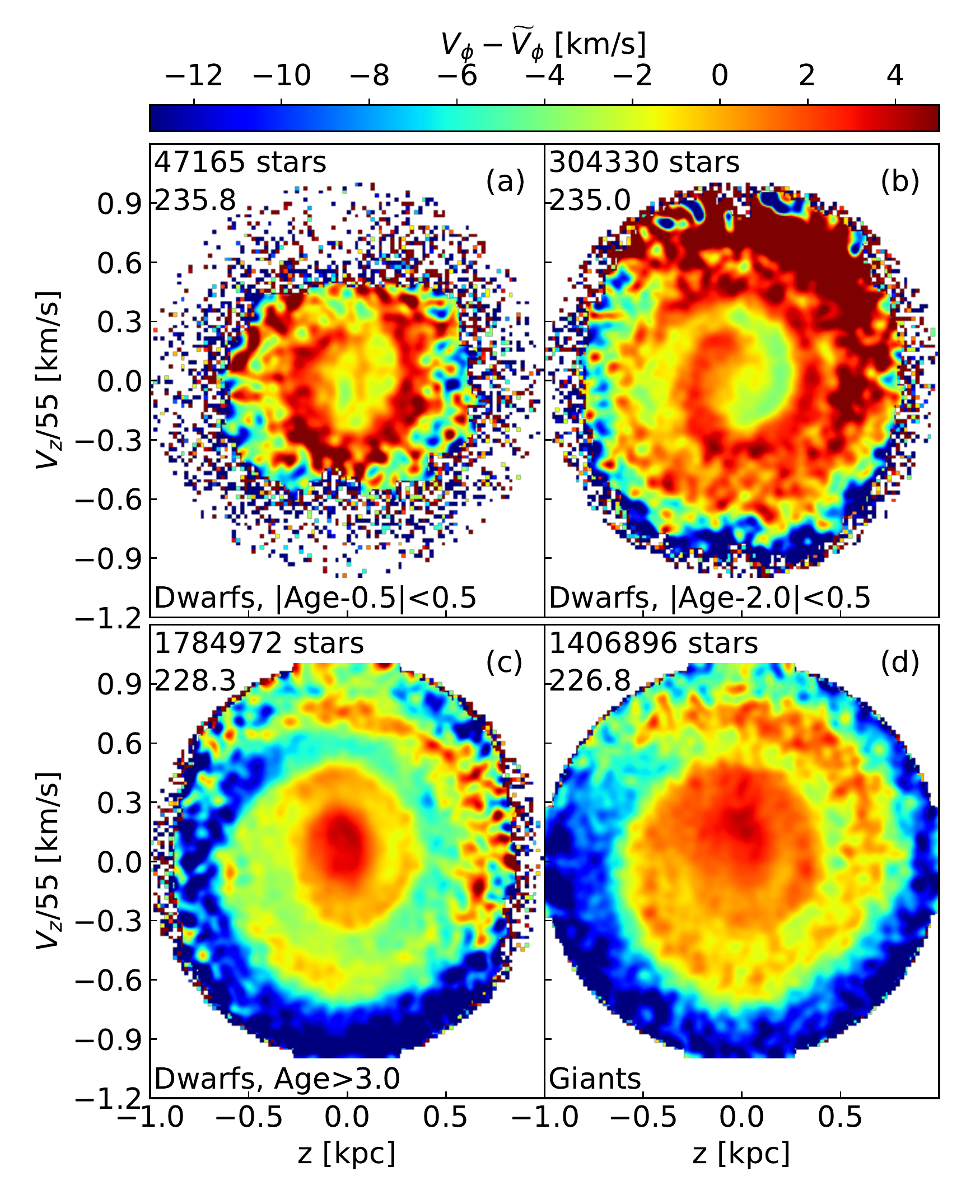}
\caption{
An analysis of the \Gaia\ photometry over the GALAH volume in the $zV_z$ plane showing crude stellar ages for (a-c) dwarfs 
and (d) giants. Stars younger than about 3 Gyr emphasise the `hook' at the centre of the phase spiral (a,b). This feature closely resembles \autoref{f:gaia_actions}b for which $J_R < 0.01$, i.e. younger stars from the local disc are on more circular orbits. The old dwarfs (c) and giants (d) resemble \autoref{f:gaia_actions}a for which $J_R > 0.01$, i.e. the older stars are on more elliptic orbits. 
\label{f:gaia_age_zVz}}
\end{figure}

\subsection{Slicing by ages}
\label{s:ages}
The \Gaia\ DR2 data release provides accurate photometry in three bands that we exploit here, i.e. the super-broadband $G$ filter, and for the split bands \GBP\ (blue) and \GRP\ (red) \citep{Jordi2010}. We use the \Gaia\ colour ($\GBP-\GRP$) to estimate crude photometric stellar ages.
\Gaia\ also provides a reddening estimate; we use these to demonstrate some problems but do not use them in the final analysis for two reasons. First, they are not available for all stars.
Secondly, the velocity dispersion of the bluest stars proves larger when we use dereddened colours than the raw colour, which indicates that dereddening contaminates a `blue' sample with older, actually redder, stars.

To calibrate the age-colour relation, we use {\tt Galaxia} to generate a mock catalog for all stars in the GALAH volume and from them derive \Gaia\ magnitudes and colours. {\tt Galaxia} uses Bayesian estimation with a scheme that takes into account the
initial mass function, the star formation rate and a sophisticated treatment for dust extinction \citep{Sharma2011,Sharma2017}. In the left panels of
\autoref{f:cmd_gaia_galaxia},
the resulting raw distribution closely matches the \Gaia\ data over the same volume. The dust-corrected photometry matches less well because the \Gaia\ photometry with dust corrections represent a biassed sample that is not accounted for in the {\tt Galaxia} model.

In \autoref{f:age_vs_bprp}, we derive
from {\tt Galaxia} the centres and widths of the distributions of age at a given colour. The profiles are given for both the extinction-corrected and uncorrected colours, both of which show a monotonic increase in age with colour up to $\GBP-\GRP$ $\approx$ 0.75 (3 Gyr). Beyond here, the \Gaia\ colour is a poor tracer of age. The separation of giants and dwarfs is shown in \autoref{f:cmd_gaia_galaxia}. We select dwarfs as $M_G<3.5\:(\GBP-\GRP-0.15)$, where $M_G$ is the absolute magnitude derived using $G$-band photometry and the parallax-derived distance.

In \autoref{f:gaia_age_zVz}, we show the $\langle V_{\phi}(z,V_z)\rangle$ map in the $zV_z$ plane for dwarfs and giants. The dwarfs are further split up into three different age slices. The vertical extent increases with age due to the age-velocity dispersion relation. The giants and the oldest dwarfs have a similar distribution in age but we suspect the the oldest stars are restricted in $z$ as compared to giants because of the selection function limit ($G<14$): red dwarfs are intrinsically fainter and harder to detect at high $z$.

For the dwarfs, the spiral exhibits a systematic trend with age. The inner `hook' of the phase spiral (a,b) is stronger and clearer for stars younger than 3 Gyr; for older stars (c,d), the centre is more filled in.
The 16 and 84 percentile age range for giants is about 1 to 9 Gyr, with a median age of 3.4 Gyr. The old dwarfs and giants (c,d) have a similar distribution in the $zV_z$ plane although the outer phase spiral for giants (d) is more diffuse, presumably because they arise from a larger volume in $(R,z)$.
For younger dwarfs (\autoref{f:gaia_age_zVz}a,b), the inner hook closely resembles that of stars with $J_R<0.01$ (\autoref{f:gaia_actions}b) as expected as these are stars on more circular orbits.

This raises an important question. Were the younger stars perturbed or kicked into the spiral pattern after they were born, or did they form from gas which was perturbed into the spiral pattern? If the former, this suggests
that it is more difficult to excite a spiral pattern within an old, dynamically hot population. Conversely, if the latter applies, this may mean the pattern is stronger and more coherent in the gas phase. Given that collisionless stars and gas respond slightly differently to a perturbation in the potential, the two scenarios may lead to different phase distributions for the younger stars. 

In a forthcoming paper, we test these ideas using an N-body simulation that has a disc continuously forming stars while interacting with the Sgr dwarf. Future HI surveys should look for the kinematic signatures of corrugation waves in the local disk.
With sufficient information on transverse motions, clumpy gas could even exhibit some of the characteristics of the phase spiral. We return to this point in Sec. 8.2.

\subsection{Stellar migration}

In Section~\ref{s:gross_struct}, we presented preliminary evidence for stellar migration based on metal-rich stars in the $\alpha$-poor disc being limited to small values in $V_z$ and $J_z$. The important point here is that stellar migration, as distinct from other scattering processes, is only efficient for stars close to the plane where the spiral arm resonances operate \citep{Sellwood2002}. The newly furnished analysis in this section provides an even stronger case because efficient stellar 
migration heavily favours stars on circular orbits, which we also observe for the same stars.

In an important study, \citet{Nieva2012} examine the elemental abundances of the local B star population and diffuse ISM. These stars reflect the local gas-phase abundances in the recent past ($\lesssim 50$ Myr). They establish a high level of chemical homogeneity across the young population, in agreement with the gas phase \citep{Sofia2001}. The authors
demonstrate that the Sun is too metal-rich for our neighbourhood and must have migrated from its birthplace near\footnote{In response to the question raised by \citet{BlandHawthorn2010}, this argues that we should be looking inwards for members of the solar family, i.e. stars born in the same star cluster.} $R\sim 5$ kpc.

In \autoref{f:gaia_galah_maps3}c,d, we compare metal-rich and metal-poor stars in the $zV_z$ plane specifically 
for the $\alpha$-poor stars. The inner part of the phase spiral is dominated by the metal-rich stars ([Fe/H] $>$ 0.1). In \autoref{f:gaia_actions}f and \autoref{f:gaia_age_zVz}a,b, we see a remarkable correlation. The innermost spiral is seen most clearly in dwarfs younger than 3 Gyr, and these stars are on near circular orbits consistent with their age. Their guiding radii lie inside of the Solar Circle. Thus, the metal-rich stars in GALAH must also have migrated outwards from the inner disc where they were formed over the past few billion years. Consistent with that picture, most migration models \citep{Roskar2012,VeraCiro2014,Daniel2018,Minchev2018} find that stars on circular orbits and close to the plane are those that move most easily, and this is what we see \citep[cf.][]{Solway2012}.

It is therefore interesting to consider the origin of metal-rich stars that are {\it not} on circular orbits in the solar neighbourhood. Such stars are observed to exist in the GALAH sample within both the $\alpha$-rich and $\alpha$-poor disks (\autoref{f:gaia_actions_feh}).
These stars appear to be older than the circularized population and conceivably require another transport process (see Sec. 3.1), or a more nuanced understanding of the \cite{Sellwood2002} proposal.

A star in a near-circular orbit is captured in a spiral-arm corotation resonance \citep{Daniel2015} and moves back and forward along the Jacobi integral line
$E_J = E - \Omega.L_z$  tangent to the circular orbit curve in the ($E,L_z$)
plane \citep[see Fig. 1 in][]{Sellwood2002}.
The star is released from this line when the spiral arm fades and the resonance weakens.  
If the star is trapped in a strong corotation resonance, then it can
move far from the circular orbit curve along the Jacobi integral line. If it
is released far from the circular orbit curve, then it is now in a non-circular orbit and will not get picked up by later spiral arm corotation resonances.  Its radial migration history has ended.  

On the other hand, if the star is released close to the circular orbit line,
though at a changed $L_z$, then it is again in a near-circular orbit and can be
picked up by later corotation resonances.  Hence the radial migration proceeds through a
series of small steps in $L_z$ from one near-circular orbit to the next.  If this sequence of
near-circular orbits is broken, the star is deposited in an eccentric orbit and stops migrating. Thus stars can migrate far in many steps but only if they never stray far from circular orbits.

We therefore expect some metal-rich stars that have migrated out from the
inner galaxy to be in eccentric orbits.  They are the ones whose radial migration was derailed after a few steps by being released into
an eccentric orbit.  A succession of weak spiral arms would favour migration via small steps between near-circular orbits.  One strong spiral arm could break the migration for some of the stars.

If this is right, then contrary to the prevailing paradigm, radial migration can lead
to some heating, but at the expense of terminating the migration for the
heated star. The energy for the heating comes from the spiral arms
which in turn feed on the Galactic differential rotation. A star captured
from a near-circular orbit and released far from the circular orbit curve in ($E,L_z$)
has acquired an increased epicyclic amplitude (radial action) and a
large change (positive or negative) in its angular momentum.  There
could be stars from the inner galaxy that now have $L_z > 1$ and large epicyclic amplitudes.

One mechanism that does not rely on transience was proposed by \citet{Minchev10a} and \citet{Minchev11a}: this quasi-chaotic, unruly process results from interference between resonances from multiple rotating patterns, e.g., the central bar and the spiral arms which are known to rotate with different pattern speeds (cf.~\citealt{Brunetti11a}). \citet{Jilkova12a} and \citet{Quillen18a} investigate this resonance overlap and find that outward migration is possible but relatively inefficient. External influences can also drive radial migration, in particular, radial in-plane orbiting galaxies that come close enough to strongly perturb the disk \citep{Quillen09a}. Thus, the ratio of metal-rich stars on radial compared to circular orbits in the local neighbourhood is a powerful constraint on the different transport mechanisms at play.

\section{$V_R$ and the velocity ellipsoid}
\label{s:v_ellip}

\begin{figure}
\centerline{\includegraphics[width=.95\hsize]{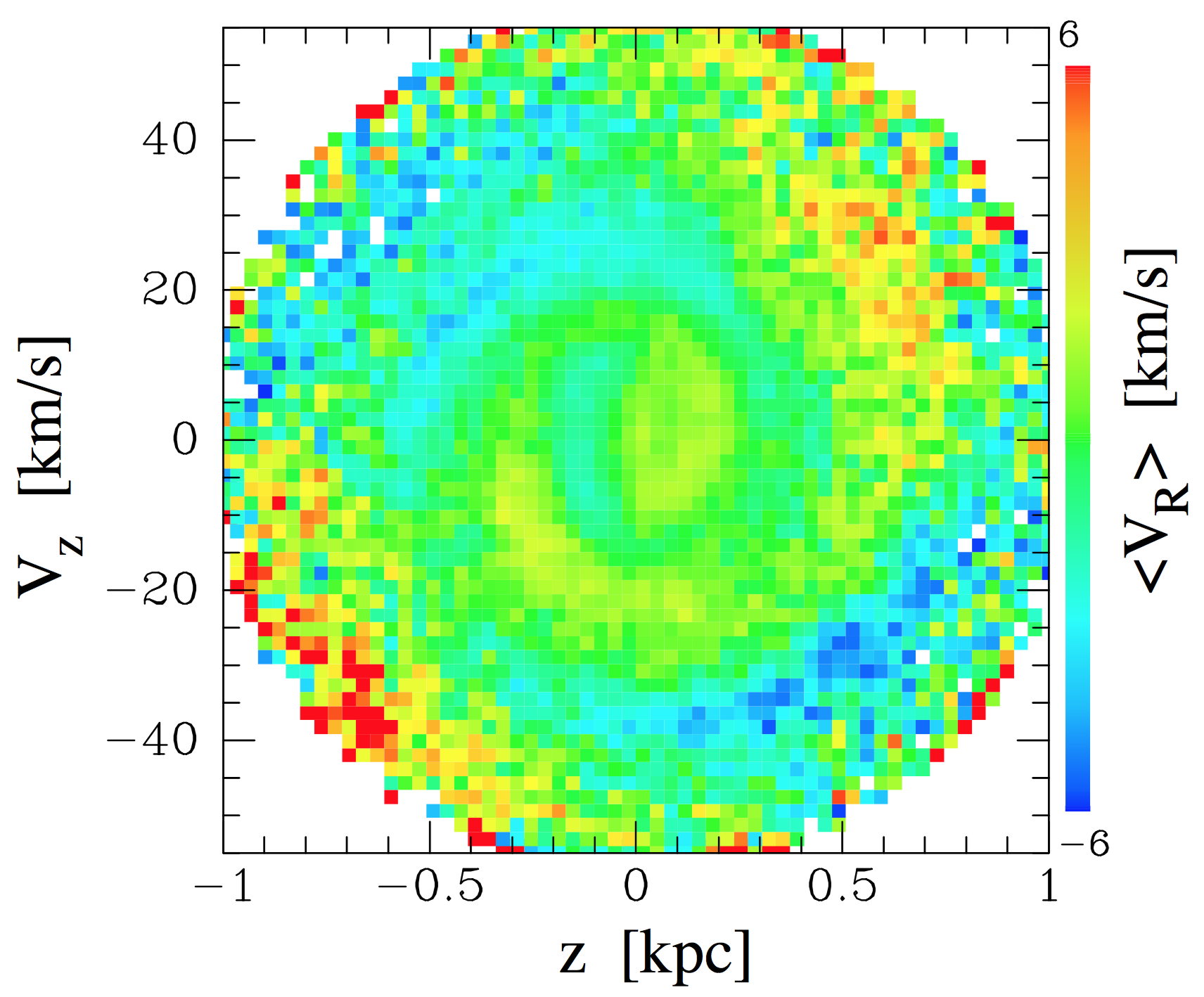}}
\vskip 0.4cm
\centerline{\includegraphics[width=.95\hsize]{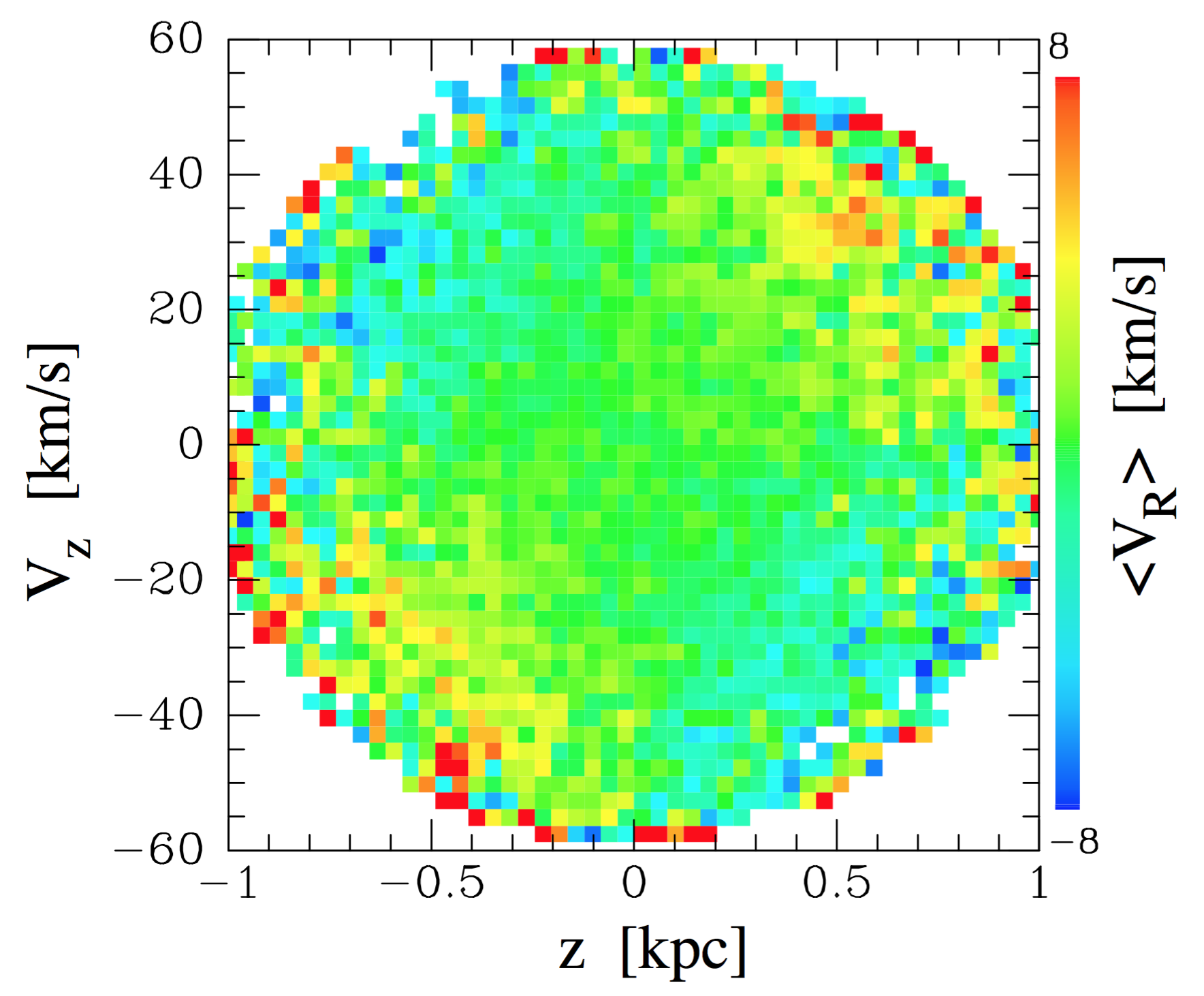}}
\caption{Top: $\ex{V_R}$ plotted in the $zV_z$ phase plane over the original volume explored by \citet{antoja2018}. A quadrupole pattern (blue top left and bottom right, red bottom left and top right) dominates the outer portion of the circle. Towards the centre, the spiral is clearly visible. Bottom: The simulated $zV_z$ plane in a
realistic axisymmetric model Galaxy (with no perturbing force) using 10$^6$ particles with sampling to match the \Gaia\ selection function. The same quadrupole is apparent and arises because of the tilt of the velocity ellipsoid.
}\label{f:jjbVR}
\end{figure}

\subsection{Tilt of the velocity ellipsoid}

The top panel of \autoref{f:jjbVR} presents $\ex{V_R}$ in the $zV_z$ phase plane as originally noted by \cite{antoja2018}.
On the largest scale, a quadrupole pattern is evident, with $\ex{V_R}$ becoming small at top left and bottom right, and large at top right and bottom left. The bottom panel shows the structure of the $zV_z$ plane 
in a realistic axisymmetric model Galaxy: the same quadrupole is evident, so the quadrupole is expected in a fully phase-mixed Galaxy.

The lower panel in \autoref{f:jjbVR} was created using the machinery described in \cite{Binney2018} and the {\tt AGAMA} software package \citep{Vasiliev2018}. The dark matter, bulge and stellar halo were each assigned a distribution function (DF) of the form $f(J)$ introduced by \cite{Posti2018}. The thin and thick discs were assigned DFs $f(J)$ with forms that will be presented in an upcoming paper (Binney \& Vasiliev 2019, in preparation). 
For given values of the parameters in the DFs, the model's potential was solved for iteratively after adding the potential of the gas disc assumed by \cite{Piffl2014}. The parameters in the DF were fitted to the \Gaia\ DR2 sample via pseudo-data created by Monte-Carlo sampling the model using a selection function that declines exponentially with distance from the Sun with a scale length of 500 pc, which is a simple approximation to the selection function of the RVS sample. From these samples, velocity histograms were computed at 35 locations around the Sun and matched to the corresponding histograms for stars with parallax errors of 20\% or less. These Monte Carlo samples are used to construct the plot of $\langle V_R\rangle$ in the $zV_z$ plane shown here.

\begin{figure}
\includegraphics[width=0.48\textwidth]{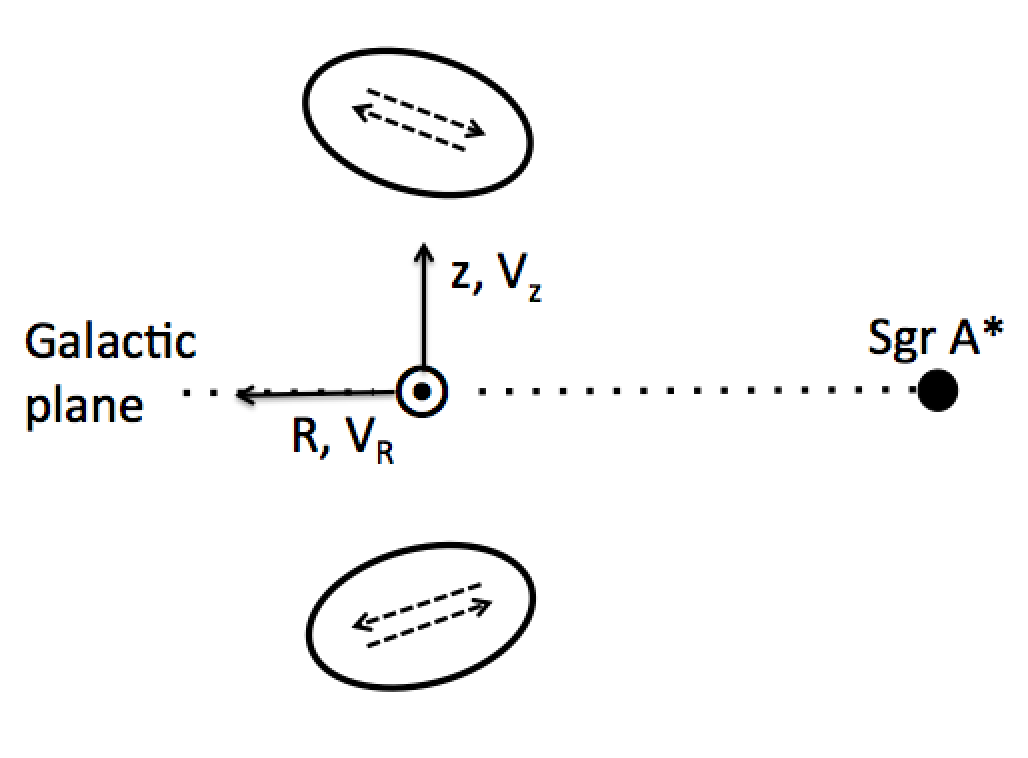}
  \caption{(Left) The adopted coordinate frame in Galactic cylindrical coordinates shown with respect to the Sun's position. The tilted velocity ellipsoids above and below the Galactic plane are also indicated. The dashed arrows show the direction of the fastest motions within the ellipsoid. When viewed from the Sun, the velocity ellipsoids produce a characteristic quadrupole $\mp\pm$ pattern seen in the 
$\ex{V_R(z,V_z)}$ plane from the GALAH and \Gaia\ data.
  }
  \label{f:v_ellip}
\end{figure}

\autoref{f:v_ellip} explains the connection between the quadrupole and the well-known tilt of the velocity ellipsoid as one goes above
or below the plane: at $z=0$, the long axis of the velocity ellipsoid lies in the Galactic plane. Off the plane, the long axis tips
almost as much as is required for it to continue to point towards the
Galactic centre \citep{Siebert2008,Binney2014}. As a consequence, at
locations above the plane, when a star is moving upwards (so it contributes
to the
upper-right quadrant in \autoref{f:jjbVR}), it is more likely to be moving
outwards than inwards and averaging over stars we get red hues in
\autoref{f:jjbVR}. Conversely, at locations below the plane, an upwards
moving star (contributing to the upper left quadrant of \autoref{f:jjbVR}) is more likely to be moving outwards than inwards, and overall we have blue hues. Hence the quadrupole shown by
\autoref{f:jjbVR} in the \Gaia\ DR2 stars is a novel signature of the velocity ellipsoid's tilt. It should have been anticipated but seems not to have been.

\subsection{The spiral in $V_R$}

\begin{figure}
\centering \includegraphics[width=0.48\textwidth]
{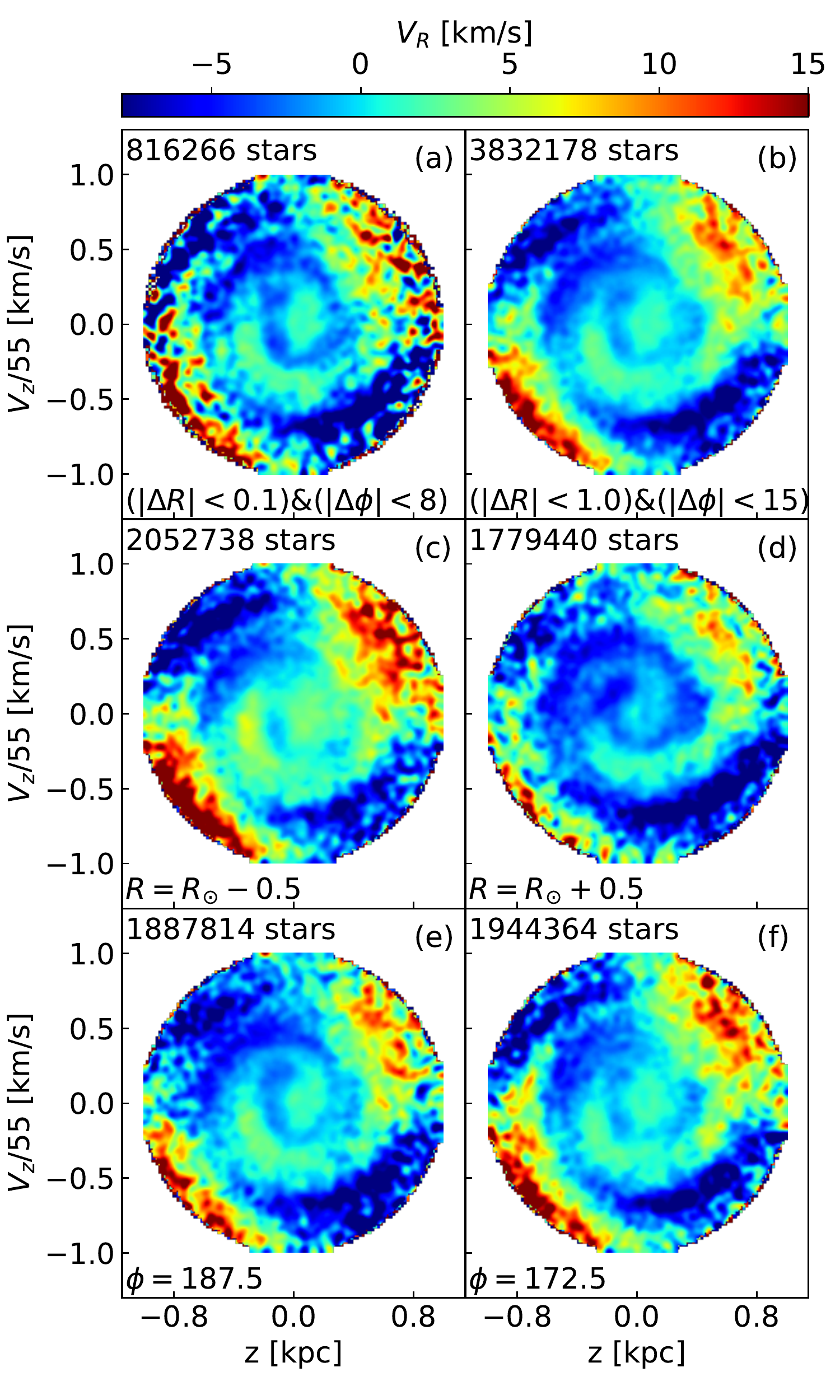}
\caption{Map of median $V_{R}$ in $(z,V_z)$ plane for stars in the solar neighborhood using data from \Gaia\  DR2. The sample definitions are the same as in \autoref{f:gaia_maps3}. The quadrupole defined by the sign changes in each quadrant are due to the tilt of the velocity ellipsoid (see the sketch in \autoref{f:v_ellip}). Note that a weak phase spiral pattern is visible in the center. The pattern is stronger for the $R=\Rsolar-0.5$ case than for the $R=\Rsolar+0.5$ case. ($\phi$ is quoted in degrees.)
\label{f:gaia_map_vr}}
\end{figure}

The most prominent feature of \autoref{f:jjbVR} is a spiral observed in $\ex{V_R(z,V_z)}$ 
that is broadly similar to that in $\ex{V_\phi}$ (\autoref{f:gaia_maps4} and
\autoref{f:jjbJplane}). In detail, the spirals differ: near the centre the $V_R$ spiral is less tightly wound than the $V_\phi$ spiral.
In \autoref{f:gaia_map_vr}, we dissect the plane along the same lines as \autoref{f:gaia_maps3}.
Once again, the $V_R$ spiral pattern in Antoja's original volume (a) is recovered in the 20$\times$ larger GALAH volume in (b). Interestingly, the inner spiral does not vary greatly with location, either in radius (c,d) or in azimuth (e,f). But
the quadrupole pattern is substantially stronger over the inner disc compared to the outer disc; there is no gradient in azimuth. These are real variations as reflected in the matched star counts in each panel.

\begin{figure}
\centering \includegraphics[width=0.48\textwidth]
{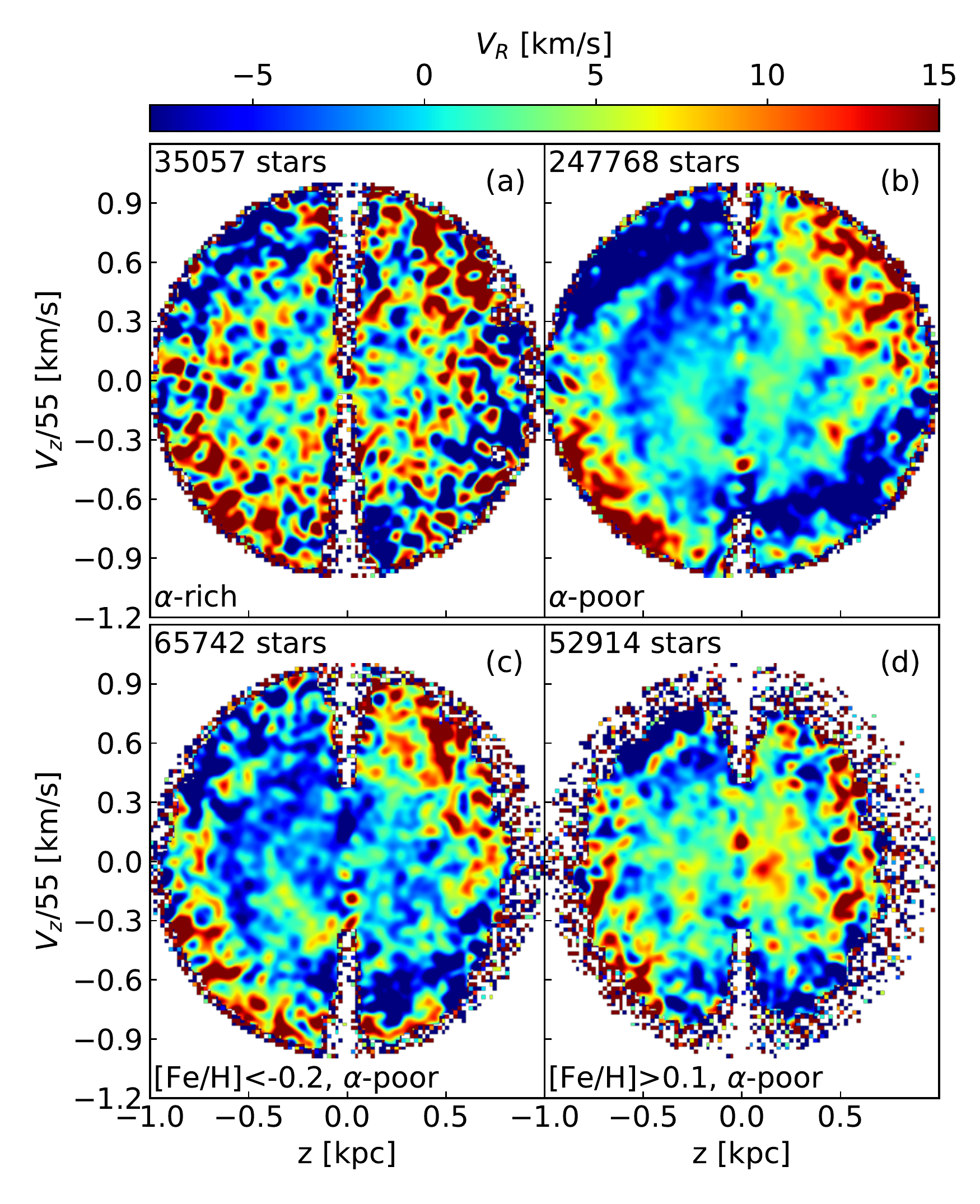}
\caption{Map of median $V_{R}$ in  $(z,V_z)$ plane for stars in the solar neighborhood using data from GALAH and \Gaia\  DR2. The sample definitions are the same as in \autoref{f:gaia_galah_maps3}.
}
\label{f:gaia_galah_map_vr}
\end{figure}

In \autoref{f:gaia_galah_map_vr}, we use the GALAH data to dissect the phase
spiral in $V_R$ by chemistry.  The quadrupole pattern is evident in all
panels and tends to obscure the phase spiral. The latter is most evident in
the panel for the summed $\alpha$-poor disc (b). Thus the spiral in $\ex{V_R}$ is manifested in the same populations as the spiral in
$\ex{V_\phi}$ as we would expect if it is simply another aspect of a common
dynamical phenomenon \citep{Binney2018}.

\section{Dynamical interpretation}
\label{s:dynamics}

\subsection{Departures from equilibrium}

In Sections \ref{s:spiral_phase} and \ref{s:v_ellip}, we encountered
abundant evidence that the phase spiral is present over a wide extent in radius and involves stars of many ages and
chemical compositions. Hence it definitely does not arise through phase
mixing of stars born in a massive starburst \citep{Candlish2014}.
Rather it is an extension to the $zV_z$ plane of one of the most important discoveries
in {\it Hipparcos} data: the detection by \cite{Dehnen1998} of clumps in the $UV$
plane. The traditional explanation of such `star streams' is that they are
dissolving star clusters, but from the work of \cite{Dehnen1998} and
\cite{Famaey2005}, it became clear that the clumps in the $UV$ plane were
heterogeneous in age and metallicity so they could not comprise stars that
formed together, but must have been swept up by some dynamical process. The
GALAH data establish that the stars that make up the $zV_z$ spiral do not
have a common origin but have been likewise caught up in a dynamical process.

In addition to the in-plane clumps, 
early evidence for departures from equilibrium came from the Galactic outer disc. Here, warps, rings and ripples are observed \citep[e.g.][]{Newberg2002,RochaPinto2004,Newberg2007} and these have been interpreted in terms of corrugations and waves triggered by a passing satellite \citep[e.g.][]{Quillen09a,Purcell2011,Widrow2012,Gomez2013,delaVega2015}.
These also manifest as kinematic \citep[][]{Widrow2012,Williams2013,Carlin2013} and density asymmetries \citep[][]{Yanny2013,Slater2014,Xu2015}
above and below the Galactic plane. While most authors conclude that the wave-like patterns can be triggered by a massive perturber like the Sgr dwarf \citep[cf.][]{Elmegreen1995}, none of the models predicted the ``phase spiral'' discovered by \citet{antoja2018} that are a natural consequence of the disturbance, as we discuss.

\subsection{Perturber models}

To a first approximation, the Galactic disc is an axisymmetric equilibrium
structure. {\it The data now at hand force us to move on from this starting point
to consider the effects on non-axisymmetric and non-stationary perturbations.}

The earlier perturber models provided the context for interpreting the \Gaia\ discovery.
\citet{antoja2018} show examples of an initial clump of stars in the $zV_z$
plane winding up into a phase spiral as they move in a toy anharmonic
potential $\Phi(z)=az^2+bz^4$. \cite{Binney2018} pointed out that this model
does not explain why the spiral is weak in a plot of the density of
Gaia DR2 stars in the $zV_z$ plane but emerges clearly when $\ex{V_\phi}$ is
plotted. They argued that the keys to understanding this phenomenon are (i)
that $\Omega_z$ is a strong function of $L_z$ as well as of $J_z$ so stars are
sorted by $L_z$ as they progress around the $zV_z$ plane, and (ii) that when
an intruder such as the Sgr dwarf crosses the plane, oscillations are
initialised in both the in-plane and perpendicular directions. That is, in the Antoja model, the observed spiral is the result of {\it two independent but synchronised oscillations}.

\cite{Binney2018} presented a toy model of the impact of an intruder crossing
the disc. They made the model tractable by using the impulse approximation to
compute the disturbance that the intruder causes.  While recognising that the
impulse approximation would in this case be invalid, they argued that errors
introduced by it are smaller than those introduced by neglect of the
perturbations to the Galaxy's gravitational potential that arise as the disc
responds to its initial stimulus. Notwithstanding its weaknesses, their toy
model reproduced for the first time spirals in $\ex{V_\phi}$ and $\ex{V_R}$
with plausible parameters for the mass and location of the intruder.

The deficiencies of the perturbative treatment 
cannot be addressed until there is a major breakthrough in the theory of disc dynamics.
It is intuitively clear that if we could solve the equations of linearised
perturbation for a self-gravitating stellar disc, at least two wave modes
would emerge: in one mode the disturbance would be largely parallel to the
plane and be associated with spiral structure, and in the other mode the
perturbation would be largely perpendicular to the plane and would be
associated with warps and corrugation waves.

\subsection{Revisiting disc dynamics}
After decades of frustration, we now have in the work of 
\cite{GoldreichDLB1965}, \citet{ToomreGpV1969,Toomre1981},
\cite{SellwoodC2014} and \cite{Fouvry2015} a convincing theory of the
dynamics of razor-thin discs. Noise from any source, including Poisson noise,
generates leading spiral waves, which are swing amplified near their
corotation resonance and subsequently absorbed by Landau damping at a Lindblad
resonance.  The disc is heated by the absorption in an annulus that is
typically very narrow, and as a consequence the impedance of the disc to
propagating spiral waves is caused to vary on small scales. When a wave
propagating from its corotation to its Lindblad resonance subsequently hits
such a region, it is partially reflected back to corotation to be
re-amplified. Gradually, as the disc ages and the number of these narrow regions
grows, significant fractions of swing-amplified waves are reflected by some
feature back to corotation to be re-amplified before they can reach their
Lindblad resonance and be absorbed. Hence the disc's responsiveness to
stimulation by noise steadily grows until the disc becomes simply unstable.
At that point the spiral structure becomes an O(1) phenomenon and a bar forms.

The picture just described certainly marks a significant step forward
in understanding galaxies, but it falls short of what is required to address
the data we now have because it is confined to razor-thin discs. It is clear
that the fundamentally in-plane mode must involve $V_z$ in addition to
$V_\phi$ because stars will be drawn down to regions of high density \citep{Masset1997}. That
is, a propagating spiral arm will force oscillations perpendicular to the
disc that satisfy the symmetry condition
$\ex{V_z(-z)}=-\ex{V_z(z)}$. These motions will remain conjectural until the theory of
spiral structure has been extended from razor-thin discs, in which vertical
motion is impossible, to discs of finite thickness; this extension  proves
extremely difficult \citep{FouvryPichon2017}.

The available formalism relating to the second kind of mode, corrugation waves, is even
more primitive than the current theory of spiral structure because it involves
neglecting epicyclic oscillations in addition to taking the disc to be razor
thin \citep{HunterT1969}. Hence we really have very little idea what a
proper theory of corrugation waves would look like. We do, however, know that their $z$-motions
would satisfy the symmetry condition $\ex{V_z(-z)}=\ex{V_z(z)}$ and they will almost
certainly involve $V_\phi$ in addition to $V_z$ because warps are all about
torques exerted by one ring on another.

In these circumstances, the natural thing to do is to resort to N-body
simulation. As we shall see, modelling the \Gaia\ DR2 data in this way is
extremely challenging because one needs to achieve high resolution in the
small part of the 6D phase space in which the spiral is detected, while at
the same time resolving the dynamic (live) dark halo, which we expect to participate in the
excitations under study \citep{Binney1998} and extends to beyond $100\kpc$, sufficiently to
prevent it becoming an important source of artificial Poisson noise.

We now revisit recent work on Sgr's interaction with the Galactic disc by carrying out a new suite of simulations. Our goal is to understand how the phase spiral can inform us of when this event happened, and how the disturbance was able to propagate through the disc. We consider the different disc response to a purely impulsive interaction (high speed, hyperbolic orbit) and the multi-crossing ``trefoil'' orbit of the Sgr dwarf. 

\begin{figure*}
\includegraphics[width=14cm]{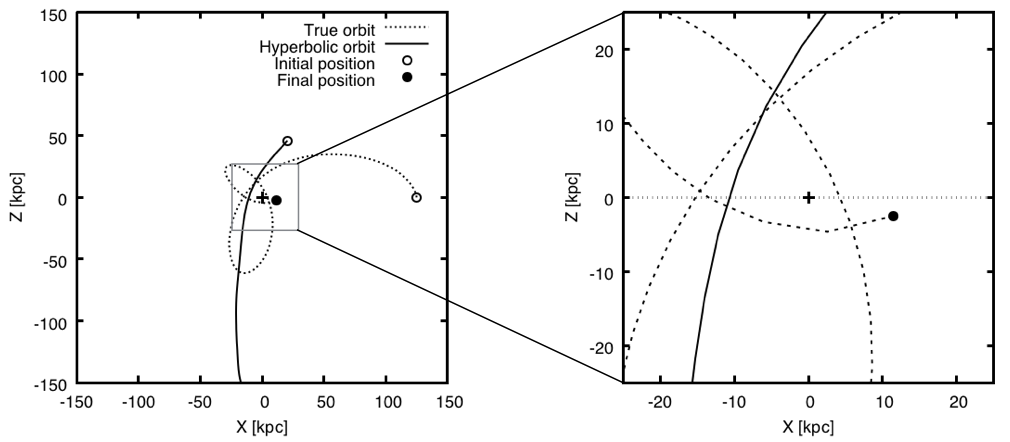}
\caption{The projection of the polar orbits for the intruders with different mass (hyperbolic vs. true) into the ($x,y$) plane seen from the NGP. The cross indicates the Galactic Centre. The dashed line shows the complex `trefoil' orbit of Sgr over the past 2.3 Gyr, now accepted across many studies since \citet{Law2005}. Sgr (shown as a filled circle) crossed the disc about 420 Myr ago at $R=13$ kpc and is due to transit again in about 50 Myr. The inset magnifies the central region. For more details, see \citet{TepperGarcia2018}.} 
\label{f:orbits}
\end{figure*}

\begin{table}
\begin{center}
\caption{Galaxy model parameters. Column headers are as follows: $M_{\rm t}$ := total mass ($10^{9}$ \Msun); $r_{\rm s}$ := scalelength (kpc); $r_{\rm tr}$ :=  truncation radius (kpc); $N_{\rm p}$ := number of particles ($10^{5}$).}
\label{t:Galaxy}
\begin{tabular}{lccccc}
\hline
\hline
 				& Profile	& $M_{\rm t}$ 	&  $r_{\rm s}$ 		& $r_{\rm tr}$ 	& $N_{\rm p}$  ~\\
\hline
Galaxy\\
\hline
DM halo			& H			& $10^3$		& 38.4		& 250	& 	10	~\\
Bulge			& H			& 9				& 0.7		& 4		&	3	~\\
Thick disc		& MN		& 20			& 5.0$^{a}$	& 20	&	6	~\\
Thin disc		& Exp/Sech	& 28			& 5.0$^{b}$	& 20	&	10 ~\\
\end{tabular}
\end{center}
\begin{list}{}{}
\item {\em Notes}. H := \citet{her90a} profile; MN := \citet{miy75a} profile; Exp := radial exponential profile.; Sech := vertical ${\rm sech^2} z$ profile.\\
$^{a}$scaleheight set to 0.5 kpc.\\
$^{b}$scaleheight set to 0.3 kpc.
 \end{list}
\end{table}
\begin{table}
\begin{center}
\caption{Overview of intruder models. Column headers are as follows: $M_{\rm
tot}$ := total mass ($10^{9}$ \Msun); $M_{\rm tid}$ := tidal mass ($10^{9}$
\Msun); $r_{\rm tr}$ := truncation radius (kpc); $N_{\rm p}$ := number of
particles ($10^{5}$). The last column gives the approximate initial orbital
speed ($\!\Kms$). See the notes below the table for more information.}
\label{t:intruder}
\begin{tabular}{lccccc}
\hline
\hline
Model						& $M_{\rm tot}$ 	& $M_{\rm tid}$ 	& $r_{\rm tr}$ 	& $N_{\rm p}$  & $v_0$~\\
\hline
K (high mass, Sgr orbit)				& 100				& 90				& 60			& 5				& 80	\\
L (intermed. mass, Sgr orbit)		& 50				& 40				& 45			& 5				& 80	\\
M (low mass, Sgr orbit)					& 10				& 7					& 25			& 5				& 80	\\
R (high mass, one transit)				& 100				& 60				& 24			& 2				& 370	\\
S (intermed. mass, one transit)		& 50				& 30				& 19			& 1				& 360	\\
T (low mass, one transit)				& 10				& 5					& 12			& 1				& 350	\\
\end{tabular}
\end{center}
\end{table}

\begin{figure*}
\centering \includegraphics[width=0.98\textwidth]
{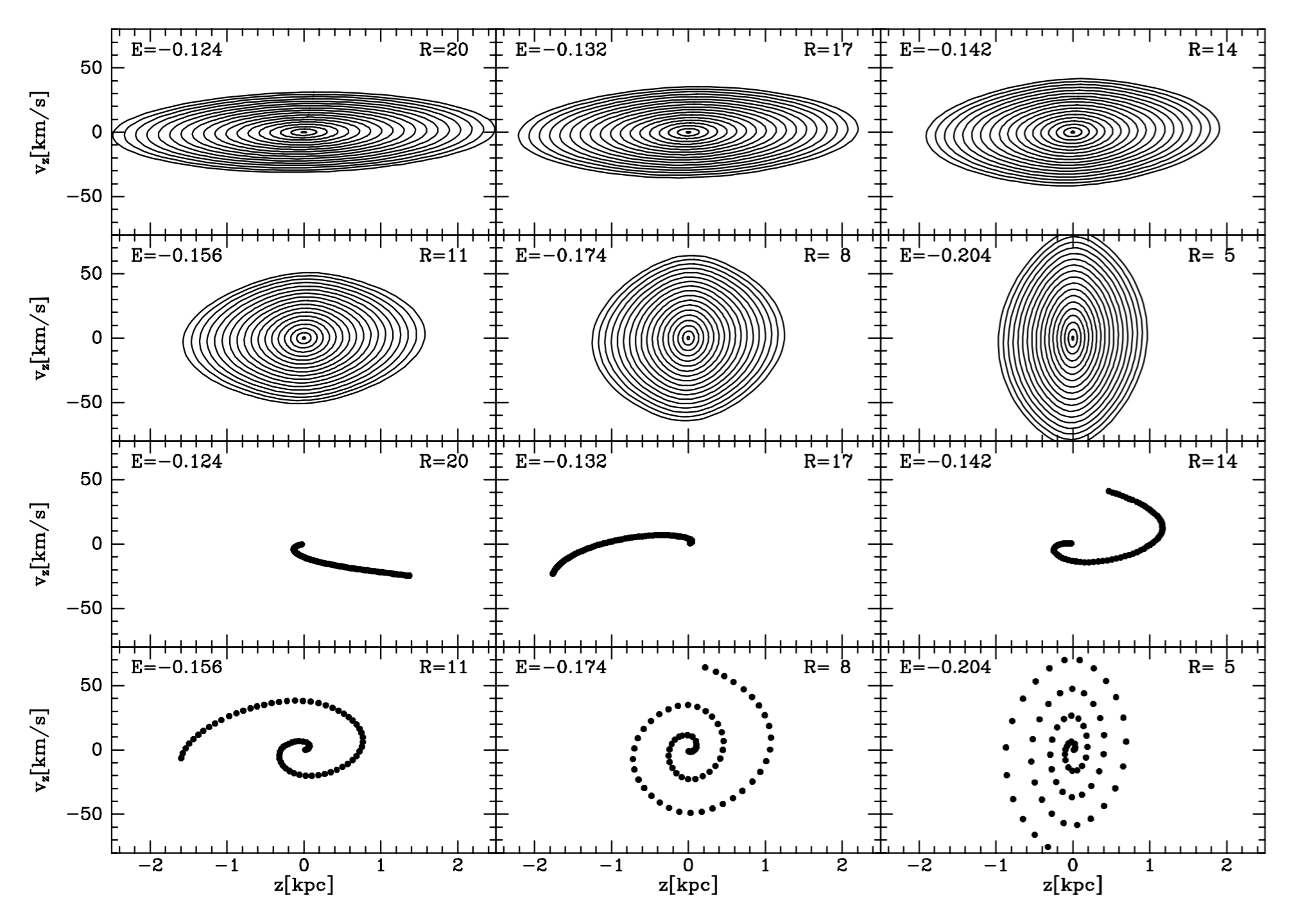}
\caption{
The top 6 panels show surfaces of section in the vertical phase plane ($z, V_z$) at six different radii ($R = 20$, 17, 14, 11, 8, 5 kpc) which are chosen to match the radial extent of our new disc simulations and the \Gaia/GALAH study. The panels were computed with the {\tt AGAMA} software package \citep{Vasiliev2018} adopting the Galactic potential from \citet{Piffl2014}. The amplitude of the outer ellipse and the phase spiral at $R=8$ kpc are chosen to match the \Gaia\ data. In every panel, the outermost ellipse has the same area ($=2\pi J_z$) and the orbital energies $E$ are indicated. The $J_z$ values for each concentric ellipse moving outwards are: $0.008, 0.20, 0.65, 1.35, 2.31, 3.52, 4.99, 6.71, 8.69, 10.93, 13.42, 16.17, 19.17, 22.43, 25.94, 29.71, 33.73, 38.01$ kpc km s$^{-1}$.
In the outer disc, the ellipses elongate in $z$ because stars travel farther and slower in the weaker disc potential. All orbits have radial action $J_R = 0.01$ kpc$^2$ Myr$^{-1}$ = 9.78 kpc km s$^{-1}$ and $J_\phi$ for the circular orbit at the quoted radius.
The bottom 6 panels coincide with the panels above indicated by the total energies; these reveal the impact of phase wrapping after 250 Myr where all points are initially lined up along the $z<0$, $V_z=0$ axis.
Across the inner disc, the stronger disc gravity leads to faster vertical oscillations which result in faster phase wrapping.
\label{f:binney}}
\end{figure*}

\section{N-body simulations}
\label{s:models}

In a recent paper, \citet{TepperGarcia2018} examine the impact of Sgr on the
Galaxy with an N-body, hydrodynamical simulation that has the unique feature
of including a gas component demanded by the resolved star-formation history
of the dwarf. This work, which used the adaptive mesh refinement (AMR)
gravito-hydrodynamics code {\tt Ramses} \citep[version 3.0 of the code
described by][]{tey02a}, emphasized that the number of disc crossings and the
timing of Sgr's orbit depend on the rate at which Sgr loses dark and baryonic
mass. Here we re-run these models without the gas component using an updated
Galaxy model (Table~\ref{t:Galaxy}) with extended ranges of intruder mass and
orbit parameters (Table~\ref{t:intruder}). Our simulations contain almost
three million particles, of which $1.6$ million are disc stars and one
million are dark-matter particles. 

We include a simulation of an unperturbed disc to emphasize the long-term stability of our models.
In addition to the realistic Sgr orbit models with their multiple crossings,
we also simulate intruders crossing the disc along (unrealistic) hyperbolic orbits to study the disc's response to a strong, one-time impulsive transit.
The face-on projection of
both orbits shown in \autoref{f:orbits} suggests that with the realistic
orbit successive crossings may influence the coherence and longevity of the
phase spiral. 

Appendix~\ref{s:movies} summarises the content of movies of these simulations that are available at \url{http://www.physics.usyd.edu.au/galah_exp/sp}.
There are movies to show both configuration space and phase space. We practice an open policy on set-up files for all of our N-body $+$ MHD studies to encourage cross checking, and encourage others to do the same.
\label{foo:code}

Below we provide plots in the $zV_z$ phase plane  for particles that are widely
distributed in $L_z$. \autoref{f:binney} helps us to understand the basic structure
of these plots. Its top
six panels show curves of constant $J_z$ in the $zV_z$ plane
for orbits with circular angular momentum and small radial action at radii that decrease from $R=20\kpc$ at top left to $R=5\kpc$ at bottom right. The values of $J_z$ for which curves are plotted are the same in each panel. As
one moves inwards, the curves become steadily more elongated vertically,
while their area, which is equal to $2\pi J_z$, remains constant because the set of $J_z$ values is the same in every panel. These orbit
traces stretch along the $V_z$ axis while shrinking along the $z$ axis in
consequence of growth in the vertical restoring force $K_z$ as $R$ decreases
and the surface density of the disc increases.  The lower six panels of
\autoref{f:binney} show the extent of phase wrapping in a given time
($250\Myr$). As $R$ shrinks, $\Omega_z$ increases and an initially radially-directed straight line in the phase plane wraps more tightly in a given time.

\subsection{Set up}

The simulations' axisymmetric initial conditions were assigned by the technique of
\citet[][]{spr05c} as implemented in the {\sc dice} code \citep[][]{per14c}.
This technique imposes $\sigma_R=\sigma_z$ and in parts of the disc
$\sigma_R$ fell below the value that makes 
\citet[][]{too64a}'s stability parameter
\[
Q_{\star} \equiv  {\sigma_R \, \kappa  \over 3.36 \, G \,
\Sigma_{\star} } = 1.5,
\]
 where $\kappa$ and $\Sigma_{\star}$ are the epicyclic frequency and stellar
surface density.  Where $Q_\star<1.5$, we increased $\sigma_R$ (but not
$\sigma_\phi$) to ensure $Q_{\star}\ge1.5$ everywhere. Since the initial
conditions are for a system slightly out of equilibrium, each simulation was
evolved for roughly 4 Gyr before being disturbed.  We refer to the model
after this relaxation phase as the `unperturbed' model (Model P).  See
Table~\ref{t:Galaxy} for details of this model Galaxy, 

We disturbed this model in six simulations. The intruders had masses of 1, 5
or $10\times10^{10}\msun$ and comprised a $4\times10^8\msun$ stellar system
embedded in a dark halo. Both the stellar system and the dark halo were
truncated Hernquist spheres. Their scale radii were $0.85$ and $10\kpc$,
respectively. The stellar system was truncated at $2.5\kpc$ while the
truncation radius of the dark halo is listed in
Table~\ref{t:intruder}. A simulation with each of these masses was started
with the intruder at $(R,z)=(21,45)\kpc$ on a hyperbolic orbit of
eccentricity $e=1.3$ and pericentre distance $10\kpc$.  For each intruder
mass, a second simulation was started with the intruder located $125\kpc$ from
the Galactic centre on a `trefoil' orbit of the type that roughly fits
observations of the Sgr dwarf \citep[see][and
\autoref{f:orbits}]{TepperGarcia2018}.  The  intruder's tidal
radius was set appropriate to a Galactocentric distance of $50\kpc$ in the
case of a hyperbolic orbit and a distance $125\kpc$ in the case of a trefoil
otbit. The simulations with trefoil orbits
are labelled K, L and M, while those with hyperbolic orbits are labelled R, S
and T. In the most realistic simulation (Model K), the first and second
pericentric passages occurred $\sim 2.5$ and $\sim 1\Gyr$ ago, while the last
passage occurred 420 Myr ago, consistent with observations
\citep[][]{iba97a}.

\subsection{Hyperbolic encounters}\label{s:hyperbol}

In simulations R, S and T with an intruder on a hyperbolic orbit, at $t\sim95\Myr$
the disc moves up towards the approaching intruder and its centre of mass
experiences a recoil. 
\autoref{f:thor1} and \autoref{f:thor2} show, for the high- and
intermediate-mass intruders respectively, the disc in real space (panels a
and b) and in the $zV_z$ phase plane (lower panels) at this time. Numerous
signs of disequilibrium are evident in all panels of both figures. In the top
panels, we see clear $m=1$ asymmetry, including a couple of distinct spiral
arms, and variations in $\ex{V_\phi}$ that have much larger amplitude than
those found in the unperturbed simulation P1. These variations seem to be
associated wih corrugation waves moving through the disc.  In the middle,
phase-plane, panels, wisps can be identified that might be part of phase
spirals. It is remarkable that there are such features since these figures
include particles irrespective of their azimuth or Galactocentric radius, and
it is to be expected that different patterns at widely differing azimuths and
radii would wash each other out more completely than they do. Plots that are
restricted in radius and azimuth lack the resolution required to trace the
spiral.

By $t\sim 130\Myr$, as the intruder crosses the plane at $R\sim13\kpc$, the
entire disc has been shaken. By $t\sim180\Myr$, the interaction has generated
a spiral arm and a strong warp in the outer disc that precesses around the
disc \citep[cf.][]{Gomez2015}. The disc does not begin to fall back down
towards the receding intruder until after $t\sim 400\Myr$.  The disc is strongly
forced by the intruder's tidal field for only $\sim100\Myr$ but its response
persists for the 2 Gyr duration of the simulation.

\subsection{Realistic orbit}

Contemporary models agree that Sgr initially crossed the disc along a
trajectory perpendicular ($i\sim90^\circ$) to the Galactic plane 
\citep[e.g.][]{Law2005,Purcell2015}, but at late times, as the orbit became
circularised by dynamical friction \citep[e.g.][]{Jiang2000}, the trajectory
evolved to one that makes a smaller angle with the disc ($i\lesssim
30^\circ$). As a result of both this change in inclination and the shrinking of
the orbit's semi-major axis, passages through the disc became steadily less
impulsive. The most recent passage occurred at a radius of about $R\approx
13\kpc$.

The left-hand panels of \autoref{f:thor3} show for Model K the real-space (top two
panels) and the phase plane (bottom panel) at $t=1.77\Gyr$, which is $30\Myr$ before the
high-mass intruder crossed the plane. The right-hand panels show the same spaces $90\Myr$ after the transit. In the bottom left panel showing the phase
plane prior to the passage, a phase spiral can be discerned that has
vanished from the bottom right panel for the moment after the passage. Closer
examination of the data plotted in the bottom left panel of \autoref{f:thor3}
reveals three distinct phase spiral patterns arising in three radial bins
($R=17, 15, 12\kpc$) and their axis ratios vary in line with the predictions
of \autoref{f:binney}. The movie K3 shows that a spiral re-appears as the
simulation progresses, so our simulations are consistent with the observed
spiral being generated $\sim400\Myr$ ago as \cite{antoja2018} and
\cite{Binney2018} suggested.

We can gain insight into the total (rather than stellar) mass of the Sgr
dwarf by combining the simulations with \Gaia\  DR2.  The low-mass intruder on a
realistic orbit (Model M) barely ruffles the disc. The high-mass intruder
(Model K), by contrast, produces features with $\zmax \lesssim 5\kpc$ and
$\Vzmax \lesssim 50\Kms$ that exceed the scale of the \Gaia\  features. The
intermediate-mass intruder produces disequilibria of about the required
amplitude at $R=\Rsolar$. This is consistent with the finding of
\cite{Binney2018}, who generated a realistic spiral with an intruder of mass
$2\times10^{10}\msun$ and an impact parameter of $R=20$ kpc.

\section{Conclusions} \label{sec:conclusions}

\subsection{Summary}
We have used the data from the second releases of the \Gaia\ and GALAH
surveys to examine the Galactic discs in a sphere of radius $\sim1\kpc$
around the Sun. The GALAH survey allows us to divide the disc into its two
fundamental components, the
$\alpha$-rich and the $\alpha$-poor discs.  Traditionally these have been
called the thick and thin discs but it is now clear that this terminology can
be confusing. 

The $\alpha$-rich disc is old and largely confined to within
$R_0$. We have shown that its velocity-dispersion tensor has a bias towards
$V_z$ that is unique in the Galaxy. Although the $\alpha$-rich disc must have
formed rather quickly (and before the $\alpha$-poor disc started to form), it
has a complex internal structure, with its more metal-poor stars being on
highly eccentric, low-angular-momentum orbits.

The $\alpha$-poor disc is the accumulation of up to $10\:\Gyr$ of gradually
declining star formation. Its history has bequeathed it a complex internal
structure that differs from that of the $\alpha$-rich disc in two key
respects: (i) stars with less angular momentum tend to be more rather than
less metal-rich; (ii) its stars have larger rather than smaller in-plane
velocity dispersions than the dispersion perpendicular to the plane.

Following the discovery by \cite{antoja2018} of a phase spiral when $\ex{V_\phi}$
is plotted in the $zV_z$ phase plane, we have used \Gaia\ and GALAH to examine
this plane closely. In this plane, stars move on ovals whose area is
proportional to the vertical action $J_z$. Hence $\alpha$-rich stars are
widely dispersed in the $zV_z$ plane whereas $\alpha$-poor stars become less
widely dispersed as [Fe/H] decreases. On account of asymmetric drift,
$\ex{V_\phi}$ decreases as one moves away from the centre of the $zV_z$
plane. On top of this systematic decrease, a spiral is evident in
$\ex{V_\phi}$ that can be fitted with remarkable accuracy by a Archimedean
spiral. Whereas \cite{antoja2018} extracted the spiral from \Gaia\  data for
stars that lie in thin cylindrical shell around the Sun, we find that a
spiral remains prominent and remarkably invariant when this volume is greatly enlarged and shifted in
radius or in azimuth. These findings suggest that the spiral is a global
feature of the disc as it would be if it were excited by a massive halo substructure, particularly in view of the extensive earlier work on warps and ripples in the outer disc.

While the general form of the phase spiral is conserved across the different slices (correcting for the variable aspect ratio), the GALAH data reveal changes in the ``contrast'' (i.e. how well defined the spiral is against the background) in sections along the spiral when the data are sliced by stellar age, action, elemental abundance or location.
The inner spiral is better defined with stars
younger than 3 Gyr and more metal rich ([Fe/H]$>$0.1). In Section 4.5, we discussed this result in the context of stellar migration.

The outer spiral is better defined in older, more metal-poor, $\alpha$-poor stars. There are variations in the phase spiral contrast (coherence) with azimuth and radius. The spiral is hard to discern in the plot for $\alpha$-rich stars. In Section 6.1, we discussed these results in the context of a disc corrugation that is mostly confined to the flaring $\alpha$-poor disc. The spiral can be seen in LAMOST data, but these data do not allow detailed study because the phase spiral has low contrast and the latest data release does not include $\FeA$.
The spiral is clearest in stars with less than the median value of $J_R$. It is somewhat tighter in stars with smaller $L_z$ as one expects from the tendency of the frequency $\Omega_z$ of motion in the $zV_z$ plane to increase as $L_z$ decreases.

\begin{figure*}
\includegraphics[width=0.8\textwidth]{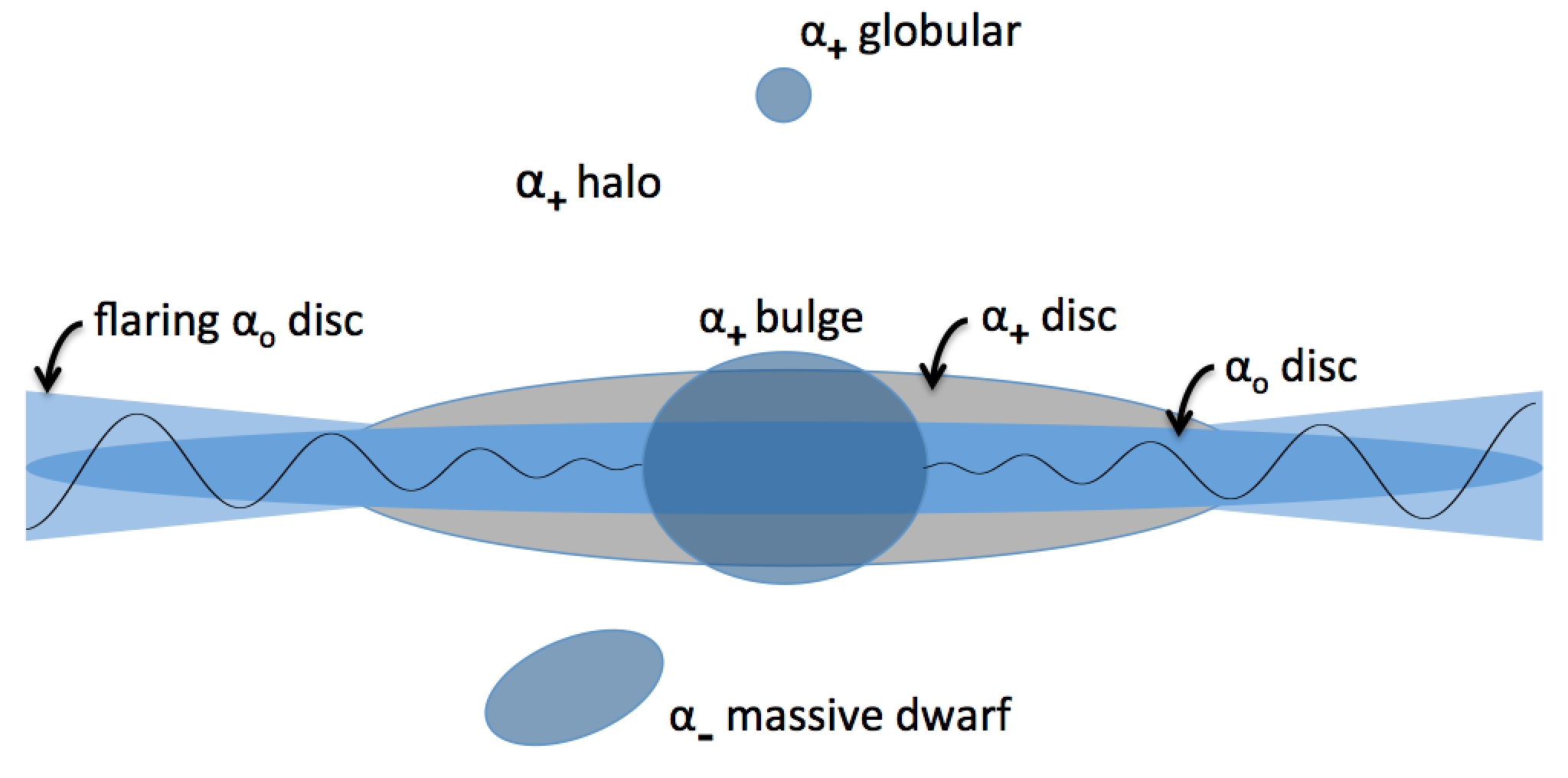}
\caption {A sketch of the corrugated disc that brings together much of what is learnt from the GALAH/\Gaia\ analysis. We also illustrate the use of the new $\alpha$ notation discussed in Sec. 2.1. 
The wave propagation gives rise to the phase spiral. As models show (Sec. 6.1), the corrugation amplitudes grow in $\vert z\vert$ with increasing radius. This can be understood in the context of the increasing vertical phase amplitude with $R$ in 
\autoref{f:binney}. Thus, the corrugation is mostly confined to the $\alpha_{\rm o}$ disc at all radii. This behaviour is also observed in external galaxies (see Sec. 8.2).
\label{f:disc_sketch_ripple}}
\end{figure*}

\subsection{The Galactic disc revisited}

In \autoref{f:disc_sketch_ripple}, we provide a sketch of the corrugated disc that we believe to be consistent with the work above. This simple schematic is remarkably successful in explaining most of the features we see in the GALAH/\Gaia\ analysis. Recall that the outer part of the spiral (large $r_{\rm ps}$) is strongest in old (\autoref{f:gaia_age_zVz}c,d), metal poor stars (\autoref{f:gaia_galah_maps3}c). It is also more prominent at larger Galactocentric radius (\autoref{f:gaia_maps3}c,d), and more evident in the $\alpha$-poor disc compared to the $\alpha$-rich disc (\autoref{f:gaia_galah_maps3}a,b). It also appears stronger at lower $\phi$ (\autoref{f:gaia_maps3}e,f) presumably because these corrugations are not purely radial waves, but exhibit a degree of spirality. We also note that it is stronger for $L_z > 1$, consistent with being more prominent outside of the Solar Circle (\autoref{f:gaia_actions}c). That so many independent data can be marshalled in this way demonstrates the power of combining \Gaia\ data with detailed stellar chemistry from ground-based surveys.

A clear example of a corrugated disc observed in both gas and stars is seen in the ``superthin'' Sd galaxy IC 2233 at a distance of about 10 Mpc \citep{Matthews2008b,Matthews2008a}. The undulations are seen in the spatial offsets of the gas and young stars, and kinematically in the cool and warm gas. There are parallels with what we observe in the Galaxy. Interestingly, the undulation effect is weaker in the old stellar population. Moreover, the wavelength ($\approx 7$ kpc) and amplitude ($\approx 150$ pc) of the corrugation \citep[see Fig. 1 in][]{Matthews2008a} are remarkably similar to the same scalelengths observed in the 
Galaxy \citep[cf. Fig. 18 in][]{Xu2015}, {\it and} they determine that the corrugation amplitude grows with increasing radius, as we do for the Galaxy. At a fixed oscillation energy, the increasing 
sinusoidal amplitude of the oscillation with radius, and its anti-correlation with the sinusoidal velocity amplitude \citep[][]{Matthews2008a}, are well explained by what we learnt from \autoref{f:binney}. 

Do such corrugations exist in the \HI\ gas of the Milky Way? As far as we can tell, this has not been found locally in recent surveys \citep{Kalberla2009} but there may be evidence for such behaviour at other radii \citep[e.g.][]{Lockman2002}. It is tempting to suggest that the tilt of the widely-discussed ``Gould's belt'' of star formation \citep{Zari2018} in the local disc is somehow indicative of a local disc corrugation, by analogy with IC 2233 and other galaxies listed by \cite{Matthews2008a}.

\subsection{Dynamical timescale}

{\it The inferred dynamical timescale for the phase spiral depends critically on the adopted disc potential.} In our final analysis, we adopt the disc potential of \citet{Piffl2014} because this was constrained to fit the well-established, vertical density profile through the Sun's position \citep{Gilmore1983}.
When angle-action coordinates are used to plot the $zV_z$ phase plane, a spiral is evident that coincides well with the curve formed after $390\Myr$ by points that start from a common value of the angle variable $\theta_z$. In contrast, when we adopt the Galactic potential from \citet{McMillan2011}, the inferred timescale is 30\% higher (515 Myr). This suggests that 
the spiral is a relic of a disturbance the disc experienced about 400-500 Myr ago, in line with the numerical simulations.

In Section 5.1, a plot of $\ex{V_R}$ in the $zV_z$ plane proves to be a superposition of two
features: a spiral that is very similar to that seen in the plot for $V_\phi$
superposed on a quadrupole pattern that we have traced to the morphology of
orbits in the $Rz$ meridional plane and the well known tilting of the velocity ellipsoid as one moves away from the plane. The results summarised here establish that the spiral is not associated with the dispersal of
stars from their natal location but, like the structure {\it Hipparcos} revealed in
the $UV$ plane, has arisen through some dynamical process disturbing the Galaxy's equilibrium. 

A strong case has been made by \cite{antoja2018} and \cite{Binney2018} that
the spiral is a consequence of the tidal pull of the Galaxy by a halo
substructure, possibly the Sgr dwarf, as it crossed the plane $\sim0.5\Gyr$
ago. \cite{Binney2018} simulated this process using two undesirable
approximations: that the event was impulsive and that it does not modify the
Galaxy's gravitational field.  Currently these approximations can only be
lifted by doing a full N-body simulation of the event. 

In Section~\ref{s:models} and the Appendix, 
we show results from six such simulations. Even with three million particles, we did not achieve sufficient resolution in an equivalent GALAH volume to reveal the phase spiral locally (i.e. within a few hundred parsecs), as others have claimed with an order of magnitude more particles \citep{Laporte2018}. Instead, we populate the $zV_z$ plane with all stars in a simulation, thus effectively superposing phase planes at all azimuths and many different radii.

Remarkably, in light of this superposition, for intruders of mass
$3\times10^{10}\:\msun$ and above (stripped down from a higher mass), spiral-like ``features" that are consistent with being phase
spirals can still be identified in the simulated phase plane. These
structures lie outside $R_0$, and reach to greater distances from the plane
and larger values of $|v_z|$ than the observed spiral. An intruder of mass $10^{10}\:\msun$ stripped down to $5\times 10^{9}\:\msun$ does not perturb the disc enough to generate measurable phase-plane structure, so we conclude that the observed structure was generated by a body of greater mass. This conclusion is consistent with the findings of \cite{Binney2018} and \cite{Laporte2018}.

Our new simulations reveal the full nature of the interaction between the Galaxy and the  infalling, shredding, multi-transiting dwarf consistent with the well established ``trefoil'' orbit of the perturber. The details of this interaction are crucial. Our simulations are significant in predicting the large-scale patterns of a corrugated disc that are now revealed in the data (Khanna et al. 2019, in preparation). The simulations reveal phase spirals co-existing at different radii due to the disc corrugation, with varying $zV_z$ aspect ratios in line with \autoref{f:binney}. The simulations also reveal for the first time that the subsequent disc crossings destroy the phase spiral coherence, such that what we observe today {\it cannot} be older than the lookback time of the last disc crossing.

\subsection{Future research}

Natural directions for future work include using the distributions of
$\alpha$-rich and $\alpha$-poor stars in action-space shown in \autoref{f:gaia_actions_feh} to
build Extended Distribution Functions (EDF) for the discs \citep{Sanders2015}, and
use these EDFs to construct a self-consistent model Galaxy. Such a model
would predict the density and kinematics of the two discs throughout the
Galaxy, and thus provide a link to other spectroscopic surveys that can
separate the discs at other locations. While the Galaxy is not in equilibrium,
the work of \cite{Binney2018} shows the value of such models as frameworks
within which to model non-equilibrium features such as the phase spiral.

The wealth of evidence that we now have, that the disc near us is vertically
excited, and the strong suspicion that the Sgr dwarf (which was discovered 30 years
after the \HI\ warp!) is the prime cause of this excitation, suggests that
it is time to develop definitive models of both the warp and the dwarf's orbit.
It seems likely that this will be achieved through meticulous and innovative
N-body modelling. This work is of direct relevance to our understanding of the nature of dark matter.
If satisfactory models can be achieved, they must surely
provide conclusive evidence that dark matter, which is dominant in the
regions of interest, absorbs energy and momentum from moving massive
bodies like ordinary matter. Then we would have conclusive evidence that dark
matter comprises non-degenerate fermions
\citep{Boehm2003}
rather than being an artefact
introduced by using the wrong theory of gravity \citep{Milgrom2008} or being a bosonic
condensate of an ultra-light quantum field \cite[q.v.][]{Hui2017}.

\begin{figure*}
\includegraphics[width=8cm]{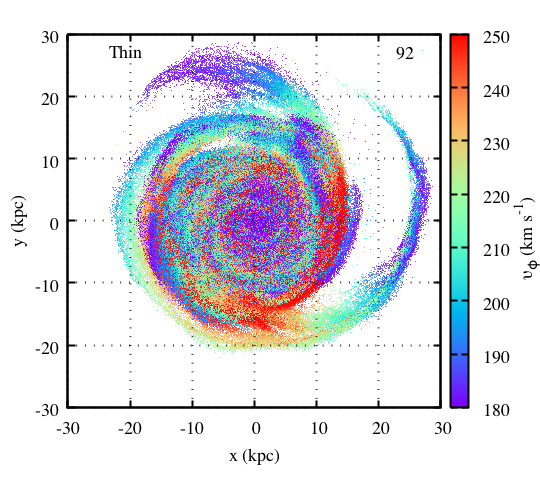}
\includegraphics[width=8cm]{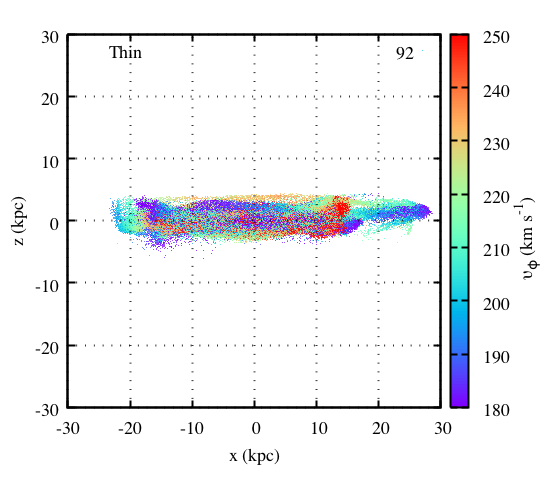}
\includegraphics[width=8cm]{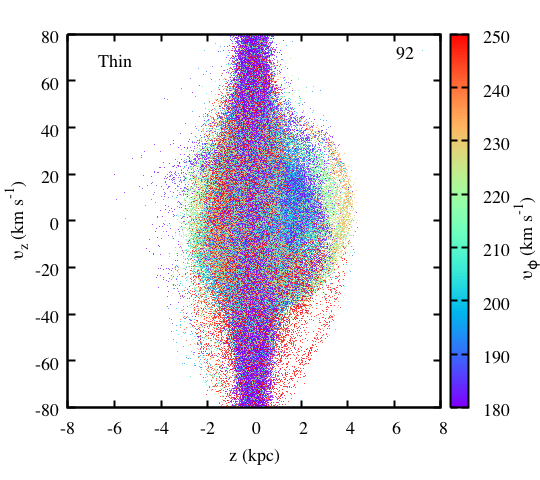}
\includegraphics[width=8cm]{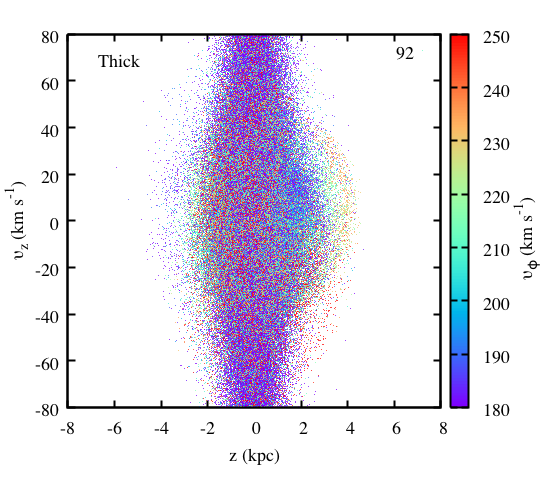}
\includegraphics[width=8cm]{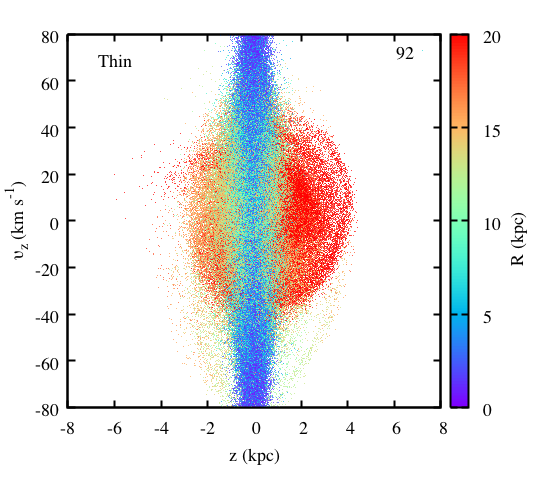}
\includegraphics[width=8cm]{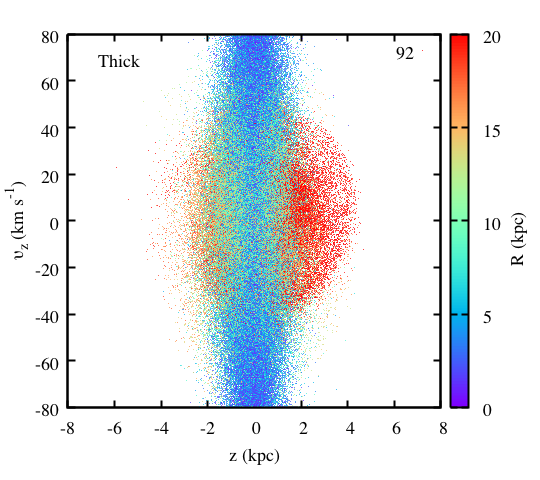}
\caption{Simulation of a high mass intruder (Model R; Table 2) on a hyperbolic orbit where only the Galaxy is shown. (a) $xy$ plane for the thin disc where the particles are colour coded with $V_\phi$ (Movie R2).  (b) $xz$ plane for the thin disc where the particles are colour coded with $V_\phi$ (Movie R2). (c) $zV_z$ phase plane for the thin disc colour coded with $V_\phi$ (Movie R3). (d) $zV_z$ phase plane for the thick disc colour coded with $V_\phi$ (Movie R3).
(e) $zV_z$ phase plane for the thin disc colour coded with $R$ (Movie R4). (f) $zV_z$ phase plane for the thick disc colour coded with $R$ (Movie R4). Note the coherent $V_\phi$ velocity structures, especially in configuration space and their interrelation across all phase planes, mostly due to strong $V_z$ and weaker $V_R$ motions. The thick disc shows the same extent and structure as the thin disc if one allows for order of magnitude fewer particles. In the vertical phase plane $V_\phi(z,V_z)$ in (c) and (e), there are coherent one-armed phase structures occurring at the same time ($t=92$ Myr) near 20 kpc (red) and 10 kpc (green).}
 \label{f:thor1}
\end{figure*}
\newpage
\begin{figure*}
\includegraphics[width=8cm]{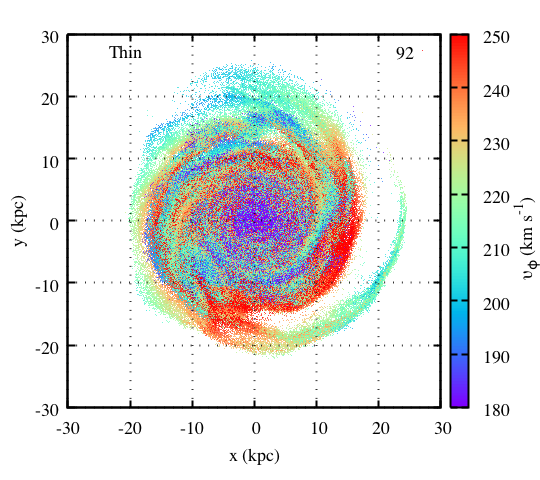}
\includegraphics[width=8cm]{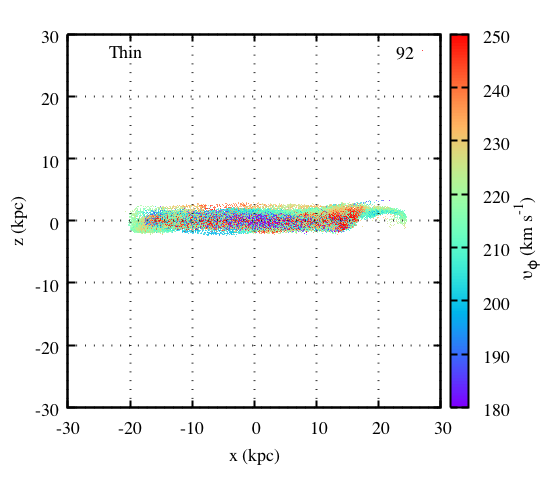}
\includegraphics[width=8cm]{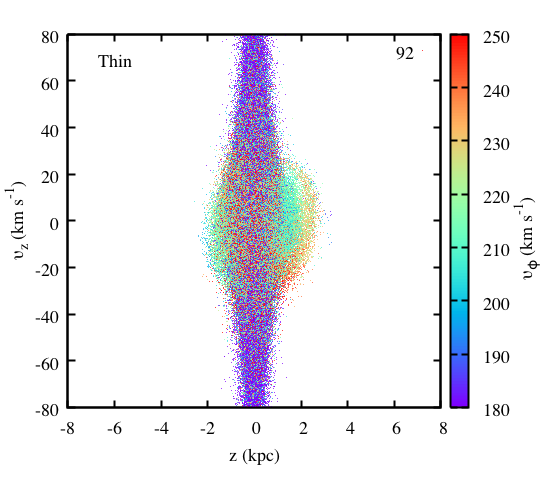}
\includegraphics[width=8cm]{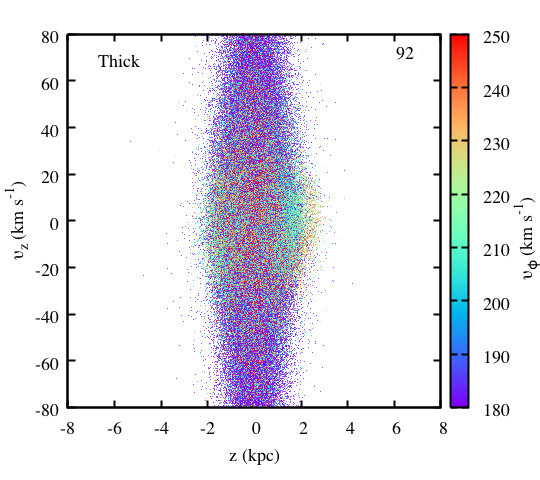}
\includegraphics[width=8cm]{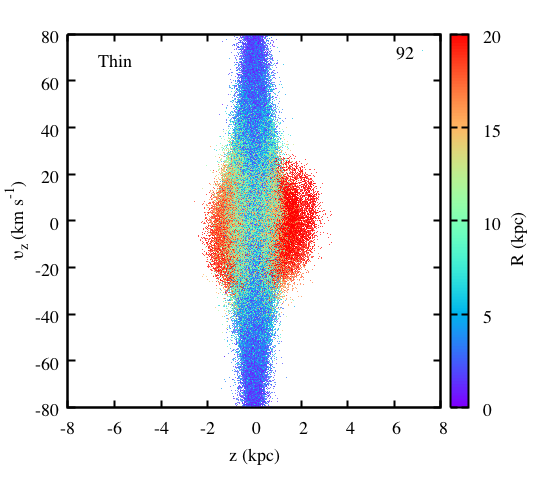}
\includegraphics[width=8cm]{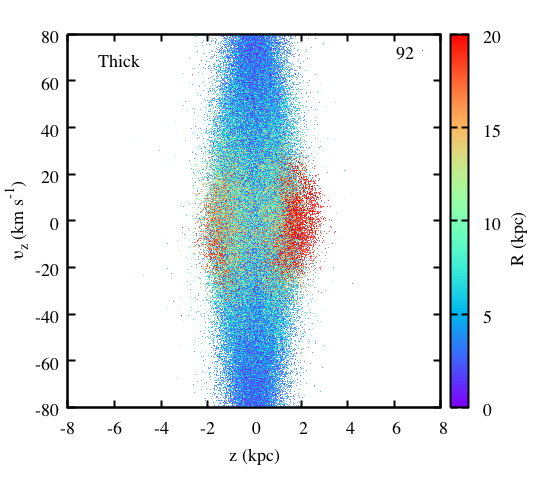}
 \caption{Simulation of an intermediate mass intruder (Model S; Table 2) on a hyperbolic orbit where only the Galaxy is shown. (a) $xy$ plane for the thin disc where the particles are colour coded with $V_\phi$ (Movie S2).  (b) $xz$ plane for the thin disc where the particles are colour coded with $V_\phi$ (Movie S2). (c) $zV_z$ phase plane for the thin disc colour coded with $V_\phi$ (Movie S3). (d) $zV_z$ phase plane for the thick disc colour coded with $V_\phi$ (Movie S3). (e) $zV_z$ phase plane for the thin disc colour coded with $R$ (Movie S4). (f) $zV_z$ phase plane for the thick disc colour coded with $R$ (Movie S4). Once again, there are coherent velocity structures across all phase planes ($t=92$ Myr). Now the physical and kinematic extent have both declined by almost a factor of two, consistent with the lower intruder mass. In (e), the one-armed phase spiral at 20 kpc is still apparent. The inner phase spiral pattern at $R=10$ kpc is no longer visible.} 
 \label{f:thor2}
\end{figure*}
\newpage
\begin{figure*}
\includegraphics[width=8cm]{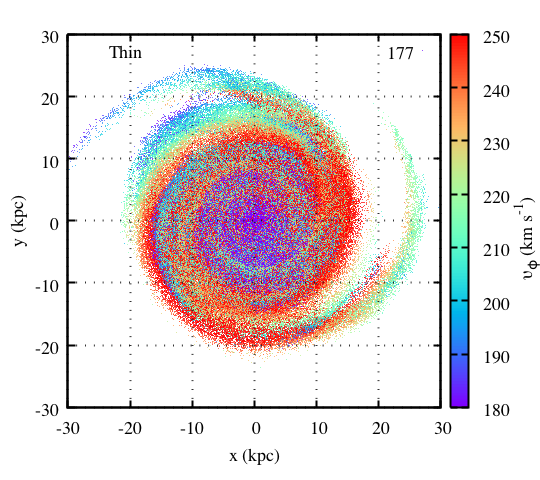}
\includegraphics[width=8cm]{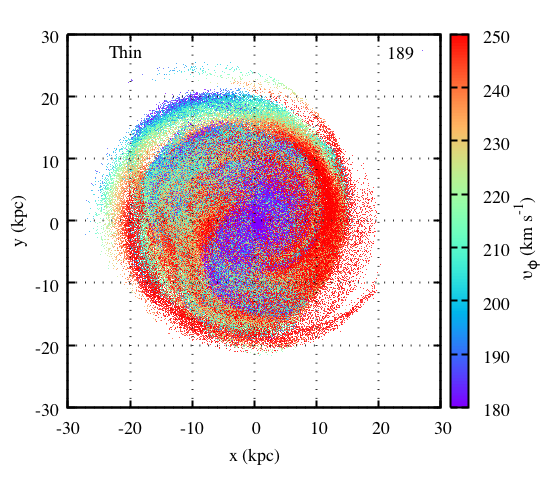}
\includegraphics[width=8cm]{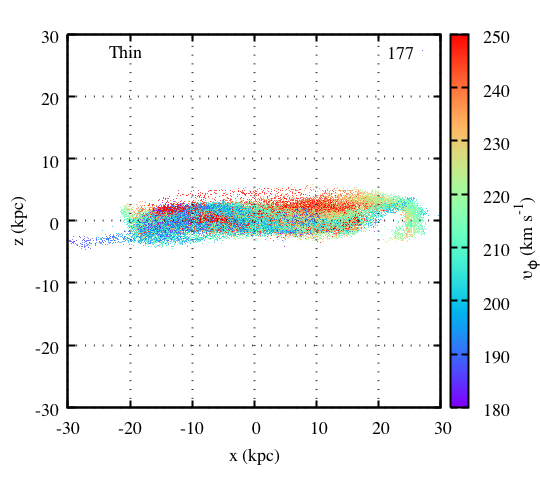}
\includegraphics[width=8cm]{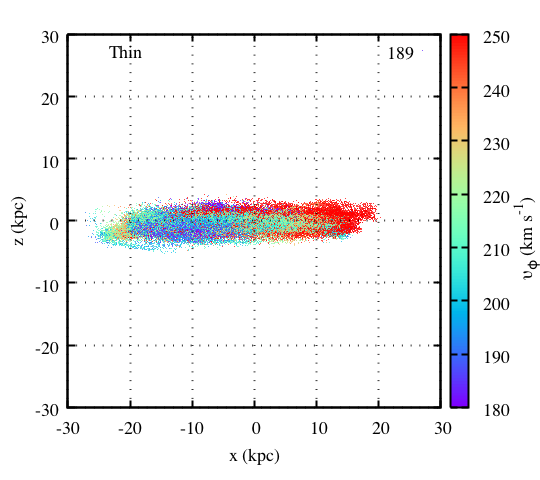}
\includegraphics[width=8cm]{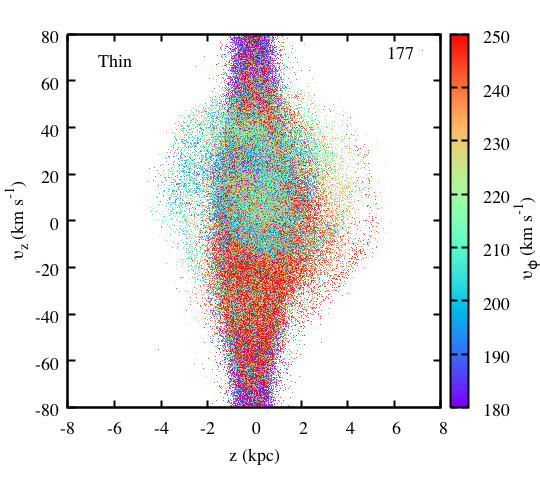}
\includegraphics[width=8cm]{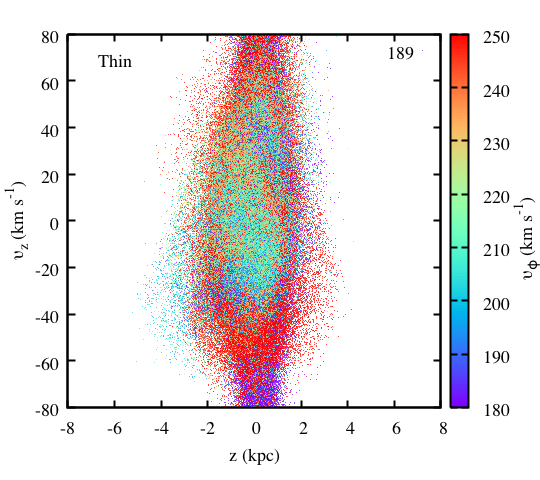}
 \caption{Simulation of a high mass intruder (Model K; Table 2) on a realistic Sgr orbit where only the Galaxy is shown. We present results for just before and just after the last disc transit which occurred at 1.8 Gyr in the simulation. (a) $xy$ plane for the thin disc 30 Myr {\it before} transit where the particles are colour coded with $V_\phi$. (b) same as in (a) but 90 Myr {\it after} the disc crossing. (c) $xz$ plane for the thin disc 30 Myr {\it before} transit where the particles are colour coded with $V_\phi$. (d) same as in (c) but 90 Myr {\it after} the disc crossing. (e) $zV_z$ plane 30 Myr {\it before} the disc crossing. (f) same as (e) but 90 Myr {\it after} the disc crossing. Note how the coherent phase space structures in (e) are wiped out in (f); the extent in $z$ is also compressed right after the disc transit such that the pattern must rebuild from scratch. Weaker related one-armed structures are seen in $V_R(z,V_z)$ as we observe in the accompanying simulations (Movie K5; Table 5).
 } 
 \label{f:thor3}
\end{figure*}

\section*{Acknowledgements}

This work has made use of data from the European Space Agency (ESA) mission
{\it \Gaia\ } (\url{https://www.cosmos.esa.int/gaia}), processed by the {\it \Gaia\ }
Data Processing and Analysis Consortium (DPAC,
\url{https://www.cosmos.esa.int/web/gaia/dpac/consortium}). 
This work is also based on data acquired from the Australian Astronomical Telescope. We acknowledge the traditional owners of the land on which the AAT stands, the Gamilaraay people, and pay our respects to elders past and present.

JBH is supported by an ARC Australian Laureate Fellowship (FL140100278) and the ARC Centre of Excellence for All Sky Astrophysics in 3 Dimensions (ASTRO-3D) through project number CE170100013. SS is funded by a Dean's University Fellowship and through JBH's Laureate Fellowship, which also supports TTG and GDS. MJH is supported by an ASTRO-3D Fellowship. JK is supported by a Discovery Project grant from the Australian Research Council (DP150104667) awarded to JBH. JJB is supported by the UK Science and Technology Facilities Council under grant number ST/N000919/1.

SB and KL acknowledge funds from the Alexander von Humboldt Foundation in the framework of the Sofja Kovalevskaja Award endowed by the Federal Ministry of Education and Research. SB, MA and KL acknowledge travel support from Universities Australia and Deutsche Akademische Austauschdienst. The research by MA, LD, JL, AMA has been supported by an Australian Research Council Laureate Fellowship to MA (grant FL110100012). LD gratefully acknowledges a scholarship from Zonta International District 24 and support from ARC grant DP160103747. LD, KF and YST are grateful for support from Australian Research Council grant DP160103747. KL acknowledges funds from the Swedish Research Council (Grant nr. 2015-00415\_3) and Marie Sklodowska Curie Actions (Cofund Project INCA 600398). SLM acknowledges support from the Australian Research Council through grant DE140100598. TZ acknowledges the financial  support  from  the  Slovenian  Research  Agency  (research core funding No. P1-0188). YST is supported by the Carnegie-Princeton Fellowship, the Chooljian Membership from the Institute for Advanced Study in Princeton and the NASA Hubble Fellowship grant HST-HF2-51425.001 awarded by the Space Telescope Science Institute.

JBH \& TTG acknowledge the Sydney Informatics Hub and the University of Sydney's high performance computing (HPC) cluster Artemis for providing the HPC resources that have contributed to the some of the research results reported within this paper. Parts of this project were undertaken with the assistance of resources and services from the National Computational Infrastructure (NCI), which is supported by the Australian Government.

We acknowledge constructive conversations with Jerry Sellwood, Elena d'Onghia, Ralph Sch\"onrich, Evgeny Vasiliev, Katherine Johnston, Ivan Minchev and Chervin Laporte. We thank Nick Prantzos at the Shanghai Galactic Archaeology meeting (November 2018) for promoting the need for improved nomenclature across the different [$\alpha$/Fe] components, Sten Hasselquist for showing us new data ahead of publication, and Lars Hernquist for suggesting the compact $\alpha$ notation in an open discussion with the GALAH team held at the University of Sydney. We are also indebted to a persistent and insightful referee.

\bibliographystyle{mnras}
\bibliography{joss}

\begin{thebibliography}{}
\makeatletter
\relax
\def\mn@urlcharsother{\let\do\@makeother \do\$\do\&\do\#\do\^\do\_\do\%\do\~}
\def\mn@doi{\begingroup\mn@urlcharsother \@ifnextchar [ {\mn@doi@}
  {\mn@doi@[]}}
\def\mn@doi@[#1]#2{\def\@tempa{#1}\ifx\@tempa\@empty \href
  {http://dx.doi.org/#2} {doi:#2}\else \href {http://dx.doi.org/#2} {#1}\fi
  \endgroup}
\def\mn@eprint#1#2{\mn@eprint@#1:#2::\@nil}
\def\mn@eprint@arXiv#1{\href {http://arxiv.org/abs/#1} {{\tt arXiv:#1}}}
\def\mn@eprint@dblp#1{\href {http://dblp.uni-trier.de/rec/bibtex/#1.xml}
  {dblp:#1}}
\def\mn@eprint@#1:#2:#3:#4\@nil{\def\@tempa {#1}\def\@tempb {#2}\def\@tempc
  {#3}\ifx \@tempc \@empty \let \@tempc \@tempb \let \@tempb \@tempa \fi \ifx
  \@tempb \@empty \def\@tempb {arXiv}\fi \@ifundefined
  {mn@eprint@\@tempb}{\@tempb:\@tempc}{\expandafter \expandafter \csname
  mn@eprint@\@tempb\endcsname \expandafter{\@tempc}}}

\bibitem[\protect\citeauthoryear{{Adibekyan}, {Sousa}, {Santos}, {Delgado
  Mena}, {Gonz{\'a}lez Hern{\'a}ndez}, {Israelian}, {Mayor}  \&
  {Khachatryan}}{{Adibekyan} et~al.}{2012}]{Adibekyan2012}
{Adibekyan} V.~Z.,  {Sousa} S.~G.,  {Santos} N.~C.,  {Delgado Mena} E.,
  {Gonz{\'a}lez Hern{\'a}ndez} J.~I.,  {Israelian} G.,  {Mayor} M.,
  {Khachatryan} G.,  2012, \mn@doi [\aap] {10.1051/0004-6361/201219401}, \href
  {http://adsabs.harvard.edu/abs/2012A%26A...545A..32A} {545, A32}

\bibitem[\protect\citeauthoryear{{Antoja} et~al.,}{{Antoja}
  et~al.}{2018}]{antoja2018}
{Antoja} T.,  et~al., 2018, \mn@doi [\nat] {10.1038/s41586-018-0510-7}, \href
  {http://adsabs.harvard.edu/abs/2018Natur.561..360A} {561, 360}

\bibitem[\protect\citeauthoryear{{Aumer}, {Binney}  \& {Sch{\"o}nrich}}{{Aumer}
  et~al.}{2016}]{AumerBS2016}
{Aumer} M.,  {Binney} J.,   {Sch{\"o}nrich} R.,  2016, \mn@doi [\mnras]
  {10.1093/mnras/stw1639}, \href
  {http://ukads.nottingham.ac.uk/abs/2016MNRAS.462.1697A} {462, 1697}

\bibitem[\protect\citeauthoryear{{Barden} et~al.,}{{Barden}
  et~al.}{2010}]{barden2010}
{Barden} S.~C.,  et~al., 2010, in Ground-based and Airborne Instrumentation for
  Astronomy III. p. 773509, \mn@doi{10.1117/12.856103}

\bibitem[\protect\citeauthoryear{{Bensby}, {Feltzing}  \&
  {Lundstr{\"o}m}}{{Bensby} et~al.}{2003}]{Bensby2003}
{Bensby} T.,  {Feltzing} S.,   {Lundstr{\"o}m} I.,  2003, \mn@doi [\aap]
  {10.1051/0004-6361:20031213}, \href
  {http://adsabs.harvard.edu/abs/2003A%26A...410..527B} {410, 527}

\bibitem[\protect\citeauthoryear{{Bensby}, {Alves-Brito}, {Oey}, {Yong}  \&
  {Mel{\'e}ndez}}{{Bensby} et~al.}{2011}]{Bensby2011}
{Bensby} T.,  {Alves-Brito} A.,  {Oey} M.~S.,  {Yong} D.,   {Mel{\'e}ndez} J.,
  2011, \mn@doi [\apjl] {10.1088/2041-8205/735/2/L46}, \href
  {http://adsabs.harvard.edu/abs/2011ApJ...735L..46B} {735, L46}

\bibitem[\protect\citeauthoryear{{Bensby}, {Feltzing}  \& {Oey}}{{Bensby}
  et~al.}{2014}]{Bensby2014}
{Bensby} T.,  {Feltzing} S.,   {Oey} M.~S.,  2014, \mn@doi [\aap]
  {10.1051/0004-6361/201322631}, \href
  {http://adsabs.harvard.edu/abs/2014A%26A...562A..71B} {562, A71}

\bibitem[\protect\citeauthoryear{{Bergemann} et~al.,}{{Bergemann}
  et~al.}{2018}]{Bergemann2018}
{Bergemann} M.,  et~al., 2018, \mn@doi [\nat] {10.1038/nature25490}, \href
  {http://adsabs.harvard.edu/abs/2018Natur.555..334B} {555, 334}

\bibitem[\protect\citeauthoryear{{Binney} \& {Sch\"onrich}}{{Binney} \&
  {Sch\"onrich}}{2018}]{Binney2018}
{Binney} J.,  {Sch\"onrich} R.,  2018, preprint, \href
  {http://adsabs.harvard.edu/abs/2018arXiv180709819B} {} (\mn@eprint {arXiv}
  {1807.09819})

\bibitem[\protect\citeauthoryear{{Binney} \& {Tremaine}}{{Binney} \&
  {Tremaine}}{2008}]{Binney2008}
{Binney} J.,  {Tremaine} S.,  2008, {Galactic Dynamics: Second Edition}.
Princeton University Press

\bibitem[\protect\citeauthoryear{{Binney}, {Jiang}  \& {Dutta}}{{Binney}
  et~al.}{1998}]{Binney1998}
{Binney} J.,  {Jiang} I.-G.,   {Dutta} S.,  1998, \mn@doi [\mnras]
  {10.1046/j.1365-8711.1998.01595.x}, \href
  {http://adsabs.harvard.edu/abs/1998MNRAS.297.1237B} {297, 1237}

\bibitem[\protect\citeauthoryear{{Binney} et~al.,}{{Binney}
  et~al.}{2014}]{Binney2014}
{Binney} J.,  et~al., 2014, \mn@doi [\mnras] {10.1093/mnras/stt2367}, \href
  {http://adsabs.harvard.edu/abs/2014MNRAS.439.1231B} {439, 1231}

\bibitem[\protect\citeauthoryear{{Bland-Hawthorn} \&
  {Gerhard}}{{Bland-Hawthorn} \& {Gerhard}}{2016}]{blandhawthorn2016a}
{Bland-Hawthorn} J.,  {Gerhard} O.,  2016, \mn@doi [\araa]
  {10.1146/annurev-astro-081915-023441}, \href
  {http://adsabs.harvard.edu/abs/2016ARA%26A..54..529B} {54, 529}

\bibitem[\protect\citeauthoryear{{Bland-Hawthorn}, {Krumholz}  \&
  {Freeman}}{{Bland-Hawthorn} et~al.}{2010}]{BlandHawthorn2010}
{Bland-Hawthorn} J.,  {Krumholz} M.~R.,   {Freeman} K.,  2010, \mn@doi [\apj]
  {10.1088/0004-637X/713/1/166}, \href
  {http://adsabs.harvard.edu/abs/2010ApJ...713..166B} {713, 166}

\bibitem[\protect\citeauthoryear{{Bland-Hawthorn}, {Kos}, {Betters}, {De
  Silva}, {O'Byrne}, {Patterson}  \& {Leon-Saval}}{{Bland-Hawthorn}
  et~al.}{2017}]{BlandHawthorn2017}
{Bland-Hawthorn} J.,  {Kos} J.,  {Betters} C.~H.,  {De Silva} G.,  {O'Byrne}
  J.,  {Patterson} R.,   {Leon-Saval} S.~G.,  2017, \mn@doi [Optics Express]
  {10.1364/OE.25.015614}, \href
  {http://adsabs.harvard.edu/abs/2017OExpr..2515614B} {25, 15614}

\bibitem[\protect\citeauthoryear{{B{\oe}hm} \& {Fayet}}{{B{\oe}hm} \&
  {Fayet}}{2004}]{Boehm2003}
{B{\oe}hm} C.,  {Fayet} P.,  2004, \mn@doi [Nuclear Physics B]
  {10.1016/j.nuclphysb.2004.01.015}, \href
  {http://adsabs.harvard.edu/abs/2004NuPhB.683..219B} {683, 219}

\bibitem[\protect\citeauthoryear{{Bovy}}{{Bovy}}{2015}]{Bovy2015}
{Bovy} J.,  2015, \mn@doi [\apjs] {10.1088/0067-0049/216/2/29}, \href
  {http://adsabs.harvard.edu/abs/2015ApJS..216...29B} {216, 29}

\bibitem[\protect\citeauthoryear{{Brunetti}, {Chiappini}  \&
  {Pfenniger}}{{Brunetti} et~al.}{2011}]{Brunetti11a}
{Brunetti} M.,  {Chiappini} C.,   {Pfenniger} D.,  2011, \mn@doi [\aap]
  {10.1051/0004-6361/201117566}, \href
  {http://adsabs.harvard.edu/abs/2011A%26A...534A..75B} {534, A75}

\bibitem[\protect\citeauthoryear{{Buder} et~al.,}{{Buder}
  et~al.}{2018}]{Buder2018a}
{Buder} S.,  et~al., 2018, preprint, \href
  {http://adsabs.harvard.edu/abs/2018arXiv180406041B} {} (\mn@eprint {arXiv}
  {1804.06041})

\bibitem[\protect\citeauthoryear{{Candlish}, {Smith}, {Fellhauer}, {Gibson},
  {Kroupa}  \& {Assmann}}{{Candlish} et~al.}{2014}]{Candlish2014}
{Candlish} G.~N.,  {Smith} R.,  {Fellhauer} M.,  {Gibson} B.~K.,  {Kroupa} P.,
   {Assmann} P.,  2014, \mn@doi [\mnras] {10.1093/mnras/stt2166}, \href
  {http://adsabs.harvard.edu/abs/2014MNRAS.437.3702C} {437, 3702}

\bibitem[\protect\citeauthoryear{{Carlin} et~al.,}{{Carlin}
  et~al.}{2013}]{Carlin2013}
{Carlin} J.~L.,  et~al., 2013, \mn@doi [\apjl] {10.1088/2041-8205/777/1/L5},
  \href {http://adsabs.harvard.edu/abs/2013ApJ...777L...5C} {777, L5}

\bibitem[\protect\citeauthoryear{{Chiappini}, {Matteucci}  \&
  {Romano}}{{Chiappini} et~al.}{2001}]{Chiappini2001}
{Chiappini} C.,  {Matteucci} F.,   {Romano} D.,  2001, \mn@doi [\apj]
  {10.1086/321427}, \href {http://adsabs.harvard.edu/abs/2001ApJ...554.1044C}
  {554, 1044}

\bibitem[\protect\citeauthoryear{{Daniel} \& {Wyse}}{{Daniel} \&
  {Wyse}}{2015}]{Daniel2015}
{Daniel} K.~J.,  {Wyse} R.~F.~G.,  2015, \mn@doi [\mnras]
  {10.1093/mnras/stu2683}, \href
  {http://adsabs.harvard.edu/abs/2015MNRAS.447.3576D} {447, 3576}

\bibitem[\protect\citeauthoryear{{Daniel} \& {Wyse}}{{Daniel} \&
  {Wyse}}{2018}]{Daniel2018}
{Daniel} K.~J.,  {Wyse} R.~F.~G.,  2018, \mn@doi [\mnras]
  {10.1093/mnras/sty199}, \href
  {http://adsabs.harvard.edu/abs/2018MNRAS.476.1561D} {476, 1561}

\bibitem[\protect\citeauthoryear{{De Silva} et~al.,}{{De Silva}
  et~al.}{2015}]{DeSilva2015}
{De Silva} G.~M.,  et~al., 2015, \mn@doi [\mnras] {10.1093/mnras/stv327}, \href
  {http://adsabs.harvard.edu/abs/2015MNRAS.449.2604D} {449, 2604}

\bibitem[\protect\citeauthoryear{{Dehnen}}{{Dehnen}}{1998}]{Dehnen1998}
{Dehnen} W.,  1998, \mn@doi [\aj] {10.1086/300364}, \href
  {http://ukads.nottingham.ac.uk/abs/1998AJ....115.2384D} {115, 2384}

\bibitem[\protect\citeauthoryear{{Deng} et~al.,}{{Deng}
  et~al.}{2012}]{Deng2012}
{Deng} L.-C.,  et~al., 2012, \mn@doi [Research in Astronomy and Astrophysics]
  {10.1088/1674-4527/12/7/003}, \href
  {http://adsabs.harvard.edu/abs/2012RAA....12..735D} {12, 735}

\bibitem[\protect\citeauthoryear{{Elmegreen}, {Sundin}, {Kaufman}, {Brinks}  \&
  {Elmegreen}}{{Elmegreen} et~al.}{1995}]{Elmegreen1995}
{Elmegreen} B.~G.,  {Sundin} M.,  {Kaufman} M.,  {Brinks} E.,   {Elmegreen}
  D.~M.,  1995, \mn@doi [\apj] {10.1086/176375}, \href
  {http://adsabs.harvard.edu/abs/1995ApJ...453..139E} {453, 139}

\bibitem[\protect\citeauthoryear{{Famaey}, {Jorissen}, {Luri}, {Mayor}, {Udry},
  {Dejonghe}  \& {Turon}}{{Famaey} et~al.}{2005}]{Famaey2005}
{Famaey} B.,  {Jorissen} A.,  {Luri} X.,  {Mayor} M.,  {Udry} S.,  {Dejonghe}
  H.,   {Turon} C.,  2005, \mn@doi [\aap] {10.1051/0004-6361:20041272}, \href
  {http://ukads.nottingham.ac.uk/abs/2005A%26A...430..165F} {430, 165}

\bibitem[\protect\citeauthoryear{{Fouvry}, {Pichon}, {Magorrian}  \&
  {Chavanis}}{{Fouvry} et~al.}{2015}]{Fouvry2015}
{Fouvry} J.~B.,  {Pichon} C.,  {Magorrian} J.,   {Chavanis} P.~H.,  2015,
  \mn@doi [\aap] {10.1051/0004-6361/201527052}, \href
  {http://ukads.nottingham.ac.uk/abs/2015A%26A...584A.129F} {584, A129}

\bibitem[\protect\citeauthoryear{{Fouvry}, {Pichon}, {Chavanis}  \&
  {Monk}}{{Fouvry} et~al.}{2017}]{FouvryPichon2017}
{Fouvry} J.-B.,  {Pichon} C.,  {Chavanis} P.-H.,   {Monk} L.,  2017, \mn@doi
  [\mnras] {10.1093/mnras/stx1625}, \href
  {http://ukads.nottingham.ac.uk/abs/2017MNRAS.471.2642F} {471, 2642}

\bibitem[\protect\citeauthoryear{{Freeman} \& {Bland-Hawthorn}}{{Freeman} \&
  {Bland-Hawthorn}}{2002}]{freeman2002}
{Freeman} K.,  {Bland-Hawthorn} J.,  2002, \mn@doi [\araa]
  {10.1146/annurev.astro.40.060401.093840}, \href
  {http://adsabs.harvard.edu/abs/2002ARA%26A..40..487F} {40, 487}

\bibitem[\protect\citeauthoryear{{Freeman} \& {Bland-Hawthorn}}{{Freeman} \&
  {Bland-Hawthorn}}{2008}]{freeman2008}
{Freeman} K.,  {Bland-Hawthorn} J.,  2008, in {Kodama} T.,  {Yamada} T.,
  {Aoki} K.,  eds,  Astronomical Society of the Pacific Conference Series Vol.
  399, Panoramic Views of Galaxy Formation and Evolution. p.~439

\bibitem[\protect\citeauthoryear{{Fuhrmann}}{{Fuhrmann}}{1998}]{Fuhrmann1998}
{Fuhrmann} K.,  1998, \aap, \href
  {http://adsabs.harvard.edu/abs/1998A%26A...338..161F} {338, 161}

\bibitem[\protect\citeauthoryear{{Gaia Collaboration} et~al.,}{{Gaia
  Collaboration} et~al.}{2018a}]{eyer2018}
{Gaia Collaboration} et~al., 2018a, preprint, \href
  {http://adsabs.harvard.edu/abs/2018arXiv180409382G} {} (\mn@eprint {arXiv}
  {1804.09382})

\bibitem[\protect\citeauthoryear{{Gaia Collaboration} et~al.,}{{Gaia
  Collaboration} et~al.}{2018b}]{Brown2018}
{Gaia Collaboration} et~al., 2018b, \mn@doi [\aap]
  {10.1051/0004-6361/201833051}, \href
  {http://adsabs.harvard.edu/abs/2018A%26A...616A...1G} {616, A1}

\bibitem[\protect\citeauthoryear{{Gaia Collaboration} et~al.,}{{Gaia
  Collaboration} et~al.}{2018c}]{babusiaux2018}
{Gaia Collaboration} et~al., 2018c, \mn@doi [\aap]
  {10.1051/0004-6361/201832843}, \href
  {http://adsabs.harvard.edu/abs/2018A%26A...616A..10G} {616, A10}

\bibitem[\protect\citeauthoryear{{Gaia Collaboration} et~al.,}{{Gaia
  Collaboration} et~al.}{2018d}]{helmi2018}
{Gaia Collaboration} et~al., 2018d, \mn@doi [\aap]
  {10.1051/0004-6361/201832698}, \href
  {http://adsabs.harvard.edu/abs/2018A%26A...616A..12G} {616, A12}

\bibitem[\protect\citeauthoryear{{Gilmore} \& {Reid}}{{Gilmore} \&
  {Reid}}{1983}]{Gilmore1983}
{Gilmore} G.,  {Reid} N.,  1983, \mn@doi [\mnras] {10.1093/mnras/202.4.1025},
  \href {http://ukads.nottingham.ac.uk/abs/1983MNRAS.202.1025G} {202, 1025}

\bibitem[\protect\citeauthoryear{{Gilmore} et~al.,}{{Gilmore}
  et~al.}{2012}]{Gilmore2012}
{Gilmore} G.,  et~al., 2012, The Messenger, \href
  {http://adsabs.harvard.edu/abs/2012Msngr.147...25G} {147, 25}

\bibitem[\protect\citeauthoryear{{Goldreich} \& {Lynden-Bell}}{{Goldreich} \&
  {Lynden-Bell}}{1965}]{GoldreichDLB1965}
{Goldreich} P.,  {Lynden-Bell} D.,  1965, \mn@doi [\mnras]
  {10.1093/mnras/130.2.125}, \href
  {http://ukads.nottingham.ac.uk/abs/1965MNRAS.130..125G} {130, 125}

\bibitem[\protect\citeauthoryear{{G{\'o}mez}, {Minchev}, {O'Shea}, {Beers},
  {Bullock}  \& {Purcell}}{{G{\'o}mez} et~al.}{2013}]{Gomez2013}
{G{\'o}mez} F.~A.,  {Minchev} I.,  {O'Shea} B.~W.,  {Beers} T.~C.,  {Bullock}
  J.~S.,   {Purcell} C.~W.,  2013, \mn@doi [\mnras] {10.1093/mnras/sts327},
  \href {http://adsabs.harvard.edu/abs/2013MNRAS.429..159G} {429, 159}

\bibitem[\protect\citeauthoryear{{G{\'o}mez}, {Besla}, {Carpintero},
  {Villalobos}, {O'Shea}  \& {Bell}}{{G{\'o}mez} et~al.}{2015}]{Gomez2015}
{G{\'o}mez} F.~A.,  {Besla} G.,  {Carpintero} D.~D.,  {Villalobos} {\'A}.,
  {O'Shea} B.~W.,   {Bell} E.~F.,  2015, \mn@doi [\apj]
  {10.1088/0004-637X/802/2/128}, \href
  {http://adsabs.harvard.edu/abs/2015ApJ...802..128G} {802, 128}

\bibitem[\protect\citeauthoryear{{Gravity Collaboration} et~al.,}{{Gravity
  Collaboration} et~al.}{2018}]{Abuter2018}
{Gravity Collaboration} et~al., 2018, \mn@doi [\aap]
  {10.1051/0004-6361/201833718}, \href
  {http://adsabs.harvard.edu/abs/2018A%26A...615L..15G} {615, L15}

\bibitem[\protect\citeauthoryear{{Guiglion} et~al.,}{{Guiglion}
  et~al.}{2015}]{Guiglion2015}
{Guiglion} G.,  et~al., 2015, \mn@doi [\aap] {10.1051/0004-6361/201525883},
  \href {http://adsabs.harvard.edu/abs/2015A%26A...583A..91G} {583, A91}

\bibitem[\protect\citeauthoryear{{Hasselquist} et~al.,}{{Hasselquist}
  et~al.}{2017}]{Hasselquist2017}
{Hasselquist} S.,  et~al., 2017, \mn@doi [\apj] {10.3847/1538-4357/aa7ddc},
  \href {http://adsabs.harvard.edu/abs/2017ApJ...845..162H} {845, 162}

\bibitem[\protect\citeauthoryear{{Hawkins}, {Jofr{\'e}}, {Masseron}  \&
  {Gilmore}}{{Hawkins} et~al.}{2015}]{Hawkins2015}
{Hawkins} K.,  {Jofr{\'e}} P.,  {Masseron} T.,   {Gilmore} G.,  2015, \mn@doi
  [\mnras] {10.1093/mnras/stv1586}, \href
  {http://adsabs.harvard.edu/abs/2015MNRAS.453..758H} {453, 758}

\bibitem[\protect\citeauthoryear{{Hayden} et~al.,}{{Hayden}
  et~al.}{2015}]{Hayden2015}
{Hayden} M.~R.,  et~al., 2015, \mn@doi [\apj] {10.1088/0004-637X/808/2/132},
  \href {http://adsabs.harvard.edu/abs/2015ApJ...808..132H} {808, 132}

\bibitem[\protect\citeauthoryear{{Hayden}, {Recio-Blanco}, {de Laverny},
  {Mikolaitis}  \& {Worley}}{{Hayden} et~al.}{2017}]{Hayden2017}
{Hayden} M.~R.,  {Recio-Blanco} A.,  {de Laverny} P.,  {Mikolaitis} S.,
  {Worley} C.~C.,  2017, \mn@doi [\aap] {10.1051/0004-6361/201731494}, \href
  {http://adsabs.harvard.edu/abs/2017A%26A...608L...1H} {608, L1}

\bibitem[\protect\citeauthoryear{{Hayden}, {Bland-Hawthorn}, {Sharma}  \&
  {GALAH team}}{{Hayden} et~al.}{2019}]{Hayden2019}
{Hayden} M.,  {Bland-Hawthorn} J.,  {Sharma} S.,   {GALAH team} 2019, preprint
  (\mn@eprint {arXiv} {in preparation})

\bibitem[\protect\citeauthoryear{{Haywood}, {Di Matteo}, {Lehnert}, {Katz}  \&
  {G{\'o}mez}}{{Haywood} et~al.}{2013}]{Haywood2013}
{Haywood} M.,  {Di Matteo} P.,  {Lehnert} M.~D.,  {Katz} D.,   {G{\'o}mez} A.,
  2013, \mn@doi [\aap] {10.1051/0004-6361/201321397}, \href
  {http://adsabs.harvard.edu/abs/2013A%26A...560A.109H} {560, A109}

\bibitem[\protect\citeauthoryear{{Hernquist}}{{Hernquist}}{1990}]{her90a}
{Hernquist} L.,  1990, \mn@doi [\apj] {10.1086/168845}, \href
  {http://adsabs.harvard.edu/abs/1990ApJ...356..359H} {356, 359}

\bibitem[\protect\citeauthoryear{{Hui}, {Ostriker}, {Tremaine}  \&
  {Witten}}{{Hui} et~al.}{2017}]{Hui2017}
{Hui} L.,  {Ostriker} J.~P.,  {Tremaine} S.,   {Witten} E.,  2017, \mn@doi
  [\prd] {10.1103/PhysRevD.95.043541}, \href
  {http://ukads.nottingham.ac.uk/abs/2017PhRvD..95d3541H} {95, 043541}

\bibitem[\protect\citeauthoryear{{Hunter} \& {Toomre}}{{Hunter} \&
  {Toomre}}{1969}]{HunterT1969}
{Hunter} C.,  {Toomre} A.,  1969, \mn@doi [\apj] {10.1086/149908}, \href
  {http://ukads.nottingham.ac.uk/abs/1969ApJ...155..747H} {155, 747}

\bibitem[\protect\citeauthoryear{{Ibata}, {Wyse}, {Gilmore}, {Irwin}  \&
  {Suntzeff}}{{Ibata} et~al.}{1997}]{iba97a}
{Ibata} R.~A.,  {Wyse} R.~F.~G.,  {Gilmore} G.,  {Irwin} M.~J.,   {Suntzeff}
  N.~B.,  1997, \mn@doi [\aj] {10.1086/118283}, \href
  {http://adsabs.harvard.edu/abs/1997AJ....113..634I} {113, 634}

\bibitem[\protect\citeauthoryear{{Jiang} \& {Binney}}{{Jiang} \&
  {Binney}}{2000}]{Jiang2000}
{Jiang} I.-G.,  {Binney} J.,  2000, \mn@doi [\mnras]
  {10.1046/j.1365-8711.2000.03311.x}, \href
  {http://adsabs.harvard.edu/abs/2000MNRAS.314..468J} {314, 468}

\bibitem[\protect\citeauthoryear{{J{\'\i}lkov{\'a}}, {Carraro}, {Jungwiert}  \&
  {Minchev}}{{J{\'\i}lkov{\'a}} et~al.}{2012}]{Jilkova12a}
{J{\'\i}lkov{\'a}} L.,  {Carraro} G.,  {Jungwiert} B.,   {Minchev} I.,  2012,
  \mn@doi [\aap] {10.1051/0004-6361/201117347}, \href
  {https://ui.adsabs.harvard.edu/#abs/2012A&A...541A..64J} {541, A64}

\bibitem[\protect\citeauthoryear{{Jordi} et~al.,}{{Jordi}
  et~al.}{2010}]{Jordi2010}
{Jordi} C.,  et~al., 2010, \mn@doi [\aap] {10.1051/0004-6361/201015441}, \href
  {http://adsabs.harvard.edu/abs/2010A%26A...523A..48J} {523, A48}

\bibitem[\protect\citeauthoryear{{Kalberla} \& {Kerp}}{{Kalberla} \&
  {Kerp}}{2009}]{Kalberla2009}
{Kalberla} P.~M.~W.,  {Kerp} J.,  2009, \mn@doi [\araa]
  {10.1146/annurev-astro-082708-101823}, \href
  {http://adsabs.harvard.edu/abs/2009ARA%26A..47...27K} {47, 27}

\bibitem[\protect\citeauthoryear{{Kos} et~al.,}{{Kos} et~al.}{2017}]{kos2017}
{Kos} J.,  et~al., 2017, \mn@doi [\mnras] {10.1093/mnras/stw2064}, \href
  {http://adsabs.harvard.edu/abs/2017MNRAS.464.1259K} {464, 1259}

\bibitem[\protect\citeauthoryear{{Kos} et~al.,}{{Kos} et~al.}{2018a}]{Kos2018b}
{Kos} J.,  et~al., 2018a, \mn@doi [\mnras] {10.1093/mnras/sty2171}, \href
  {http://adsabs.harvard.edu/abs/2018MNRAS.480.5242K} {480, 5242}

\bibitem[\protect\citeauthoryear{{Kos} et~al.,}{{Kos} et~al.}{2018b}]{Kos2018c}
{Kos} J.,  et~al., 2018b, \mn@doi [\mnras] {10.1093/mnras/sty2175}, \href
  {http://adsabs.harvard.edu/abs/2018MNRAS.480.5475K} {480, 5475}

\bibitem[\protect\citeauthoryear{{Laporte}, {G{\'o}mez}, {Besla}, {Johnston}
  \& {Garavito-Camargo}}{{Laporte} et~al.}{2018}]{Laporte2018}
{Laporte} C.~F.~P.,  {G{\'o}mez} F.~A.,  {Besla} G.,  {Johnston} K.~V.,
  {Garavito-Camargo} N.,  2018, \mn@doi [\mnras] {10.1093/mnras/stx2146}, \href
  {http://adsabs.harvard.edu/abs/2018MNRAS.473.1218L} {473, 1218}

\bibitem[\protect\citeauthoryear{{Law}, {Johnston}  \& {Majewski}}{{Law}
  et~al.}{2005}]{Law2005}
{Law} D.~R.,  {Johnston} K.~V.,   {Majewski} S.~R.,  2005, \mn@doi [\apj]
  {10.1086/426779}, \href {http://adsabs.harvard.edu/abs/2005ApJ...619..807L}
  {619, 807}

\bibitem[\protect\citeauthoryear{{Li} et~al.,}{{Li} et~al.}{2017}]{Li2017}
{Li} T.~S.,  et~al., 2017, \mn@doi [\apj] {10.3847/1538-4357/aa7a0d}, \href
  {http://adsabs.harvard.edu/abs/2017ApJ...844...74L} {844, 74}

\bibitem[\protect\citeauthoryear{{Lockman}}{{Lockman}}{2002}]{Lockman2002}
{Lockman} F.~J.,  2002, in {Taylor} A.~R.,  {Landecker} T.~L.,   {Willis}
  A.~G.,  eds,  Astronomical Society of the Pacific Conference Series Vol. 276,
  Seeing Through the Dust: The Detection of HI and the Exploration of the ISM
  in Galaxies. p.~107 (\mn@eprint {} {astro-ph/0203210})

\bibitem[\protect\citeauthoryear{{Lynden-Bell}}{{Lynden-Bell}}{1967}]{LyndenBell1967}
{Lynden-Bell} D.,  1967, \mn@doi [\mnras] {10.1093/mnras/136.1.101}, \href
  {http://adsabs.harvard.edu/abs/1967MNRAS.136..101L} {136, 101}

\bibitem[\protect\citeauthoryear{{Majewski} et~al.,}{{Majewski}
  et~al.}{2017}]{Majewski2017}
{Majewski} S.~R.,  et~al., 2017, \mn@doi [\aj] {10.3847/1538-3881/aa784d},
  \href {http://adsabs.harvard.edu/abs/2017AJ....154...94M} {154, 94}

\bibitem[\protect\citeauthoryear{{Martell} et~al.,}{{Martell}
  et~al.}{2017}]{Martell2017}
{Martell} S.~L.,  et~al., 2017, \mn@doi [\mnras] {10.1093/mnras/stw2835}, \href
  {http://adsabs.harvard.edu/abs/2017MNRAS.465.3203M} {465, 3203}

\bibitem[\protect\citeauthoryear{{Masseron} \& {Gilmore}}{{Masseron} \&
  {Gilmore}}{2015}]{Masseron2015}
{Masseron} T.,  {Gilmore} G.,  2015, \mn@doi [\mnras] {10.1093/mnras/stv1731},
  \href {http://adsabs.harvard.edu/abs/2015MNRAS.453.1855M} {453, 1855}

\bibitem[\protect\citeauthoryear{{Masset} \& {Tagger}}{{Masset} \&
  {Tagger}}{1997}]{Masset1997}
{Masset} F.,  {Tagger} M.,  1997, \aap, \href
  {http://adsabs.harvard.edu/abs/1997A%26A...322..442M} {322, 442}

\bibitem[\protect\citeauthoryear{{Matthews} \& {Uson}}{{Matthews} \&
  {Uson}}{2008a}]{Matthews2008b}
{Matthews} L.~D.,  {Uson} J.~M.,  2008a, \mn@doi [\aj]
  {10.1088/0004-6256/135/1/291}, \href
  {http://adsabs.harvard.edu/abs/2008AJ....135..291M} {135, 291}

\bibitem[\protect\citeauthoryear{{Matthews} \& {Uson}}{{Matthews} \&
  {Uson}}{2008b}]{Matthews2008a}
{Matthews} L.~D.,  {Uson} J.~M.,  2008b, \mn@doi [\apj] {10.1086/592086}, \href
  {http://adsabs.harvard.edu/abs/2008ApJ...688..237M} {688, 237}

\bibitem[\protect\citeauthoryear{{McMillan}}{{McMillan}}{2011}]{McMillan2011}
{McMillan} P.~J.,  2011, \mn@doi [\mnras] {10.1111/j.1365-2966.2011.18564.x},
  \href {http://adsabs.harvard.edu/abs/2011MNRAS.414.2446M} {414, 2446}

\bibitem[\protect\citeauthoryear{{Milgrom} \& {Sanders}}{{Milgrom} \&
  {Sanders}}{2008}]{Milgrom2008}
{Milgrom} M.,  {Sanders} R.~H.,  2008, \mn@doi [\apj] {10.1086/529119}, \href
  {http://ukads.nottingham.ac.uk/abs/2008ApJ...678..131M} {678, 131}

\bibitem[\protect\citeauthoryear{{Minchev} \& {Famaey}}{{Minchev} \&
  {Famaey}}{2010}]{Minchev10a}
{Minchev} I.,  {Famaey} B.,  2010, \mn@doi [\apj]
  {10.1088/0004-637X/722/1/112}, \href
  {https://ui.adsabs.harvard.edu/#abs/2010ApJ...722..112M} {722, 112}

\bibitem[\protect\citeauthoryear{{Minchev}, {Famaey}, {Combes}, {Di Matteo},
  {Mouhcine}  \& {Wozniak}}{{Minchev} et~al.}{2011}]{Minchev11a}
{Minchev} I.,  {Famaey} B.,  {Combes} F.,  {Di Matteo} P.,  {Mouhcine} M.,
  {Wozniak} H.,  2011, \mn@doi [\aap] {10.1051/0004-6361/201015139}, \href
  {http://adsabs.harvard.edu/abs/2011A\%26A...527A.147M} {527, A147}

\bibitem[\protect\citeauthoryear{{Minchev} et~al.,}{{Minchev}
  et~al.}{2018}]{Minchev2018}
{Minchev} I.,  et~al., 2018, \mn@doi [\mnras] {10.1093/mnras/sty2033}, \href
  {http://adsabs.harvard.edu/abs/2018MNRAS.481.1645M} {481, 1645}

\bibitem[\protect\citeauthoryear{{Miyamoto} \& {Nagai}}{{Miyamoto} \&
  {Nagai}}{1975}]{miy75a}
{Miyamoto} M.,  {Nagai} R.,  1975, \pasj, \href
  {http://adsabs.harvard.edu/abs/1975PASJ...27..533M} {27, 533}

\bibitem[\protect\citeauthoryear{{Newberg} et~al.,}{{Newberg}
  et~al.}{2002}]{Newberg2002}
{Newberg} H.~J.,  et~al., 2002, \mn@doi [\apj] {10.1086/338983}, \href
  {http://adsabs.harvard.edu/abs/2002ApJ...569..245N} {569, 245}

\bibitem[\protect\citeauthoryear{{Newberg}, {Yanny}, {Cole}, {Beers}, {Re
  Fiorentin}, {Schneider}  \& {Wilhelm}}{{Newberg} et~al.}{2007}]{Newberg2007}
{Newberg} H.~J.,  {Yanny} B.,  {Cole} N.,  {Beers} T.~C.,  {Re Fiorentin} P.,
  {Schneider} D.~P.,   {Wilhelm} R.,  2007, \mn@doi [\apj] {10.1086/521068},
  \href {http://adsabs.harvard.edu/abs/2007ApJ...668..221N} {668, 221}

\bibitem[\protect\citeauthoryear{{Nieva} \& {Przybilla}}{{Nieva} \&
  {Przybilla}}{2012}]{Nieva2012}
{Nieva} M.-F.,  {Przybilla} N.,  2012, \mn@doi [\aap]
  {10.1051/0004-6361/201118158}, \href
  {http://adsabs.harvard.edu/abs/2012A%26A...539A.143N} {539, A143}

\bibitem[\protect\citeauthoryear{{Perret}, {Renaud}, {Epinat}, {Amram},
  {Bournaud}, {Contini}, {Teyssier}  \& {Lambert}}{{Perret}
  et~al.}{2014}]{per14c}
{Perret} V.,  {Renaud} F.,  {Epinat} B.,  {Amram} P.,  {Bournaud} F.,
  {Contini} T.,  {Teyssier} R.,   {Lambert} J.-C.,  2014, \mn@doi [\aap]
  {10.1051/0004-6361/201322395}, \href
  {http://adsabs.harvard.edu/abs/2014A%26A...562A...1P} {562, A1}

\bibitem[\protect\citeauthoryear{{Perryman} et~al.,}{{Perryman}
  et~al.}{2001}]{perryman2001}
{Perryman} M.~A.~C.,  et~al., 2001, \mn@doi [\aap]
  {10.1051/0004-6361:20010085}, \href
  {http://adsabs.harvard.edu/abs/2001A%26A...369..339P} {369, 339}

\bibitem[\protect\citeauthoryear{{Piffl} et~al.,}{{Piffl}
  et~al.}{2014}]{Piffl2014}
{Piffl} T.,  et~al., 2014, \mn@doi [\mnras] {10.1093/mnras/stu1948}, \href
  {http://adsabs.harvard.edu/abs/2014MNRAS.445.3133P} {445, 3133}

\bibitem[\protect\citeauthoryear{{Posti}, {Helmi}, {Veljanoski}  \&
  {Breddels}}{{Posti} et~al.}{2018}]{Posti2018}
{Posti} L.,  {Helmi} A.,  {Veljanoski} J.,   {Breddels} M.~A.,  2018, \mn@doi
  [\aap] {10.1051/0004-6361/201732277}, \href
  {http://adsabs.harvard.edu/abs/2018A%26A...615A..70P} {615, A70}

\bibitem[\protect\citeauthoryear{{Prusti} et~al.,}{{Prusti}
  et~al.}{2016}]{prusti2016}
{Prusti} T.,  et~al., 2016, \mn@doi [\aap] {10.1051/0004-6361/201629272}, \href
  {http://adsabs.harvard.edu/abs/2016A%26A...595A...1G} {595, A1}

\bibitem[\protect\citeauthoryear{{Purcell}, {Bullock}, {Tollerud}, {Rocha}  \&
  {Chakrabarti}}{{Purcell} et~al.}{2011a}]{Purcell2011}
{Purcell} C.~W.,  {Bullock} J.~S.,  {Tollerud} E.~J.,  {Rocha} M.,
  {Chakrabarti} S.,  2011a, \mn@doi [\nat] {10.1038/nature10417}, \href
  {http://adsabs.harvard.edu/abs/2011Natur.477..301P} {477, 301}

\bibitem[\protect\citeauthoryear{{Purcell}, {Bullock}, {Tollerud}, {Rocha}  \&
  {Chakrabarti}}{{Purcell} et~al.}{2011b}]{Purcell2015}
{Purcell} C.~W.,  {Bullock} J.~S.,  {Tollerud} E.~J.,  {Rocha} M.,
  {Chakrabarti} S.,  2011b, \mn@doi [\nat] {10.1038/nature10417}, \href
  {http://adsabs.harvard.edu/abs/2011Natur.477..301P} {477, 301}

\bibitem[\protect\citeauthoryear{{Quillen}, {Minchev}, {Bland-Hawthorn}  \&
  {Haywood}}{{Quillen} et~al.}{2009}]{Quillen09a}
{Quillen} A.~C.,  {Minchev} I.,  {Bland-Hawthorn} J.,   {Haywood} M.,  2009,
  \mn@doi [\mnras] {10.1111/j.1365-2966.2009.15054.x}, \href
  {https://ui.adsabs.harvard.edu/#abs/2009MNRAS.397.1599Q} {397, 1599}

\bibitem[\protect\citeauthoryear{{Quillen}, {Nolting}, {Minchev}, {De Silva}
  \& {Chiappini}}{{Quillen} et~al.}{2018}]{Quillen18a}
{Quillen} A.~C.,  {Nolting} E.,  {Minchev} I.,  {De Silva} G.,   {Chiappini}
  C.,  2018, \mn@doi [\mnras] {10.1093/mnras/sty125}, \href
  {https://ui.adsabs.harvard.edu/#abs/2018MNRAS.475.4450Q} {475, 4450}

\bibitem[\protect\citeauthoryear{{Reid} \& {Brunthaler}}{{Reid} \&
  {Brunthaler}}{2004}]{Reid2004}
{Reid} M.~J.,  {Brunthaler} A.,  2004, \mn@doi [\apj] {10.1086/424960}, \href
  {http://adsabs.harvard.edu/abs/2004ApJ...616..872R} {616, 872}

\bibitem[\protect\citeauthoryear{{Rocha-Pinto}, {Majewski}, {Skrutskie},
  {Crane}  \& {Patterson}}{{Rocha-Pinto} et~al.}{2004}]{RochaPinto2004}
{Rocha-Pinto} H.~J.,  {Majewski} S.~R.,  {Skrutskie} M.~F.,  {Crane} J.~D.,
  {Patterson} R.~J.,  2004, \mn@doi [\apj] {10.1086/424585}, \href
  {http://adsabs.harvard.edu/abs/2004ApJ...615..732R} {615, 732}

\bibitem[\protect\citeauthoryear{{Ro{\v s}kar}, {Debattista}, {Quinn}  \&
  {Wadsley}}{{Ro{\v s}kar} et~al.}{2012}]{Roskar2012}
{Ro{\v s}kar} R.,  {Debattista} V.~P.,  {Quinn} T.~R.,   {Wadsley} J.,  2012,
  \mn@doi [\mnras] {10.1111/j.1365-2966.2012.21860.x}, \href
  {http://adsabs.harvard.edu/abs/2012MNRAS.426.2089R} {426, 2089}

\bibitem[\protect\citeauthoryear{{Russell} \& {Dopita}}{{Russell} \&
  {Dopita}}{1992}]{Russell1992}
{Russell} S.~C.,  {Dopita} M.~A.,  1992, \mn@doi [\apj] {10.1086/170893}, \href
  {http://adsabs.harvard.edu/abs/1992ApJ...384..508R} {384, 508}

\bibitem[\protect\citeauthoryear{{Sanders} \& {Binney}}{{Sanders} \&
  {Binney}}{2015}]{Sanders2015}
{Sanders} J.~L.,  {Binney} J.,  2015, \mn@doi [\mnras] {10.1093/mnras/stv578},
  \href {http://ukads.nottingham.ac.uk/abs/2015MNRAS.449.3479S} {449, 3479}

\bibitem[\protect\citeauthoryear{{Sch{\"o}nrich} \& {Binney}}{{Sch{\"o}nrich}
  \& {Binney}}{2009}]{SchoenrichB2009}
{Sch{\"o}nrich} R.,  {Binney} J.,  2009, \mn@doi [\mnras]
  {10.1111/j.1365-2966.2009.14750.x}, \href
  {http://ukads.nottingham.ac.uk/abs/2009MNRAS.396..203S} {396, 203}

\bibitem[\protect\citeauthoryear{{Sch{\"o}nrich} \& {Dehnen}}{{Sch{\"o}nrich}
  \& {Dehnen}}{2018}]{Schoenrich2018}
{Sch{\"o}nrich} R.,  {Dehnen} W.,  2018, \mn@doi [\mnras]
  {10.1093/mnras/sty1256}, \href
  {http://adsabs.harvard.edu/abs/2018MNRAS.478.3809S} {478, 3809}

\bibitem[\protect\citeauthoryear{{Sch{\"o}nrich} \& {McMillan}}{{Sch{\"o}nrich}
  \& {McMillan}}{2017}]{SchoenrichM2017}
{Sch{\"o}nrich} R.,  {McMillan} P.~J.,  2017, \mn@doi [\mnras]
  {10.1093/mnras/stx093}, \href
  {http://ukads.nottingham.ac.uk/abs/2017MNRAS.467.1154S} {467, 1154}

\bibitem[\protect\citeauthoryear{{Sch{\"o}nrich}, {Binney}  \&
  {Dehnen}}{{Sch{\"o}nrich} et~al.}{2010}]{Schoenrich2010}
{Sch{\"o}nrich} R.,  {Binney} J.,   {Dehnen} W.,  2010, \mn@doi [\mnras]
  {10.1111/j.1365-2966.2010.16253.x}, \href
  {http://adsabs.harvard.edu/abs/2010MNRAS.403.1829S} {403, 1829}

\bibitem[\protect\citeauthoryear{{Sellwood} \& {Binney}}{{Sellwood} \&
  {Binney}}{2002}]{Sellwood2002}
{Sellwood} J.~A.,  {Binney} J.~J.,  2002, \mn@doi [\mnras]
  {10.1046/j.1365-8711.2002.05806.x}, \href
  {http://adsabs.harvard.edu/abs/2002MNRAS.336..785S} {336, 785}

\bibitem[\protect\citeauthoryear{{Sellwood} \& {Carlberg}}{{Sellwood} \&
  {Carlberg}}{2014}]{SellwoodC2014}
{Sellwood} J.~A.,  {Carlberg} R.~G.,  2014, \mn@doi [\apj]
  {10.1088/0004-637X/785/2/137}, \href
  {http://ukads.nottingham.ac.uk/abs/2014ApJ...785..137S} {785, 137}

\bibitem[\protect\citeauthoryear{{Sharma}}{{Sharma}}{2017}]{Sharma2017}
{Sharma} S.,  2017, \mn@doi [\araa] {10.1146/annurev-astro-082214-122339},
  \href {http://adsabs.harvard.edu/abs/2017ARA%26A..55..213S} {55, 213}

\bibitem[\protect\citeauthoryear{{Sharma}, {Bland-Hawthorn}, {Johnston}  \&
  {Binney}}{{Sharma} et~al.}{2011}]{Sharma2011}
{Sharma} S.,  {Bland-Hawthorn} J.,  {Johnston} K.~V.,   {Binney} J.,  2011,
  \mn@doi [\apj] {10.1088/0004-637X/730/1/3}, \href
  {http://adsabs.harvard.edu/abs/2011ApJ...730....3S} {730, 3}

\bibitem[\protect\citeauthoryear{{Sharma} et~al.,}{{Sharma}
  et~al.}{2014}]{Sharma2014}
{Sharma} S.,  et~al., 2014, \mn@doi [\apj] {10.1088/0004-637X/793/1/51}, \href
  {http://adsabs.harvard.edu/abs/2014ApJ...793...51S} {793, 51}

\bibitem[\protect\citeauthoryear{{Sharma} et~al.,}{{Sharma}
  et~al.}{2018}]{Sharma2018}
{Sharma} S.,  et~al., 2018, \mn@doi [\mnras] {10.1093/mnras/stx2582}, \href
  {http://adsabs.harvard.edu/abs/2018MNRAS.473.2004S} {473, 2004}

\bibitem[\protect\citeauthoryear{{Sheinis} et~al.,}{{Sheinis}
  et~al.}{2015}]{Sheinis2015}
{Sheinis} A.,  et~al., 2015, \mn@doi [Journal of Astronomical Telescopes,
  Instruments, and Systems] {10.1117/1.JATIS.1.3.035002}, \href
  {http://adsabs.harvard.edu/abs/2015JATIS...1c5002S} {1, 035002}

\bibitem[\protect\citeauthoryear{{Siebert} et~al.,}{{Siebert}
  et~al.}{2008}]{Siebert2008}
{Siebert} A.,  et~al., 2008, \mn@doi [\mnras]
  {10.1111/j.1365-2966.2008.13912.x}, \href
  {http://adsabs.harvard.edu/abs/2008MNRAS.391..793S} {391, 793}

\bibitem[\protect\citeauthoryear{{Slater} et~al.,}{{Slater}
  et~al.}{2014}]{Slater2014}
{Slater} C.~T.,  et~al., 2014, \mn@doi [\apj] {10.1088/0004-637X/791/1/9},
  \href {http://adsabs.harvard.edu/abs/2014ApJ...791....9S} {791, 9}

\bibitem[\protect\citeauthoryear{{Sofia} \& {Meyer}}{{Sofia} \&
  {Meyer}}{2001}]{Sofia2001}
{Sofia} U.~J.,  {Meyer} D.~M.,  2001, \mn@doi [\apjl] {10.1086/321715}, \href
  {http://adsabs.harvard.edu/abs/2001ApJ...554L.221S} {554, L221}

\bibitem[\protect\citeauthoryear{{Solway}, {Sellwood}  \&
  {Sch{\"o}nrich}}{{Solway} et~al.}{2012}]{Solway2012}
{Solway} M.,  {Sellwood} J.~A.,   {Sch{\"o}nrich} R.,  2012, \mn@doi [\mnras]
  {10.1111/j.1365-2966.2012.20712.x}, \href
  {http://ukads.nottingham.ac.uk/abs/2012MNRAS.422.1363S} {422, 1363}

\bibitem[\protect\citeauthoryear{{Spagna}, {Lattanzi}, {Re Fiorentin}  \&
  {Smart}}{{Spagna} et~al.}{2010}]{Spagna2010}
{Spagna} A.,  {Lattanzi} M.~G.,  {Re Fiorentin} P.,   {Smart} R.~L.,  2010,
  \mn@doi [\aap] {10.1051/0004-6361/200913538}, \href
  {http://ukads.nottingham.ac.uk/abs/2010A%26A...510L...4S} {510, L4}

\bibitem[\protect\citeauthoryear{{Springel}, {Di Matteo}  \&
  {Hernquist}}{{Springel} et~al.}{2005}]{spr05c}
{Springel} V.,  {Di Matteo} T.,   {Hernquist} L.,  2005, \mn@doi [\mnras]
  {10.1111/j.1365-2966.2005.09238.x}, \href
  {http://adsabs.harvard.edu/abs/2005MNRAS.361..776S} {361, 776}

\bibitem[\protect\citeauthoryear{{Steinmetz} et~al.,}{{Steinmetz}
  et~al.}{2006}]{Steinmetz2006}
{Steinmetz} M.,  et~al., 2006, \mn@doi [\aj] {10.1086/506564}, \href
  {http://adsabs.harvard.edu/abs/2006AJ....132.1645S} {132, 1645}

\bibitem[\protect\citeauthoryear{{Tepper-Garc{\'{\i}}a} \&
  {Bland-Hawthorn}}{{Tepper-Garc{\'{\i}}a} \&
  {Bland-Hawthorn}}{2018}]{TepperGarcia2018}
{Tepper-Garc{\'{\i}}a} T.,  {Bland-Hawthorn} J.,  2018, \mn@doi [\mnras]
  {10.1093/mnras/sty1359}, \href
  {http://adsabs.harvard.edu/abs/2018MNRAS.478.5263T} {478, 5263}

\bibitem[\protect\citeauthoryear{{Tepper-Garc\'ia}, {Bland-Hawthorn}, {Binney},
  {Sharma}  \& {GALAH team}}{{Tepper-Garc\'ia} et~al.}{2019}]{TepperGarcia2019}
{Tepper-Garc\'ia} T.,  {Bland-Hawthorn} J.,  {Binney} J.,  {Sharma} S.,
  {GALAH team} 2019, preprint (\mn@eprint {arXiv} {in preparation})

\bibitem[\protect\citeauthoryear{{Teyssier}}{{Teyssier}}{2002}]{tey02a}
{Teyssier} R.,  2002, \mn@doi [\aap] {10.1051/0004-6361:20011817}, \href
  {http://adsabs.harvard.edu/abs/2002A%26A...385..337T} {385, 337}

\bibitem[\protect\citeauthoryear{{Toomre}}{{Toomre}}{1964}]{too64a}
{Toomre} A.,  1964, \mn@doi [\apj] {10.1086/147861}, \href
  {http://adsabs.harvard.edu/abs/1964ApJ...139.1217T} {139, 1217}

\bibitem[\protect\citeauthoryear{{Toomre}}{{Toomre}}{1969}]{ToomreGpV1969}
{Toomre} A.,  1969, \mn@doi [\apj] {10.1086/150250}, \href
  {http://ukads.nottingham.ac.uk/abs/1969ApJ...158..899T} {158, 899}

\bibitem[\protect\citeauthoryear{{Toomre}}{{Toomre}}{1981}]{Toomre1981}
{Toomre} A.,  1981, in {Fall} S.~M.,  {Lynden-Bell} D.,  eds, Structure and
  Evolution of Normal Galaxies. pp 111--136

\bibitem[\protect\citeauthoryear{{Vasiliev}}{{Vasiliev}}{2018}]{Vasiliev2018}
{Vasiliev} E.,  2018, {AGAMA: Action-based galaxy modeling framework},
  Astrophysics Source Code Library (\mn@eprint {ascl} {1805.008})

\bibitem[\protect\citeauthoryear{{Vera-Ciro}, {D'Onghia}, {Navarro}  \&
  {Abadi}}{{Vera-Ciro} et~al.}{2014}]{VeraCiro2014}
{Vera-Ciro} C.,  {D'Onghia} E.,  {Navarro} J.,   {Abadi} M.,  2014, \mn@doi
  [\apj] {10.1088/0004-637X/794/2/173}, \href
  {http://adsabs.harvard.edu/abs/2014ApJ...794..173V} {794, 173}

\bibitem[\protect\citeauthoryear{{Widrow}, {Gardner}, {Yanny}, {Dodelson}  \&
  {Chen}}{{Widrow} et~al.}{2012}]{Widrow2012}
{Widrow} L.~M.,  {Gardner} S.,  {Yanny} B.,  {Dodelson} S.,   {Chen} H.-Y.,
  2012, \mn@doi [\apjl] {10.1088/2041-8205/750/2/L41}, \href
  {http://adsabs.harvard.edu/abs/2012ApJ...750L..41W} {750, L41}

\bibitem[\protect\citeauthoryear{{Williams} et~al.,}{{Williams}
  et~al.}{2013}]{Williams2013}
{Williams} M.~E.~K.,  et~al., 2013, \mn@doi [\mnras] {10.1093/mnras/stt1522},
  \href {http://adsabs.harvard.edu/abs/2013MNRAS.436..101W} {436, 101}

\bibitem[\protect\citeauthoryear{{Wittenmyer} et~al.,}{{Wittenmyer}
  et~al.}{2018}]{Wittenmyer2018}
{Wittenmyer} R.~A.,  et~al., 2018, \mn@doi [\aj] {10.3847/1538-3881/aaa3e4},
  \href {http://adsabs.harvard.edu/abs/2018AJ....155...84W} {155, 84}

\bibitem[\protect\citeauthoryear{{Xu}, {Newberg}, {Carlin}, {Liu}, {Deng},
  {Li}, {Sch{\"o}nrich}  \& {Yanny}}{{Xu} et~al.}{2015}]{Xu2015}
{Xu} Y.,  {Newberg} H.~J.,  {Carlin} J.~L.,  {Liu} C.,  {Deng} L.,  {Li} J.,
  {Sch{\"o}nrich} R.,   {Yanny} B.,  2015, \mn@doi [\apj]
  {10.1088/0004-637X/801/2/105}, \href
  {http://adsabs.harvard.edu/abs/2015ApJ...801..105X} {801, 105}

\bibitem[\protect\citeauthoryear{{Yanny} \& {Gardner}}{{Yanny} \&
  {Gardner}}{2013}]{Yanny2013}
{Yanny} B.,  {Gardner} S.,  2013, \mn@doi [\apj] {10.1088/0004-637X/777/2/91},
  \href {http://adsabs.harvard.edu/abs/2013ApJ...777...91Y} {777, 91}

\bibitem[\protect\citeauthoryear{{Zari}, {Hashemi}, {Brown}, {Jardine}  \& {de
  Zeeuw}}{{Zari} et~al.}{2018}]{Zari2018}
{Zari} E.,  {Hashemi} H.,  {Brown} A.~G.~A.,  {Jardine} K.,   {de Zeeuw} P.~T.,
   2018, preprint, \href {http://adsabs.harvard.edu/abs/2018arXiv181009819Z} {}
  (\mn@eprint {arXiv} {1810.09819})

\bibitem[\protect\citeauthoryear{{Zwitter} et~al.,}{{Zwitter}
  et~al.}{2018}]{zwitter2018}
{Zwitter} T.,  et~al., 2018, preprint, \href
  {http://adsabs.harvard.edu/abs/2018arXiv180406344Z} {} (\mn@eprint {arXiv}
  {1804.06344})

\bibitem[\protect\citeauthoryear{{de la Vega}, {Quillen}, {Carlin},
  {Chakrabarti}  \& {D'Onghia}}{{de la Vega} et~al.}{2015}]{delaVega2015}
{de la Vega} A.,  {Quillen} A.~C.,  {Carlin} J.~L.,  {Chakrabarti} S.,
  {D'Onghia} E.,  2015, \mn@doi [\mnras] {10.1093/mnras/stv2055}, \href
  {http://adsabs.harvard.edu/abs/2015MNRAS.454..933D} {454, 933}

\makeatother
\end{thebibliography}

\appendix
\section{Movies}\label{s:movies}
\subsection{Reference model: isolated Galaxy}

In movie
P1,\footnote{\url{http://www.physics.usyd.edu.au/galah_exp/sp
}
\label{l:website}} we witness the long-term evolution of the isolated Galaxy
model (Model P) summarised in Table~\ref{t:Galaxy}. The frames are shown with time steps of $\Delta t = 10$ Myr that are short enough to trace most stellar
orbits reliably.  The disc settles to an equilibrium configuration within a
few hundred Myr. This can be seen from the settling to a cold, thin disc in
the edge-on and face-on projections, with a constant vertical scaleheight.
The disc generates low-level flocculent (transient) spiral perturbations consistent with
the intrinsic numerical and spatial resolution, but individual stellar orbits
confirm that the coarse-grained potential is well behaved. The simulation
includes a thick disc which is shown only in the vertical phase plane. Since
the thick disc is older and more metal poor than the thin disc, including two
discs allows us to study the predicted age and metallicity dependence of the
cohent spiral. This is the base model we use to study the perturbation
induced by an intruder.

In movie P2, we show the same configuration space for movie P1, but now each
star is colour-coded by its azimuthal velocity $V_\phi$ in the plane of the
disc.  The velocity field reflects the underlying rotation curve consistent
with the total gravitational potential. This becomes relevant when comparing
to the perturbed cases below. There are small variations across the disc
consistent with the stochastic spiral perturbations.

In movie P3, the evolution of the thin and thick discs is shown in the
vertical phase plane ($z,V_z$) coded by the azimuthal velocity $V_\phi$, i.e.
$V_\phi (z,V_z)$. There is no statistical averaging $\langle V_\phi\rangle$
over the population as carried out for the \Gaia\ data (e.g.
\autoref{f:gaia_maps4}). These are projections of the {\it entire} disc population except that we have removed all points inside of $R=5$ kpc. Otherwise, the phase plane would be dominated by the inner disc with its higher $V_z$
motions compressed in $z$. The different vertical thicknesses are evident and
they remain fairly constant as the disc evolves. Each star moves clockwise
around ($z,V_z$)=(0,0) as it oscillates in the disc potential with an angular
frequency $\Omega_z$. No coherent patterns emerge because the phases are
randomised over $2\pi$.

In movie P4, we show the evolution of the thin and thick discs in the
vertical phase plane ($z,V_z$) coded by the radial distance $R$, i.e. $R (z,V_z)$. No significant structural evolution is evident once the disc has settled.

\subsection{Perturbed model: hyperbolic orbit}

Movies T1, S1 and R1 show the interaction of the Galaxy with the low-mass,
intermediate-mass and high-mass intruders respectively, each on a
hyperbolic orbit (shown as a filled red circle) crossing the disc at about
$R=13\kpc$. At $t\sim95\Myr$, the disc moves up towards the approaching
intruder and its centre of mass experiences a recoil.
By the time the
intruder has transited the disc plane, the entire disc has responded to the
perturber. By $t=180\Myr$, the interaction has excited a spiral arm and a strong warp in the outer disc that precesses in the plane around the centre
of mass \citep[cf.][]{Gomez2015}.  The upwards momentum of the disc does not
reverse until after $t=400$ Myr.  The strong forcing by the perturber is
active for less than 100 Myr but the disc response persists for the 2 Gyr
duration of the movie. For reference, the results are also shown in
\autoref{f:thor1} and \autoref{f:thor2} respectively at a single timestep
($t=900$ Myr).

In \autoref{f:thor1} (R2) and \autoref{f:thor2} (S2), when compared to the isolated Galaxy model (movie P2), the azimuthal velocity field $V_\phi$ shows
systematic variations due to the low-amplitude bending waves (corrugations or
wrinkles) propagating through the disc. The kinematically distinct, azimuthal
`plumes' (confined in radius) arise from disc segments displaced vertically
by varying amounts, with larger displacements resulting in bigger lags with respect to the disc rotation at that radius. In other words, these kinematic
plumes are mostly associated with large $V_z$ motion and some $V_R$ motion.

In \autoref{f:thor1} (R3) and \autoref{f:thor2} (S3), these same kinematic
plumes encoded with the same $V_\phi$ velocities manifest in the ($z,V_z$)
plane. In \autoref{f:thor1} (R4) and \autoref{f:thor2} (S4) using the $R(z,V_z)$ plane, we show plumes at two different radii for a single
timestep ($t=900$ Myr). The spiral feature at 10 kpc aligns and
elongates with the disc whereas the feature in the outer disc appears more circular as expected from theory (\autoref{f:binney}). 
In the absence of
another disc-transiting event, these spiral features can be long-lived
($\gtrsim 500$ Myr) in the vertical phase plane.

There is only a finite number of kinematic plumes in the $zV_z$ plane because the corrugations have a long wavelength of order several kiloparsecs, as observed in the Galaxy \citep{Xu2015} and in external galaxies \citep{Matthews2008a}, and expected from theory \citep{Masset1997}. The aspect ratio in $zV_z$ space, the degree of winding up and the total amplitude depend on the disc location (\autoref{f:binney}).

At times, a one-armed spiral feature is clear (e.g. movies S3 and R3 during
$1.2<t<1.9$ Gyr) although tracing it to the origin in ($z,V_z$) proves to be
difficult without the action analysis (Section~\ref{s:action}) applied
directly to the models \citep{TepperGarcia2019}. This is because our sampling
(limited by $N_{\rm P}$) is not sufficient to provide contrast against the
clockwise population from the dominant underlying disc. 

In \autoref{f:thor1}, for the high-mass intruder (R3), there are several
coherent one-armed plumes in the ($z,V_z$) plane. Once again, these grow to
their maximum amplitude in \zmax\ $\approx$ 6 kpc and \Vzmax\ $\approx 60$ km
s$^{-1}$ long after the transit has occurred. This appears to be associated
with the strong vertical recoil of the disc after its reversal along the
intruder's orbit. In particular, note the coherent {\it elliptic} plume
aligned with and encircling the disc ($V_z < 0$) emerging at $t=900$ Myr and
persisting for 100 Myr. The elongation in $V_z$ compared to $z$ 
(\autoref{f:binney}) reinforces
that this feature occurs at smaller radius ($R\approx 13-15$ kpc). There are
spiral plumes down to maybe $R\approx$ 10 kpc but insufficiently populated to
give good contrast. This is a failing of the current models. Much larger simulations with an order of magnitude more particles are already under way.

In \autoref{f:thor2}, for the intermediate-mass intruder (S3), we see the
kinematic spiral has lower amplitude in both axes. It grows out of the disc,
rotates as a fixed pattern, reaching its maximum amplitude in \zmax\
$\approx$ 2.5-3 kpc and \Vzmax\ $\approx 30-40$ km s$^{-1}$.  In
\autoref{f:thor2} (S4), the vertical phase plane is encoded with the
Galactocentric radius $R$: we see that the coherent spiral patterns occur at
$R\approx 15-20$ kpc.  (These features may also be associated with the
reversal of the disc's momentum parallel to the intruder's orbit.) We note
that \zmax\ and \Vzmax\ approximately match the predicted values in
\autoref{f:binney} at $R=20$ kpc; the model effectively ``calibrates'' the
expected amplitudes at other radii, in particular the anticipated values of
\zmax\ $\approx$ 1 kpc and \Vzmax\ $\approx 50$ km s$^{-1}$ at $R\approx
\Rsolar$ \citep{antoja2018}. We see the effects of the phase spiral at lower
radii down to 10 kpc but the contrast is poor.

In \autoref{f:thor1} (R4), the strongest phase spiral signal is further out
at $R\approx 20$ kpc. The outer disc of the Galaxy must be experiencing a
strong forced oscillation, with stars confined to well-defined corrugations
or wrinkles \citep{Schoenrich2018}. This is a strong prediction of our
simulations given the low-flying orbit of Sgr over the disc. It seems
reasonable to consider that this is the explanation for the TriAndromeda and
Monoceros ``rings'' discovered by the SDSS survey towards the outer stellar
disc \citep{Xu2015,Li2017} and maybe even most of the wave-like structure, ripples
and corrugations claimed to date \citep{Schoenrich2018,Bergemann2018}.

\subsection{Perturbed model: realistic orbit}

In movie M1 (\autoref{l:website}), we present our model for the low-mass Sgr
impact along a realistic orbit \citep[see][]{TepperGarcia2018}. Contemporary
models agree that Sgr initially crossed the disc along a trajectory
perpendicular to the Galactic plane \citep[e.g.][]{Law2005,Purcell2015}. But
at late times, as the orbit became circularised by dynamical friction
\citep[e.g.][]{Jiang2000}, the trajectory evolved to be more inclined to the
disc ($i\lesssim 30^\circ$), and therefore less impulsive. The last crossing
occurred at a radius of about $R\approx 13$ kpc.

All of the movies from Model M look very similar to the unperturbed case
(Model P), with the exception of one. When we compare model M3 to the stable
model P3 for the thin disc, there is clear evidence of heating in the former
due to the action of the low-mass perturber. Such dynamical heating can
happen over the disc where the forcing frequency is out of phase with the
intrinsic disc response. This heating also occurs in the high-mass and
intermediate-mass cases but this is obscured by the dramatic plumes arising
from the disc. 

In movies L and K (\autoref{l:website}), we present our model for the
intermediate and high mass impacts respectively along essentially the same
orbit.  In \autoref{f:thor3}, we show frames from movie K4 at two different
timesteps, 30 Myr before and 90 Myr after a disc transit. This is to
emphasise how clean the phase spiral signature is right before impact, and
how it is wiped out for up to $\approx 100-150$ Myr after the impact,
reforming thereafter.  

In \autoref{f:thor3}(e), there are three distinct phase spiral patterns
arising in three radial bins ($R=17, 15, 12$ kpc). We can calibrate the
strength of the signal in \citet{antoja2018} from our numerical simulations,
confirmed by the single crossing hyperbolic models. The low-mass intruder
(Model M) barely ruffles the disc. The high-mass intruder (Model K) produces
features with $\zmax \lesssim 5$ kpc and $\Vzmax \lesssim 50$ km s $^{-1}$,
comparable to the results for the high-mass hyperbolic case (Model R). The
intermediate-mass case scales down as it did for the hyperbolic models, thus
consistent with the amplitude of the Antoja spiral at $R=\Rsolar$. 

After impact, the spiral features arise again from the ashes and are
persistent until the next impact. The fact that we see a clear phase spiral
today is consistent with the passage of time since the last crossing about
400$-$500 Myr ago (by general consensus) and our imminent disc crossing in
$\gtrsim$50 Myr. This makes the spiral phenomenon no older than $\sim 0.5$ Gyr.

\newpage
\noindent \rule{8.5cm}{1pt}

\noindent
$^{1}$Sydney Institute for Astronomy, School of Physics, A28, The University of Sydney, NSW 2006, Australia\\
$^{2}$Center of Excellence for Astrophysics in Three Dimensions (ASTRO-3D), Australia\\
$^{3}$Miller Professor, Miller Institute, University of California Berkeley, CA 94720, USA\\
$^{4}$Rudolf Peierls Centre for Theoretical Physics, Clarendon Laboratory, Oxford, OX1 3PU, UK\\
$^{5}$Research School of Astronomy \& Astrophysics, Australian National University, ACT 2611, Australia\\
$^{6}$Department of Physics and Astronomy, Macquarie University, Sydney, NSW 2109, Australia\\
$^{7}$Max Planck Institute  for Astronomy (MPIA), Koenigstuhl 17, 69117 Heidelberg, Germany\\
$^{8}$Fellow of the International Max Planck Research School for Astronomy \& Cosmic Physics at the University of Heidelberg, Germany\\
$^{9}$Monash Centre for Astrophysics, Monash University, Australia \\
$^{10}$School of Physics and Astronomy, Monash University, Australia\\
$^{11}$INAF Astronomical Observatory of Padova, 36012 Asiago, Italy \\
$^{12}$Department of Physics and Astronomy, Uppsala University, Box 516, SE-751 20 Uppsala, Sweden\\
$^{13}$School of Physics, UNSW, Sydney, NSW 2052, Australia\\
$^{14}$Department of Astronomy, Columbia University, Pupin Physics Laboratories, New York, NY 10027, USA\\
$^{15}$Center for Computational Astrophysics, Flatiron Institute, 162 Fifth Avenue, New York, NY 10010, USA\\
$^{16}$Faculty of Mathematics and Physics, University of Ljubljana, Jadranska 19, 1000 Ljubljana, Slovenia\\
$^{17}$ICRAR, The Uni. of Western Australia, 35 Stirling Highway, Crawley, WA 6009, Australia \\
$^{18}$Department of Physics and Astronomy, University of Rochester, Rochester, NY 14627, USA \\
$^{19}$Institute for Advanced Study, Princeton, NJ 08540, USA \\
$^{20}$Department of Astrophysical Sciences, Princeton University, Princeton, NJ 08544, USA \\
$^{21}$Observatories of the Carnegie Institution of Washington, 813 Santa Barbara Street, Pasadena, CA 91101, USA \\
$^{22}$Department of Physics and Astronomy, The Johns Hopkins University, Baltimore, MD 21218, USA \\

\bsp	
\label{lastpage}
\end{document}